\documentclass[12pt]{article}
\usepackage[mathscr]{eucal}
\usepackage{bm}
\usepackage[usenames, dvipsnames]{color}
\usepackage{epsfig,amsfonts}
\usepackage[body={170mm,247mm}]{geometry}
\usepackage{epstopdf}
\usepackage{graphicx}
\usepackage[margin=15pt,font=small,labelfont=bf]{caption}
\usepackage{amsmath}
\usepackage{amsthm,amssymb}
\usepackage{graphicx}
\usepackage{hhline}
\usepackage{cite}

\usepackage{psfrag}
\usepackage{enumerate}
\usepackage{tikz}
\usepackage{upgreek}
\makeatletter
\@addtoreset{equation}{section}
\makeatother
\renewcommand{\theequation}{\thesection.\arabic{equation}}

\newcounter{app}
\newcounter{sapp}[app]
\def\theapp{\Alph{app}}
\newcommand{\app}[1]{
\refstepcounter{app}{\vspace{7mm}
\noindent\Large\bf Appendix
\theapp.
 \ #1 \par \vspace{5mm}}
\setcounter{equation}{0}
\def\theequation{\Alph{app}.\arabic{equation}}}

\def\gl{g_0}

\def\be{\begin{equation}}
\def\ee{\end{equation}}
\def\bdm{\begin{displaymath}}
\def\edm{\end{displaymath}}
\def\bea{\begin{eqnarray}}
\def\eea{\end{eqnarray}}

\def\s{\sigma}
\def\sig{a}

\def\ri{{\rm i}}
\def\Mba{{\mathcal M}}
\def\Mba{{M_0}}
\def\newp{{p}}
\def\mye{{\phi}}
\def\myenergy{{\mathsf e}}
\def\myp{{P}}

\def\XXint#1#2#3{{\setbox0=\hbox{$#1{#2#3}{\int}$}
    \vcenter{\hbox{$#2#3$}}\kern-.5\wd0}}

\def\hf{{\frac{1}{2}}}
\def\tshf{{\textstyle\frac{1}{2}}}
\def\thf{{\frac{3}{2}}}

\def\Wsf{{\tt W}}

\def\Asf{{\mathsf Q}}
\def\Qbs{{\mathsf W}}

\def\Zbs{{\mathsf Z}}
\def\Abl{{\Asf}^{\rm(BL)}}
\def\Tbs{{\mathsf T}}
\def\be{\begin{equation}}
\def\ee{\end{equation}}
\def\wh{\widehat}
\def\beq{\begin{equation}}
\def\eeq{\end{equation}}

\newcommand{\rd}{\mbox{d}}
\newcommand{\re}{\mbox{e}}

\def\ds{\displaystyle}
\newcommand{\JJ}{{\mathscr J}}
\newcommand{\EE}{{E}}
\newcommand{\PP}{{P}}
\def\nop{{\mathcal N}}
\newcommand{\NN}{{\mathcal N}}

\newcommand{\RR}{{S}}

\newcommand{\Ss}[2]{{\boldsymbol S}^{(#1)}_{#2}}
\renewcommand{\s}{\sigma}
\renewcommand{\sp}{\sigma^{\prime}}
\newcommand{\spp}{\sigma^{\prime\prime}}
\newcommand{\gla}{\lambda}

\newcommand{\z}{\bar{z}}
\def\massPsi{{\boldsymbol\Psi}}
\def\confPsi{\Upsilon}
\def\confpsi{\varphi}

\begin{document}
\begin{titlepage}
{}
\vglue .5cm
{}

\begin{center}
\begin{LARGE}
{\bf Bukhvostov-Lipatov model \\}
\end{LARGE}
\vspace{.7cm}
\begin{LARGE}
{\bf and quantum-classical duality}
\end{LARGE}

\vspace{2.3cm}
\begin{large}

{\bf Vladimir V. Bazhanov$^{1}$, Sergei  L. Lukyanov$^{2,3}$\\
\bigskip
and
\\
\bigskip
  Boris A. Runov$^{1,4}$}

\end{large}

\vspace{1.cm}
$^1$Department of Theoretical Physics,\\
         Research School of Physics and Engineering,\\
    Australian National University, Canberra, ACT 2601, Australia\\\ \\
${}^{2}$NHETC, Department of Physics and Astronomy,\\
     Rutgers University,\\
     Piscataway, NJ 08855-0849, USA\\\ \\
  ${}^{3}$Kharkevich Institute for Information Transmission Problems,\\
 Moscow, 127994 Russia\\
\vspace{.2cm}
and\\
\vspace{.2cm}
${}^{4}$Institute for Theoretical and Experimental Physics,\\
Moscow, 117218, Russia\\

\end{center}
\vspace{.5cm}
\begin{center}
\centerline{\bf Abstract} \vspace{.8cm}\small

\parbox{16cm} 
{ The Bukhvostov-Lipatov model is an exactly soluble
  model of two interacting Dirac fermions in 1+1 dimensions.  The
  model describes weakly interacting instantons and anti-instantons in
  the $O(3)$ non-linear sigma model. In our previous work
  [arXiv:1607.04839] we have proposed an exact formula for the vacuum
  energy of the Bukhvostov-Lipatov model in terms of special solutions
  of the classical sinh-Gordon equation, which can be viewed as an example of a remarkable duality between integrable quantum field theories and integrable classical field theories in two dimensions. 
Here we present a complete derivation
  of this duality based on the classical inverse scattering transform
  method, traditional Bethe ansatz techniques and analytic theory of
  ordinary differential equations. 
In particular, 
we show that the Bethe ansatz equations defining the vacuum state 
of the quantum theory also define connection coefficients 
of an auxiliary linear problem for the classical sinh-Gordon equation. 
Moreover, we also present details of the
  derivation of the non-linear integral equations determining the
  vacuum energy and other spectral characteristics of the model in the
  case when the vacuum state is filled by 2-string solutions of the
  Bethe ansatz equations.}
\end{center}
\vspace{.8cm}

\vfill

\end{titlepage}
\newpage
\addtocounter{page}{1}
\tableofcontents

\section{\label{intro}Introduction}

This paper is a sequel to our previous work
\cite{Bazhanov:2016glt} 
devoted to the study of the Bukhvostov-Lipatov (BL) 
model \cite{Bukhvostov:1980sn}. 
Originally, this model was introduced 
to describe weakly interacting
instantons and anti-instantons in the $O(3)$ non-linear sigma model in
two dimensions, extending the results of 
\cite{Polyakov:1975yp,Fateev:1979dc}.
It is an exactly soluble (1+1)-dimensional QFT, containing two
interacting Dirac fermions $\varPsi_a$\ ($a=\pm$), defined by the
renormalized Lagrangian
\bea\label{Lagr1}
{\cal L}=
\sum_{a=\pm }{\bar \varPsi}_a
\big(\ri \gamma^\mu\partial_\mu-M\big){ \varPsi}_a-
g\, \big({\bar \varPsi}_+\gamma^\mu{\, \varPsi}_+\big)
\big({\bar \varPsi}_- \gamma_\mu{ \varPsi}_-\big)\,+\,\mbox{counterterms}\,.
\eea
This model provides a remarkable illustration to the idea of the exact
instanton counting. Indeed, as explained in \cite{Bazhanov:2016glt},
the model can be reformulated as a bosonic QFT with two interacting
bosons, which upon an analytic continuation into a strong coupling
regime, becomes equivalent to the originating $O(3)$ non-linear sigma
model.

Our interest to the BL model is also motivated by the study of an
intriguing correspondence between Integrable Quantum Field Theories
(IQFT) and Integrable Classical Field Theories in two dimensions,
which cannot be expected from the standard correspondence principle.
Over the past two decades this topic has been continuously developed,
as can be seen from the works \cite{Vor94, DT99b,Bazhanov:1998wj,
  Suzuki:2000fc,Bazhanov:2001xm, Bazhanov:2003ni,Fioravanti:2004cz,
  Dorey:2006an,Feigin:2007mr, Lukyanov:2010rn,Dorey:2012bx,
  Lukyanov:2013wra,Bazhanov:2013cua,Masoero:2015lga,Ito:2015nla}.
The new correspondence yields an extremely efficient extension of the
Quantum Inverse Scattering Method (QISM) \cite{Faddeev:1979} by
connecting it with well-developed techniques of the spectral theory of
ordinary differential equations (ODE) and, most importantly, the
Classical Inverse Scattering Method \cite{FT87}.
The resulting approach is usually referred to as ``ODE/IM'' or
``ODE/IQFT'' correspondence.  Ultimately, for a massive IQFT, it
allows one to describe quantum stationary states (and the
corresponding eigenvalues of quantum integrals of motion) in terms of
special singular solutions of classical integrable partial
differential equations. To this moment, the approach has already been
applied to a number of IQFT models, such as the sine(h)-Gordon model
\cite{Lukyanov:2010rn}, the Bullough-Dodd model \cite{Dorey:2012bx}
and various regimes
\cite{Lukyanov:2013wra, Bazhanov:2013cua,Bazhanov:2014joa} of the
Fateev model \cite{Fateev:1996ea}, including the ``sausage model''
\cite{Fateev:1992tk} (a one-parameter deformation of the $O(3)$
sigma-model). Despite this progress, the subject, certainly, deserves
further studies for a better understanding of the mathematical
structure of the ODE/IQFT correspondence, as well as unraveling
fundamental underlying reasons of its existence.
In this work we continue to address these problems in the case of the
BL model. As noted before, this model is integrable. Its coordinate 
Bethe ansatz solution was presented in the original BL paper
\cite{Bukhvostov:1980sn}. The corresponding
factorized scattering theory was proposed in \cite{Fateev:1996ea}.
The non-linear integral equations  for calculating the ground state
energy were derived in \cite{Saleur:1998wa}.

As in \cite{Bazhanov:2016glt} we consider the theory \eqref{Lagr1} in
a finite volume imposing twisted (quasiperiodic) boundary conditions
on the fundamental fermion fields,
\bea\label{apssspps}
{\varPsi}_\pm(t,x+R)=-\re^{2\pi\ri k_\pm}\, {\varPsi}_\pm(t,x)\ ,\ \ \ \ 
{\bar \varPsi}_\pm(t,x+R)=-\re^{-2\pi\ri k_\pm}\, {\bar \varPsi}_\pm(t,x)\ .
\eea
The pair of real numbers $(k_+,k_-)$ labels different sectors of the
theory and, therefore, one can address the problem of computing of
vacuum energy $E_{\bf k}$ in each sector.
The perturbative treatment of \eqref{Lagr1} leads to ultraviolet (UV)
divergences. In \cite{Bazhanov:2016glt} we used the renormalization
procedure which preserves the integrability of the theory.  In this
case the fermion mass $M$ could only have a finite renormalization and
the only UV divergent quantity remaining in the theory is the bulk
vacuum energy, defined as ${\cal E}=\lim_{R\to\infty} E_{\bf k}/R$.
Therefore, it is convenient to extract the (divergent) extensive part
from $E_{\bf k}$ and introduce a scaling function
\bea\label{F-def}
{\mathfrak F}(r,{\bf k})=\frac{R}{\pi}\ ( E_{\bf k}-R\ {\cal E})\,,
\eea
which is simply related to the so-called effective central
charge,\ ${\mathfrak F}=-c_{\rm eff}/6$.  Notice that it is a
dimensionless function of the dimensionless variables $r\equiv MR$ and
${\bf k}$, satisfying the normalization condition\ $
\lim_{r\to+\infty}{\mathfrak F}(r,{\bf k})=0 $.

In \cite{Bazhanov:2016glt} the scaling function \eqref{F-def} was
computed in a variety of different ways. Let us briefly review this
work and then describe new considerations contained in this
paper. First, we used {\em the renormalized Matsubara perturbation
  theory\/} (recall that we consider the finite-volume theory) to
calculate the first two non-trivial orders of the expansion of the
scaling function \eqref{F-def} in powers of the coupling constant $g$.
The physical fermion mass $M$ is normalized by the large distance
asymptotics of the scaling function. As for the coupling constant
$g$, it is convenient to trade it off the for a physically observable
parameter $\delta$, which enters the $S$-matrix of the model.
According to \cite{Fateev:1996ea} the particle spectrum contains a
fundamental quadruplet of mass $M$ whose two-particle $S$-matrix is
given by the direct product $(-S_{a_1}\otimes S_{a_2})$ of two
$U(1)$-symmetric solutions of the $S$-matrix bootstrap, with
$a_1=2-a_2=1-\delta$. Each of the factors $S_a$ coincides with the
soliton $S$-matrix in the quantum sine-Gordon theory with the
renormalized coupling constant $a$.

Further, in \cite{Bazhanov:2016glt} we also used the {\em conformal
  perturbation theory\/} and computed the small-$r$ expansion of the
scaling function \eqref{F-def} to within the forth order terms
$O(r^4)$, inclusively. Then, applying rather non-trivial integral
identities for the hypergeometric function, we discovered that the
result of these perturbative calculations can be expressed through a
particular solution of the Liouville equation.
Guided by this peculiar observation and the previous results
\cite{Lukyanov:2013wra,Bazhanov:2013cua,Bazhanov:2014joa} concerning
different regimes of the Fateev model  we presented an exact
formula for the scaling function
\bea\label{F-sinh}
{\mathfrak F}(r,{\bf k})=-{\mathfrak f}_{\rm B}\big(2r 
\cos\big({\textstyle\frac{\pi\delta}{2}}\big)\big)
-{ \frac{8}{\pi}}\  \int_{{\mathbb D}_{\rm BL}}\rd^2 w\  \sinh^2({\hat\eta})+
\sum_{i=1}^2  a_i\, \big(|k_i|-{\textstyle\frac {1}{2}}\big)^2 \ , 
\eea
where $k_\pm=k_1\pm k_2$, in terms of a real valued solution ${\hat \eta}$ of the classical
sinh-Gordon equation
\bea\label{sinh-eq}
\partial_w\partial_{\bar w}{\hat \eta}-\re^{2{ \hat \eta}}+ \re^{-2{ \hat \eta}}=0
\eea
in the domain ${\mathbb D}_{\rm BL}$.
\begin{figure}
[!ht]
\centering
\psfrag{a}{$\pi a_1$}
\psfrag{b}{$\pi a_2$}
\psfrag{c}{$r/4$}
\psfrag{w1}{$w_1$}
\psfrag{w2}{$w_2$}
\psfrag{w}{$w$}
\includegraphics[width=4.5  cm]{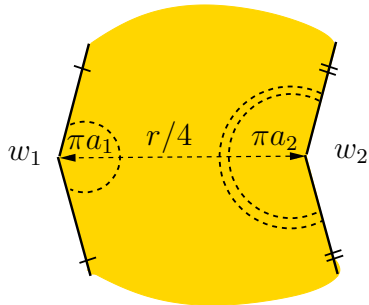}
\caption{Domain  ${\mathbb D}_{\rm BL}$ has the topology of 
a thrice-punctured 2D sphere. 
The horizontal width of  ${\mathbb D}_{\rm BL}$ is controlled by
a length of the segment $(w_1, w_2)$,  which coincides with
$r/4$. The boundaries of the strips in the upper and lower half
planes are identified, as marked.} 
\label{fig3sgsg}
\end{figure}
This domain is obtained
from two inclined half-infinite strips of the complex plane glued 
together along their infinite sides, as shown in Fig.\,\ref{fig3sgsg}.   
It has the topology of 
a two-dimensional sphere with three punctures $(w_1,w_2,\infty)$. 
At the singular
points $w=w_i\ (i=1,2)$ on the real axis 
the solution has the following  asymptotic behavior:
\bea\label{osasail}
{\hat\eta}=(2 |k_i|-1)\ \log|w-w_i|+O(1)\ \ \ \ \ \ \ \ {\rm as}\ \ \ \ \ w\to w_i\  
\eea
and
\bea\label{asym-z3}
{\hat \eta}\to 0\ \ \ \ \ {\rm as}\ \ \ \ |w|\to\infty\ .
\eea
The conical angles at the points $w_1$ and $w_2$ (see Fig.\,\ref{fig3sgsg})
are determined by the parameters 
$a_1=1-\delta$ and $a_2=1+\delta$, while the number $k_{1,2}$ are related to the twists parameters $k_\pm=k_1\pm k_2$ in \eqref{apssspps}. The quantity 
${\mathfrak f}_{\rm B}$ is the free energy of a free 1D boson, given
by 
\beq\label{free-boson}
{\mathfrak f}_{\rm B}(\beta)=\frac{\beta}{2\pi^2}\ 
\int_{-\infty}^\infty\rd\theta\ \cosh(\theta)\ \log\big(1-\re^{-\beta\cosh(\theta)}\big)\ ,
\eeq
where $\beta$ is the (dimensionless) inverse temperature.

In this paper we present a derivation of the formula \eqref{F-sinh}.
It is based on the fact that the sinh-Gordon equation is a
classical integrable equation, which can be treated by the inverse
scattering transform method. This allows one to relate the classical
conserved charges to the connection coefficients of ODE's arising  
in the auxiliary linear problem for the sinh-Gordon equation. 
It turns out that these
 connection coefficients are determined by the same Bethe ansatz
 equations that appear in the coordinate 
Bethe ansatz solution of the BL model. As a result 
eigenvalues and eigenstates of QFT integrals motion are determined by   
solutions of the classical sinh-Gordon equation.
In addition, using standard approaches of integrable QFT,  
we derive another expression for the scaling function in terms
of solutions of Non-Linear Integral Equations (NLIE) which are equivalent
to Bethe ansatz equations, but more efficient for numerical 
analysis.

The organization of the paper is as follows. The coordinate Bethe 
ansatz for the BL model is considered in Sect.\,\ref{CBA}. 
The lattice-type
regularization of the Bethe ansatz equations and associated 
particle-hole transformations are considered in
Sect.\,\ref{lat-type}. The functional
relations for the connection coefficients in the auxiliary linear problem   
for the modified sinh-Gordon equation are derived in Sect.\,\ref{MShG}. 
It is shown that these functional relations are equivalent to the Bethe ansatz
equation for the BL model.
The calculation of vacuum eigenvalues of local and
non-local integral of motion from the NLIE and their comparison against
small- and large-$R$ asymptotics derived in our previous paper
\cite{Bazhanov:2016glt}  
is given in Sect.\,\ref{secnew4}.
Appendix~\ref{app:A} contains details of the derivation of 
functional relations for the connection coefficients. Their properties
in the short distance (CFT) limit, are discussed in Appendix~\ref{app:B}.
Appendix~\ref{app:C} contains derivations of the NLIE,
and Appendix~\ref{app:D} contains
integral representations for connection coefficients.  
Appendix~\ref{app:E} 
contains details of small- and large-$R$ expansions of the vacuum
energy as well as eigenvalues of higher integrals of motion.

\section{\label{CBA} Bethe ansatz results}
In this section we review and extend the original results 
of \cite{Bukhvostov:1980sn}, concerning the 
coordinate Bethe ansatz for the BL model. 
\subsection{Coordinate Bethe ansatz}
Consider (1+1)-dimensional quantum field theory 
of two interacting Dirac fermions
$\Psi_\sig$ ($\sig=\pm$), defined by
the {\em bare\/} Lagrangian
\begin{equation}
\label{lag1}
	\mathcal{L}=\sum_{\sig=\pm}\bar{\Psi}_{\sig}
\big(\ri\gamma^{\mu}\partial_{\mu}
        -{\Mba}\big)
        \Psi_{\sig}-\gl\,\big
(\bar{\Psi}_{+}\gamma^{\mu}\Psi_{+}\big)\, \big(\bar{\Psi}_{-}
        \gamma_{\mu}\Psi_{-} \big)\, ,
\end{equation}
where $\partial_0=\partial/\partial_t$,
$\partial_1=\partial/\partial_x$ and  
$\bar{\Psi}_\sig=\Psi_\sig^\dagger\gamma_0$. 
The $\gamma$-matrices are defined as $\gamma_0=\sigma_1$,
$\gamma_1=\ri\,\sigma_2$, with $\sigma_1,\sigma_2,\sigma_3$ being the
usual Pauli matrices.
We will consider the theory in a finite volume, imposing twisted
(quasiperiodic) boundary conditions on the fermionic fields, 
\bea\label{twisted}
{\Psi}_\pm(t,x+R)=- \re^{2\pi\ri p_\pm}\, {\Psi}_\pm(t,x)\ ,\ \ \ \ 
{\bar \Psi}_\pm(t,x+R)=- \re^{-2\pi\ri p_\pm}\, {\bar \Psi}_\pm(t,x)\ .
\eea
The standard perturbation theory leads to ultraviolet
divergencies and, therefore, 
the Lagrangian \eqref{lag1} requires renormalization. Typically this
procedure is performed over the physical vacuum state of the
continuous theory. For instance, 
the renormalized Matsubara
perturbation theory (with the coordinate space cutoff)
for one- and two-loop diagrams was considered 
in our previous paper \cite{Bazhanov:2016glt}. However, 
the traditional Bethe ansatz approach is formulated in terms of
unrenormalized quantities and a bare (unphysical) vacuum state.
Therefore, in this section we will work with the unrenormalized
Lagrangian \eqref{lag1}. 
The connection of the bare fermion
mass ${\Mba}$, coupling constant $\gl$ and twist parameters $p_\pm$ 
to their 
renormalized physical counterparts will be given below (see
\eqref{rmatrix},\,\eqref{physmass},\,\eqref{asksaasu}\,\eqref{twist2}). 

We write the bispinors $\Psi_\sig(x)$ in the form 
\beq\label{bispinor}
	\Psi_{\sig}(x)=\left(\begin{array}{c}
	\psi_{a}^+(x)\\
	\psi_{a}^-(x)
	\end{array}\right)\,,
\eeq
where $\psi^+_\sig(x)$ and $\psi^-_\sig(x)$ stand for their components with
the Lorentz spin $+\hf$ and $-\hf$, respectively. 
The Hamiltonian corresponding to \eqref{lag1} reads
\begin{equation}
\label{hamilton}
	\widehat{ H}=\int \rd x\, \Big( \sum_{a=\pm}(-\ri\Psi^{\dag}_{a}\sigma_3
       \, \partial_x \Psi_{a}+{\Mba}\Psi_{a}^{\dag}\sigma_1\Psi_a) 
	+\gl\,(\Psi_{+}^{\dag}\Psi_{+})(\Psi_{-}^{\dag}\Psi_{-})
-\gl(\Psi_{+}^{\dag}\sigma_3\Psi_{+})(\Psi_{-}^{\dag}\sigma_3\Psi_{-})\Big). 
\end{equation}
Under the second quantization the quantities $\psi_{a}^{s\,\dag}(x)$
and $\psi_{a}^{s}(x)$
become creation and annihilation operators for 
a quasiparticle with the flavor $a$ and spin $s\,\hf$\   $(s=\pm)$ at
the point $x$. These operators satisfy the canonical anticommutation relations,
\beq\label{anticom}
\big\{\psi_{a_1}^{s_1}(x_1),\psi_{a_2}^{s_2\,\dagger}(x_2)\big\}=
\delta_{a_1a_2}\delta_{s_1s_2}\delta(x_1-x_2)\,,
\quad
\big\{\psi_{a_1}^{s_1}(x_1),\psi_{a_2}^{s_2}(x_2)\big\}=
\big\{\psi_{a_1}^{s_1\,\dagger}(x_1),\psi_{a_2}^{s_2\,\dagger}(x_2)\big\}=0\,.
\eeq
It is easy to check that the operators of the total number of
quasiparticles of each flavor 
\begin{equation}
\label{numop}
\widehat{\nop}_\pm=\int \rd x
\ \Psi_{\pm}^{\dag}\Psi_{\pm}^{\phantom{\dag}}\,,\qquad
[\,\widehat{\nop}_+,\,\widehat{\nop}_-\,]=
[\,\widehat{\nop}_\pm,\,\widehat{ H}\,]=0\,,
\end{equation}
commute among themselves and with the Hamiltonian
\eqref{hamilton}. 
Correspondingly, both these operators  $\widehat{\NN}_\pm$ 
are separately conserved quantities. 
In particular, this means that the stationary Schr\"odinger
equation $\widehat{ H}\,|\Phi\rangle 
=E\,|\Phi\rangle$ has solutions with a fixed total number
$\NN=\NN_++\NN_-$\ \  of quasiparticles, 
\begin{equation}
\label{wave}
	|\Phi\rangle=\sum_{a_1,{\ldots},a_{{\NN}}}
\sum_{s_1,{\ldots},s_{{\NN}}}
         \int\prod_{i=1}^{{\NN}}\rd x_i\ 
        \chi_{a_1a_2\ldots a_{\NN}}^{s_1s_2\ldots s_{\NN}}
            (x_1,x_2,\ldots, x_{{\NN}})\ 
        \psi^{s_1\,\dag}_{a_1}(x_1)\,{\cdots}\,
            \psi^{s_\NN\,\dag}_{a_{{\NN}}}(x_{{\NN}})\ |0\rangle\,. 
\end{equation}
Here $|0\rangle$ stands for the bare vacuum state, i.e., 
the state with no quasiparticles (which should not be confused with the 
physical vacuum state, obtained after filling the Dirac sea of
negative energy states). The variables $\{a_i\}$ and $\{s_i\}$, taking
values $a_i=\pm\,, s_i=\pm$, label
the flavors and the spins of created quasiparticles.
Substituting \eqref{wave} into the Schr${\rm \ddot o}$dinger equation and using the 
commutation relations \eqref{anticom}, one obtains a partial
differential equation 
\begin{equation}
\label{SE} 
	\left(\sum_{i=1}^{{\NN}}\Big(-\ri\sigma^{(i)}_{3}\frac{\partial}
{\partial x_i}+{\Mba}\sigma^{(i)}_1\Big)+\frac{\gl}{2}\,
        \sum_{j=1}^{{\NN}}\sum_{i<j}\delta(x_i-x_j)\big(1-\sigma^{(i)}_3
        \sigma^{(j)}_3\big)\big(1-\tau^{(i)}_3\tau^{(j)}_3\big)\right)
\boldsymbol{\chi}=E\,\boldsymbol{\chi} 
\end{equation}
where the ``wave function'' $\boldsymbol{\chi}$ is understood as a vector
in ${\mathbb C}^{2\NN}$ with the components given by the coefficients 
$\chi_{a_1,{\ldots},a_{{\NN}}}^{s_1,{\ldots},
s_{{\NN}}}(x_1,{\ldots},x_{{\NN}})$.
The quantities $\tau^{(i)}_k$ and
$\sigma^{(i)}_k$ \ \  $(k=1,2,3,\ i=1,\ldots,\NN)$\  denote the Pauli
matrices, acting, respectively, on the flavor and
spin indices of the $i$-th
quasiparticle. Thus, 
the diagonalization of the Hamiltonian \eqref{hamilton} is reduced
to a quantum many-body problem \eqref{SE} for ${\NN}$
fermions with a  $\delta$-function potential.  
For ${\NN}=1$ this leads to the free Dirac equations with the solution 
\begin{equation}
\label{onepart}
u_{\theta}(x,s)=\ri\, s\, \re^{\ri x\,{\Mba}\,\sinh(\theta)-\ri s\,
  \theta/2}
\end{equation}
describing a single quasiparticle with the energy $\varepsilon$ and momentum
$k$, where 
\begin{equation}
	\varepsilon=-{\Mba}\cosh (\theta)\,,\qquad k={\Mba}\sinh( \theta)\,, 
\qquad \varepsilon^2-k^2={\Mba}^2\,.
\end{equation}
The rapidity variable $\theta$ is a free parameter, which can take
complex values (e.g., for the positive energy states one should
substitute $\theta\to \ri\pi-\theta$).  

For general values of $\NN>1$, 
the problem \eqref{SE} can be solved by the Bethe ansatz. First, note that if
no coordinates $\{x_i\}$ pairwise coincide the interaction potential
vanishes and the wave function becomes a linear combination of
products of the free quasiparticle solutions \eqref{onepart}. 
Using the antisymmetry of the wave function 
\beq\label{asym}
\chi_{a_1{\ldots}a_i a_{i+1}\ldots a_{{\NN}}}^{s_1 {\ldots}
s_i s_{i+1} \ldots
s_{{\NN}}}(x_1,{\ldots},x_i,x_{i+1},\ldots,x_{{\NN}})=
-\chi_{a_1{\ldots} a_{i+1} a_{i} \ldots a_{{\NN}}}^{s_1 {\ldots}
s_{i+1} s_{i} \ldots,
s_{{\NN}}}(x_1,{\ldots},x_{i+1},x_{i},\ldots,x_{{\NN}})
\eeq
one can reduce the considerations to the case
$x_1<x_2<\cdots<x_\NN$. Then we use the ansatz 
\beq 
\chi_{a_1{\ldots}a_{{\NN}}}^{s_1{\ldots}
s_{{\NN}}}(x_1,\ldots,x_{{\NN}})\Big\vert_{x_1<x_2<\cdots<x_\NN}=
\sum_Q A^{q_1q_2\ldots q_\NN}_{a_1a_2\ldots a_\NN}\,
u_{\theta_{q_1}}(x_1,s_1)\,
u_{\theta_{q_2}}(x_2,s_2)\cdots
u_{\theta_{q_\NN}}(x_\NN,s_\NN)\label{ansatz}
\eeq
where the sum is taken over all permutations
$Q=\{q_1,q_2,\ldots,q_\NN\}$ of $\{1,2,\ldots,\NN\}$.  The last two formulae
define a solution of \eqref{SE} with the total energy and momentum
\beq\label{E-total}
\EE=-{\Mba}\,\sum_{i=1}^\NN \cosh\theta_i\,,\qquad \PP=\Mba\,\sum_{i=1}^\NN \sinh\theta_i\,,
\eeq
which is valid everywhere, except at the hyperplanes $x_i=x_j$, separating  
regions with different orderings of the coordinates.
The continuity of the wave function at these boundaries imposes
multiple linear relations between   
the coefficients $A^{q_1q_2\ldots q_\NN}_{a_1a_2\ldots a_\NN}$.
Consider, for instance, the
two-particle case $\NN=2$. Integrating \eqref{SE} over $x_2$ from
$x_1-\epsilon$ to $x_1+\epsilon$, with $\epsilon\to0$,  and using
\eqref{asym}, \eqref{ansatz},   one obtains exactly four linear relations 
\beq  
\label{smat}
	A^{12}_{a_1 a_2}=\sum_{a_1',a_2'}
\RR^{a'_1 a'_2}_{a_1a_2}(\theta_{1}-\theta_2) A^{21}_{a'_2a'_1}\ ,
\end{equation}
where the quasiparticle scattering $S$-matrix  is defined as 
\beq
\begin{array}{cc}
\ds	
\RR^{+-}_{+-}(\theta)=\RR^{-+}_{-+}(\theta)=-\frac{\sinh (\theta)}{\sinh
  (\theta-i\pi\delta)}\,,&\qquad\ds 
\RR^{-+}_{+-}(\theta)=\RR^{+-}_{-+}(\theta)=\frac{\sinh(\ri\pi\delta)}{\sinh(\theta-\ri\pi\delta)}\,,\\[.6cm]
\ds \RR^{++}_{++}(\theta)=\RR^{--}_{--}(\theta)=-1\,,
&\ds\qquad \delta=\frac{2}{\pi}\arctan(\gl)\,.
\end{array}\label{rmatrix}
\end{equation}
The key to the solvability of the problem \eqref{SE} is that this
matrix literally coincides with 
the $R$-matrix 
of the 6-vertex model, satisfying the
Yang-Baxter equation \cite{Baxter:book:1982}
\begin{equation}
\label{YBE}
	\sum_{a',b',c'}
\RR^{a'b'}_{a^{\phantom{'}}b^{\phantom{'}}}(\theta_{12})\,
\RR^{a''c'}_{a^{'\phantom{'}}c^{\phantom{'}}}(\theta_{23})\, 
\RR^{b''c''}_{b^{'\phantom{'}}c^{'\phantom{'}}}(\theta_{13})
=\sum_{a',b',c'}
\RR^{b'c'}_{b^{\phantom{'}}c^{\phantom{'}}}(\theta_{23})\,
\RR^{a'c''}_{a^{\phantom{'}}c^{'\phantom{'}}}(\theta_{13})\,
\RR^{a''b''}_{a^{'\phantom{'}}b^{'\phantom{'}}}(\theta_{12})
\end{equation}
and the inversion relation
\beq\label{IR}
\sum_{b_1,b_2}\RR_{a_1a_2}^{b_1b_2}(\theta_{12})\,\RR_{b_1b_2}^{a'_1a'_2}(\theta_{21})=
\delta_{a^{\phantom{'}}_1a'_1}\,\delta_{a^{\phantom{'}}_2a'_2}\, ,
\eeq
where $\theta_{ij}=\theta_i-\theta_j$. Note also the ``flavor
conservation''  property,
\beq\label{arrows}
\RR_{a_1a_2}^{b_1b_2}(\theta)\equiv 0\,,\qquad\mbox{if} \qquad 
a_1+a_2\not=b_1+b_2
\eeq
and a special point where
the $S$-matrix becomes proportional to the permutation matrix 
\beq\label{init}
\RR_{a_1a_2}^{b_1b_2}(0)=-\delta_{a_1b_2}\delta_{a_2b_1}\,.
\eeq
Eq.\eqref{smat} can be readily generalized for general values of\/ $\NN$. 
Let $Q=\{q_1,\ldots ,q_j,q_{j+1},\ldots , q_\NN\}$ 
and $Q'=\{q_1,\ldots ,q_{j+1},q_{j},\ldots , q_\NN\}$ be any two permutations,
differing from each other by the order of just two elements $q_j$ and
$q_{j+1}$. Then, repeating the above arguments for the case
$x_{q_j}=x_{q_{j+1}}$, one obtains an overdetermined system of
homogeneous linear equations
\beq\label{linrel}
{A}^{
{q_1\,\ldots \,q_jq_{j+1}\ldots  q_\NN}_{\phantom{|}}}_{{a_1\ldots
    \,a_ja_{j+1}\ldots a_\NN}^{\phantom{|}}}= \sum_{a'_{j}a'_{j+1}}
{\RR}^{{a'_j a'_{j+1}}_{\phantom{|}}}_{{a_ja_{j+1}}^{\phantom{|}}}
(\theta_{q_j}-\theta_{q_{j+1}})  \,
{A}^{{q_1\ldots \, q_{j+1}q_{j}\ldots  q_\NN}_{\phantom{|}}}_{{a_1\ldots\,
  a'_{j+1}a'_{j}\ldots a_\NN}^{\phantom{|}}}\,.
\eeq
Thanks to the Yang-Baxter \eqref{YBE} and inversion \eqref{IR}
relations this system is {\em consistent\/} and has non-zero solutions for 
the coefficients
$A_{a_1\,\ldots\,a_\NN}^{q_1,\,\ldots\,q_\NN}$. Namely, all of them
can be linearly expressed through their subset corresponding to one
particular permutation $Q$, e.g.,  
through the set
\beq\label{phi-vec}
\phi_{a_1\,\ldots\,a_\NN}=A_{a_1\ldots\,a_\NN}^{1\,\ldots\,\,\/\NN}\, \ \ \ \qquad (a_i=\pm)
\eeq
 to which the
system \eqref{linrel} imposes no restrictions.\footnote{
Indeed, by virtue of \eqref{YBE} and \eqref{IR} the relations \eqref{linrel} 
realize a 
representation of the permutation group on $\NN$ elements.}
At this point it is worth remembering that the numbers of
quasiparticles of each flavor are separately conserved. This means
that the vector \eqref{phi-vec} must be an eigenvector of the operator
\beq\label{nplus}
\widehat{\NN}_+ =\tshf \sum_{i=1}^\NN \,(\tau_3^{(i)}+1)\,,\qquad
\widehat{\NN}_+ \, \phi=\NN_+\,\phi\,\qquad \ (\,0\le \NN_+\le \NN\,)\, ,
\eeq
where, similarly to \eqref{SE}, 
the Pauli matrix $\tau_3^{(i)}$ acts only on the
flavor of the $i$-th quasiparticle.

\subsection{Bethe ansatz equations}
To find the remaining undetermined coefficients \eqref{phi-vec} and 
the rapidities $\{\theta_i\}$ one needs to use the periodicity 
of the wave functions
\beq\label{period}
\chi_{a_1{\ldots}a_j\ldots a_{{\NN}}}^{s_1 {\ldots}
s_j \ldots
s_{{\NN}}}(x_1,{\ldots},x_j,\ldots,x_{{\NN}})=-\re^{-2\pi \ri p_{a_j}}\,
\chi_{a_1{\ldots} a_{j}\ldots a_{{\NN}}}^{s_1 {\ldots}
s_{j} \ldots,
s_{{\NN}}}(x_1,{\ldots},x_{j}+R,\ldots,x_{{\NN}}) \qquad  (\,\forall j\,)
\eeq
implied by \eqref{twisted} and \eqref{wave}. Note that for 
this relation one assumes the following ordering  
$x_j<x_1<\cdots<x_{j-1}<x_{j+1}<\cdots<x_\NN$ of the coordinates on the circle.
Then using \eqref{ansatz} and
\eqref{linrel} one obtains 
\begin{equation}
\label{A-period}
	A^{{\,j\,\,1\,\,{\ldots}\,\backslash\kern -.30em j 
\,\ldots\,\,{\NN}}_{\phantom{|}}}_{{a_j a_1 {\ldots}
\backslash\kern -.33em a_j\ldots
 \, a_{{\NN}}}^{\phantom{|}}} 
=(-)^\NN\, \re^{-2\pi \ri p_{a_j}}\, \re^{\ri{\Mba}R\sinh\theta_j}\,
A^{{\,1\,\,{\ldots}\,\,\backslash\kern -.30em j 
\,\ldots\,\,{\NN}\,\,j}_{\phantom{|}}}_{{a_1 {\ldots}
\backslash\kern -.33em a_j\ldots
 \, a_{{\NN}}\, a_j}^{\phantom{|}}} \ ,
\end{equation}
where the crossed symbol $\backslash\kern -.33em j$ 
means that the corresponding entry is
omitted.
With the help of \eqref{linrel} and \eqref{init}
the last relation can be viewed 
as an eigenvalue
problem for a set of operators 
\beq\label{tj-def}
\widehat{T}(\theta_j)\,\phi=-\re^{-2\pi \ri p_{a_j}}\,
\re^{\ri{\Mba}R\sinh\theta_j}\,\phi\,\ \ \qquad
(j=1,\ldots,\,\NN)
\eeq
which all have the same eigenvector \eqref{phi-vec}.  
Here $\widehat{T}(\theta)$ stands for the transfer matrix of an inhomogeneous
six-vertex model with a (horizontal) field, defined as  
\begin{equation}
\begin{split}
\Vert\,\widehat{T}(\theta)\,
\Vert_{a_1 a_2\ldots a_\NN}^{a'_1 a'_2\ldots a'_\NN}=(-1)^\NN\,
\sum_{b_1,b_2,\ldots, b_\NN}\re^{2\pi\ri p_{b_1}} 
\RR_{b_1a_1}^{b_2a'_1}(\theta-\theta_1)
\RR_{b_2a_2}^{b_3a'_2}(\theta-\theta_2)\cdots
\RR_{b_\NN a_\NN}^{b_1 a'_\NN}(\theta-\theta_\NN)\,.
\end{split}
\label{tmat}
\end{equation}
As a consequence of the Yang-Baxter equation \eqref{YBE} the transfer matrices
with different values of $\theta$ commute among themselves 
$[\widehat{T}(\theta),\widehat{T}(\theta')]=0$. Moreover, due to
\eqref{arrows} 
they also
commute with the operator $\widehat{\NN}_+$, defined by \eqref{nplus}. Thus, 
all operators 
$\widehat{T}(\theta_j)$ in \eqref{tj-def} and the operator $\widehat{\NN}_+$ 
can be simultaneously diagonalized. 

We can now use celebrated results of Lieb \cite{Lieb66}
and Baxter \cite{Baxter:1971sam,Baxterbook} for the lattice 6-vertex model. 
The
eigenvalues of $\widehat{T}(\theta)$ in the sector with fixed $\NN_+$
read\footnote{Note, that 
there is an alternative (but equivalent) expression for
  $T(\theta)$ obtained from \eqref{T-eigen} by the replacement
  $p_\pm\to  p_\mp$ and $\NN_+ \to \NN_-=\NN-\NN_+$.}  
\begin{equation}
\label{T-eigen}
	T(\theta)=e^{2\pi\ri\,p_-}\prod_{k=1}^{\NN_+} 
\frac{\sinh(\theta-u_k+\frac{\ri\pi}{2}\delta)}
{\sinh(\theta-u_k-\frac{\ri\pi}{2}\delta)}
+e^{2\pi
  \ri\,p_+}
\prod_{j=1}^{{\NN}}\frac{\sinh(\theta-\theta_j)}{\sinh(\theta-\theta_j
-\ri\pi\delta)}
\prod_{k=1}^{\NN_+}\frac{\sinh(\theta-u_k-\frac{3\ri
\pi}{2}\delta)}{\sinh(\theta-u_k-\frac{i\pi}{2}\delta)}\,.
\end{equation}
This formula contains $\NN_+$ new (complex) parameters $\{u_\ell\}$,  
which are rapidities of
auxiliary ``magnons''. The latter are quasiparticles of yet another
kind, required for the diagonalization of the transfer matrix. The
new parameters $\{u_\ell\}$ are 
determined by the following Bethe Ansatz Equations (BAE)
\bea
-1&=&\ds\re^{-4\pi\ri p_1}\,\prod_{\ell'}
\frac{\sinh\big(u_\ell-u_{\ell'}+\ri \pi \delta\big)}
{\sinh\big(u_\ell-u_{\ell'}-\ri \pi \delta\big)}\ 
\prod_\JJ\  \frac{\sinh\big(u_\ell-\theta_\JJ-\frac{\ri \pi}{2} \delta\big)}
{\sinh\big(u_\ell-\theta_\JJ+\frac{\ri\pi}{2} \delta\big)}\,,\label{bae2}
\eea
where the indices $\ell,\ell'$ take $\NN_+$
different values,
whereas the index $\JJ$ take
$\NN$ different values. The parameters 
$p_1, p_2$ are defined as
\bea\label{asksaasu}
p_\pm=p_2\pm p_1\ .
\eea
Different solutions of \eqref{bae2} correspond to different eigenvalues of the transfer matrix \eqref{tmat}.

Next, in order to satisfy the boundary conditions (\ref{period}) the 
rapidities $\theta_j$ should satisfy (\ref{tj-def}). Together 
with \eqref{T-eigen} these conditions take the form
\bea
-1&=&\ds\re^{2\pi\ri(p_1-p_2)}\,
\re^{\ri \Mba R\sinh
  \theta_\JJ}\ \prod_\ell\  
\frac{\sinh\big(\theta_\JJ-u_\ell-\frac{\ri \pi}{2} \delta\big)} 
{\sinh\big(\theta_\JJ-u_\ell+\frac{\ri \pi}{2} \delta\big)}\,.
\label{bae1}
\eea
Thus, we have have obtained the system of coupled BAE
 \eqref{bae2} and \eqref{bae1} for the set of rapidities $\{u_\ell \}$
and $\{\theta_\JJ\}$.  Once they are solved, the energy and momentum of
the state can be calculated from \eqref{E-total}. To complete 
the construction the eigenvectors \eqref{ansatz} one needs to find the
corresponding eigenvector $\phi$ of the transfer matrix \eqref{tmat}
 of the six-vertex model. This part of analysis is well
 known  and can be found in \cite{Lieb66,
   Baxter:1971sam,TF79,Baxterbook}, as well as in the original BL paper
 \cite{Bukhvostov:1980sn}. We do not reproduce these results here
 since explicit expressions for the eigenvectors are not used in
 this paper.

It is worth noting again that the parameter ${\Mba}$ used
in this section is the {\em bare mass} parameter. Its relationship
with the physical fermion mass $M$ in \eqref{Lagr1} follows from the
requirement that the scaling function \eqref{F-def}, calculated from
BAE, at large distances should
decay as $\propto\exp(-M R)$. 
As shown in \cite{Bazhanov:2016glt} this is achieved if one
sets
\beq\label{physmass}
{\Mba}=M\cos\big({\textstyle\frac{\pi\delta}{2}}\big)\,.
\eeq
In what follows this relation will always be assumed. 

\subsection{The vacuum state}
It is useful to rewrite the above equations \eqref{bae2}, \eqref{bae1}
in the logarithmic form
\begin{subequations}\label{baelog}
\bea
m_\JJ&=&{\hf}+p_1-p_2+\frac{1}
{2\pi}\,r\cos(\pi\delta/2)\sinh(\theta_\JJ)\ + \sum_\ell\  {\mye}_{2\delta}(\theta_\JJ-u_\ell)
\label{baelog1}
\\[.8cm] 
{\overline m}_\ell&=&{\hf}-2p_1 -\sum_{\ell'}
{\mye}_{4\delta}(u_\ell-u_{\ell'})
+\sum_\JJ{\mye}_{2\delta}(u_\ell-\theta_\JJ)\,,\label{baelog2}
\eea
\end{subequations}
where $r=MR$, 
\beq\label{e-def}
{\mye}_\alpha(\theta)=\frac{1}{2\pi\ri}\log\left[\frac{\sinh
\big(\frac{\ri\pi}{4}
\, \alpha-\theta\big)}
{\sinh\big(\frac{\ri\pi}{4}\, \alpha+\theta\big)}\right]
\eeq
with the phase of the logarithm is chosen such that ${\mye}_\alpha(0)=0$.
The integer phases $\{m_\JJ\}$ 
and $\{{\overline m}_\ell\}$ play the r$\rm {\hat o}$le of quantum numbers, which uniquely 
characterize solutions of the BAE. 
Different solutions define different eigenstates
of the Hamiltonian. The energy of the corresponding state reads
\beq
E=-M\cos(\pi\delta/2)
\,\sum_\JJ \cosh({\theta_\JJ})\,. \label{e-lip}
\eeq
For the vacuum state the location of zeroes and the associated phases 
in \eqref{baelog} were analyzed in 
\cite{Bukhvostov:1980sn} and \cite{Bazhanov:2016glt}. First note that the
vacuum is flavor neutral. Thus, we set $\NN_+=\NN_-$ and
$\NN=2\NN_+$. We denote $\NN_+=N$, assuming it is an even number.  
To make our notations identical to those of 
\cite{Bazhanov:2016glt} we also need to shift the numbering of roots, 
assuming that the indices $\JJ$ and $\ell$ run over the values
\begin{equation}\label{set2}
	\ell \in \textstyle 
\{-\frac{N}{2}+1,-\frac{N}{2}+2,{\ldots},\frac{N}{2}\}\, ,\qquad
	\JJ \in \{-N+1,-N+2,{\ldots},N\}\, ,
\end{equation}
so that there $3N$ unknown roots altogether.  According to 
Refs.\!\cite{Bukhvostov:1980sn, Bazhanov:2016glt}
for $\delta>0$ the vacuum
roots $\{u_\ell\}$ and $\{\theta_\JJ\}$ are real and the integer
phases in \eqref{baelog} read
\beq\label{phases}
m_\JJ=\JJ\,,\qquad {\overline m}_\ell=\ell\,,\qquad \delta>0\,.
\eeq
For the relativistic QFT \eqref{lag1} 
the number of quasiparticles filling the bare vacuum state should be 
infinite. Therefore, the number $N$ plays the r${\hat{\rm o}}$le of the
ultraviolet cutoff. Standard estimates
\cite{Bukhvostov:1980sn} show
that for large $N$ the roots behave as 
\beq
r\cos(\pi\delta/2)\,\re^{\theta_\JJ}\sim2\pi\,\JJ+O(1)\,,\qquad
r\,\re^{u_\ell}\sim\textstyle
4\pi\, \ell+O(1)\,,\qquad 1\ll|\JJ|,|\ell|\ll N\,.
\eeq
This indicates that the 
products in \eqref{bae2} and \eqref{bae1} require regularization for
$N \to \infty$. Moreover, the total energy \eqref{E-total} in this limit 
diverges quadratically 
and should be renormalized by subtracting its extensive
part, as in \eqref{F-def}. The most convenient way to do this is to
use a ``lattice-type'' regularization, considered below.

\section{Lattice-type regularization and particle-hole duality\label{lat-type}}

\subsection{\label{sec:latreg}Lattice-type regularization} 
Here we consider a lattice-type regularization of
the above BAE \eqref{bae2},\,\eqref{bae1}.    
It is achieved by replacing the relativistic phase term $r\cos(\pi\delta/2)
\sinh(\theta)$ in \eqref{bae1} by a suitable lattice-type
expression
\beq
r \cos(\pi\delta/2)\,\sinh(\theta)\to N\,
{\myp}(\theta,\delta)\,,\qquad
{{\myp}(\theta,\delta)}=
{2\pi}\,{\mye}_{(1-\delta)}\big({\textstyle\hf}(\theta+\Theta)\big)
\textstyle+{2\pi}\,
{\mye}_{(1-\delta)}\big(\hf(\theta-\Theta)\big)\, ,\label{p-def}
\eeq
where ${\mye}_\alpha(\theta)$ is defined in \eqref{e-def}. 
Eq.\eqref{bae1} then becomes 
\beq\label{baelat1}
\left[\frac{{\mathrm s}\big(\theta_\JJ+\frac{\ri \pi}{2}(1-\delta)\big)}
{{\mathrm s}\big(\theta_\JJ-\frac{\ri \pi}{2}
  (1-\delta)\big)}\right]^N 
=-\ds\re^{2\pi\ri(p_1-p_2)}\,
\prod_{\ell}\  
\frac{\sinh\big(\theta_\JJ-u_\ell -\frac{\ri\pi}{2} \delta\big)} 
{\sinh\big(\theta_\JJ-u_\ell+\frac{\ri\pi}{2} \delta\big)}\,,
\eeq
where
\beq\label{phi-def}
{\mathrm s}(\theta)=\sinh\left({\textstyle\hf}(\theta+\Theta)\right)
\sinh\left({\textstyle\hf}(\theta-\Theta)\right)\,,
\eeq
or, in the logarithmic form,
\beq
m_\JJ={\hf}+p_1-p_2+
\frac{N\,\myp(\theta_\JJ,\delta)}{2\pi}\ + \sum_\ell\  {\mye}_{2\delta}(\theta_\JJ-u_\ell)\, .
\label{baelog3}
\eeq
Eqs.\eqref{bae2} and \eqref{baelog2} remain unchanged. 
The indices $\ell$ and $\JJ$
run over the same sets of integers as in \eqref{set2}.
The resulting equations look 
like typical BAE for some 1+1 dimensional solvable lattice model, where
the parameter $N$ stands for a half of the number of sites for
a periodic spin chain; the real valued parameter $\Theta$ 
controls the column inhomogeneity of Boltzmann weights,
whereas $p_1,p_2$ define twist parameters for
quasiperiodic boundary conditions. 
The energy $E_N$ of the corresponding 
lattice state can be computed by the formula 
\beq\label{energy}
E_N(\Theta,\delta) =2\,\sum_\JJ \myenergy(\theta_\JJ,\delta)\,,\qquad
\myenergy(\theta,\delta)={\mye}_{(1-\delta)}\big({\textstyle\hf}(\theta+\Theta)\big)
\textstyle-{\mye}_{(1-\delta)}\big(\hf(\theta-\Theta)\big)\,,
\eeq
which is quite natural in the context of the light-cone lattice
regularization of integrable QFT models 
(see \cite{Destri:1994bv} for the case of the
sine-Gordon model).  
The original BAE \eqref{bae1}, corresponding to the continuous model
\eqref{Lagr1}, are recovered in the limit when 
both $N$ and $\Theta$ tend to
infinity, while the scaling parameter  
\beq
r=4N\, \re^{-\Theta}
\label{qft-lim}
\eeq
is kept fixed.
 The lattice energy expression \eqref{energy} then
formally reduces to \eqref{e-lip},
\beq\label{E-relation}
N E_N(\Theta,\delta)=\frac{R E}{\pi}+2
N^2\,\big(1+\delta+O(\re^{-2\Theta})\big)\,,
\eeq
up to a diverging additive constant. 
Note that our lattice-type regularization
\eqref{p-def},\,\eqref{energy} is different from that suggested in
\cite{Saleur:1998wa}.

It is not difficult to check that the new BAE possess qualitatively the same
patterns of vacuum roots as described in the previous subsection.
First consider the case $\delta>0$. As before, repeating the arguments of
\cite{Bukhvostov:1980sn, Bazhanov:2016glt}, 
one concludes that all the roots
$\theta_\JJ$ and $u_\ell$ for the vacuum state are real and the
integer 
phases have the same assignment \eqref{phases}. The BAE
then imply that these 
roots form two ordered sets 
\beq\label{vacsol}
\textstyle\theta_{-2N+\hf}<\theta_{-2N+\thf}<\cdots<\theta_{2N-\hf}\, ,\qquad
u_{-N+\hf}<u_{-N+\thf}<\cdots<u_{N-\hf}\,.
\eeq
For large $N$ and fixed $\Theta$ their distribution,
\beq
\rho_u^{(N)}\big(u_{n+\hf}\big)=
\frac{1}{N(u_{n+1}-u_n)}\, \qquad \big(u_{n+1/2}\equiv 
{\textstyle\hf}(u_{n+1}+u_n)\,\big)\,,\label{rho-lat}
\eeq
is well approximated by the continuous density,
which is determined by the standard lattice model methods
\cite{Baxter:book:1982}, 
\beq
\rho_u(u)=\frac{1}{2\pi}\left(\frac{1}{\cosh(u+\Theta)}+
\frac{1}{\cosh(u-\Theta)}\right)\,.\label{dens1}
\eeq
Similarly for the roots $\theta_\JJ$ one has
\beq
\rho_\theta(\theta)=2\,\Re e
\Big(\rho_u\big(\theta+\tshf\ri\pi\delta\big)\Big)\,.
\label{dens2}
\eeq
For large $N$ and $\Theta$ the roots split into two cluster centered 
at $\pm\Theta$. Their asymptotic positions there can be 
estimated with the formulae 
\bea
N\, {\mathsf
  z}_\theta(\theta_\JJ)&\approx&\JJ-\hf+\frac{p_1}{a_1}+\frac{p_2}{a_2}
\qquad (1\ll |\JJ|\ll N)\nonumber\\[.2cm] 
N\, {\mathsf
  z}_u(u_\ell)&\approx&\ell-\hf+\frac{2p_2}{a_2}\qquad 
\qquad (1\ll 2|\ell|\ll N)\,,
\label{assz}
\eea
where the functions 
${\mathsf z}_\theta(\theta)$ and ${\mathsf z}_u(u)$ are given by 
\begin{subequations}\label{zcf}
\bea\label{zcf1}
{\mathsf z}_\theta(\theta)&=&\int_0^\theta
\rho_\theta(\theta')\,\rd\theta'=\frac{2}{\pi}\,
\int_{0}^{\infty}\frac{\rd \nu}{\nu} \ \sin({\nu\theta})\ 
\frac{\cos(\nu\Theta)\cosh\big(\frac{\pi\nu\delta}{2}\big)}
{\cosh\big(\frac{\pi\nu}{2}\big)}
\\[.4cm]
{\mathsf z}_u(u)&=&\int_0^u
\rho_u(u')\,\rd u'=\frac{1}{\pi}\,
\int_{0}^{\infty}\frac{\rd \nu}{\nu}\ \sin(\nu u)\ 
\frac{\cos(\nu\Theta)}
{\cosh\big(\frac{\pi\nu}{2}\big)} \,.\label{zcf2}
\eea
\end{subequations}
For large $N$ and fixed $\Theta$ the energy \eqref{energy} behaves as 
\beq
E^{(1)}_N(\Theta,\delta)=2 \sum_\JJ \myenergy(\theta_\JJ,\delta)=
N \varepsilon_\infty^{(1)}
+O(N^{-1})\ \qquad (\delta>0)\,,
\eeq
where the superscript ``$(1)$'' indicates that the vacuum is filled
by the real roots $\{\theta_\JJ\}$, which in this context are 
usually called ``1-strings''.
The constant $\varepsilon_\infty$ can be easily calculated by combining
\eqref{energy} and \eqref{dens2},
\beq\label{einfinity}
\varepsilon_\infty^{(1)}=
\varepsilon_\infty(\Theta,\delta)=\frac{2}{\pi}
\,\int_{-\infty}^{\infty}\frac{\rd \nu}{\nu}\ \sin(2\nu\Theta)\ \frac{\sinh\big(\frac{\pi\nu(1+\delta)}{2}\big)
\cosh\big(\frac{\pi\nu\delta}{2}\big)}
{\sinh\big({\pi\nu}\big)
\cosh\big(\frac{\pi\nu}{2}\big)}\,.
\eeq
In particular, for large positive  $\Theta$ one has
\beq
\varepsilon_\infty(\Theta,\delta)=(1+\delta)
-\big( {\textstyle \frac{4}{\pi}}\, \re^{-\Theta} \cos(\tshf \pi\delta)\big)^2\  \Big(\,\Theta+\tshf+
{\textstyle \frac{1}{4}}\pi \, (1+2\delta)\tan(\tshf \pi\delta)\,\Big)+O(\re^{-3\Theta})\ .
\eeq

Similar considerations apply to the case $\delta<0$. For practical
purposes it is convenient to 
always assume that the constant $\delta$ is positive, and consider
different BAE, which are obtained from \eqref{bae2} and 
\eqref{baelat1} by the negation of
$\delta$. In addition it is also convenient to simultaneously  
interchange  $p_1$ and $p_2$. Let $\{\theta_\JJ\}$
and $\{u_\ell\}$ now solve the BAE modified in this way. 
As explained in 
\cite{Bukhvostov:1980sn, Bazhanov:2016glt}, the
$\theta$-roots become complex and form 2-strings, which for large $N$
are asymptotically
approaching the values
\beq\label{two-strings}
\theta_{2\ell-\hf\pm\hf}\sim u_\ell\pm \tshf\ri\pi\delta\, ,
\eeq
where $\ell$ runs over the set \eqref{set2}. The phase assignment of
these complex roots was discussed in details in the first part of this
work \cite{Bazhanov:2016glt} (see Eqs.\,(5.10a),\,(5.10b) therein).  
The corresponding vacuum energy \eqref{energy},  
\beq
E^{(2)}_N(\Theta,\delta)
=2 \sum_\JJ \myenergy(\theta_\JJ,-\delta)\,,
\eeq
involves the sum over the 2-strings \eqref{two-strings}.   
The $u$-roots remain real.
More precisely, as we shall see in the next section, 
their positions remain unaffected by
the transformation $(\delta, p_1, p_2)\to(-\delta, p_2,p_1)$.
Then, as follows from 
\eqref{two-strings}, the asymptotic density of the 2-strings exactly coincides
with that of the $u$-roots given by \eqref{dens1}, 
which implies that for $N\to\infty$
\beq
E^{(2)}_N(\Theta,\delta)=N\,\epsilon_\infty^{(2)}+O(N^{-1})\,,\qquad
\varepsilon_\infty^{(2)}=\varepsilon_\infty(\Theta,-\delta)\,,
\label{einf2}
\eeq
where $\epsilon_\infty(\Theta,-\delta)$ is defined by 
\eqref{einfinity} with $\delta$ replaced by $-\delta$.

Remarkably, the regularized vacuum energies for the above
two cases exactly coincide 
\beq\label{identity}
E^{(1)}_N(\Theta,\delta)
-N\,\varepsilon_\infty^{(1)}=
E^{(2)}_N(\Theta,\delta)
-N\,\varepsilon_\infty^{(2)}
\eeq
for any finite values of $N$ and $\Theta$. Originally, we have noted
this from numerical calculations and then proven analytically (see
Appendix~\ref{app:D} for details). The working is essentially based on
the particle-hole duality symmetry, considered below. 

\subsection{\label{phole}Particle-hole duality transformations}
\begin{figure}[ht]
\centering
\hspace*{-1cm}
\includegraphics[width=16.8cm]{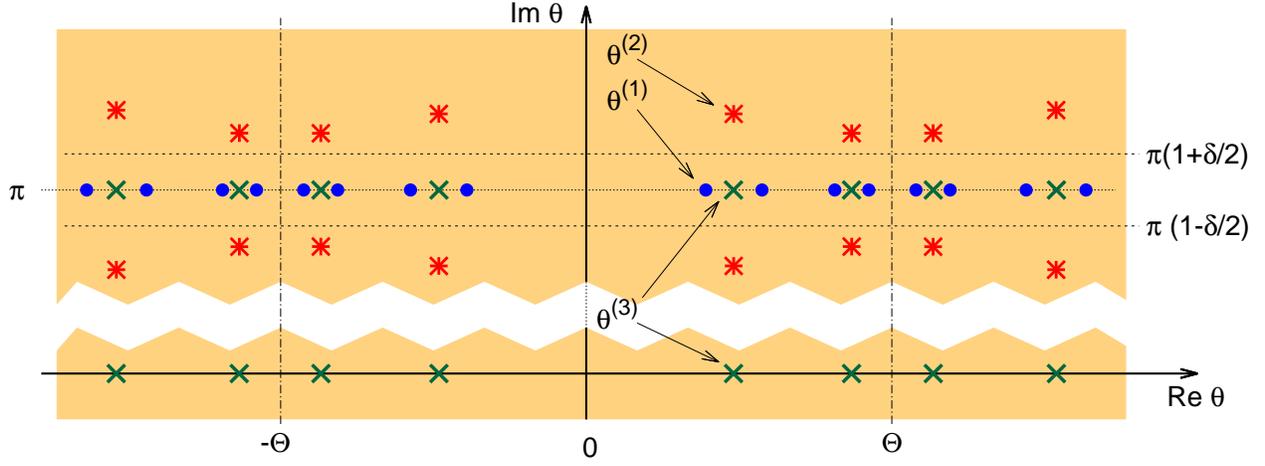}
\caption{Roots of Bethe Ansatz equations with the lattice-type
  regularization, given in Sect.\,\ref{lat-type}. The (blue) dots
  show zeroes of ${\Asf}^{(1)}(e^{\theta})$,
 the (red) asterisks show zeroes 
of ${\Asf}^{(2)}(e^{\theta})$ 
and the (green) crosses represent zeroes of ${\Asf}^{(3)}(e^{\theta})$.} 
\label{lattice_string}
\end{figure}

The lattice type BAE \eqref{bae2} and \eqref{baelat1} can be brought to
various equivalent forms by means of the so-called {\em particle-hole}
transformations which are well known for sypersymmetric solvable
lattice models \cite{Essler:1992uc, Kulish:1985bj}. We use this
symmetry to bring the BAE 
to a form convenient for a derivation of
NLIE valid for both positive and negative $\delta$ in the interval 
$-1<\delta<1$.  
Consider the vacuum solution containing real roots 
\eqref{vacsol} corresponding to $\delta>0$. 
Define two Laurent polynomials ${\Asf}_1(\lambda)$ and 
${\Asf}_3(\lambda)$ of the degree $2N$ 
in the variable $\lambda=\re^\theta$,
\beq
{\Asf}_1(\re^\theta)=\prod_\JJ \,
\sinh\big({\textstyle\hf}(\theta-\theta^{(1)}_\JJ)\big)\,,
\qquad
{\Asf}_3(\re^\theta)=\prod_{\ell} \,
\sinh\big(\theta-\theta^{(3)}_\ell\big)\,,\label{A13-def}
\eeq
whose zeroes are determined by the vacuum roots,
\beq
\theta^{(1)}_\JJ=\theta_\JJ+\ri\pi,\qquad 
\theta^{(3)}_\ell=u_\ell\, ,
\eeq
where, as before, the
indices $\ell$ and $\JJ$ run over the values \eqref{set2} (note an
$\ri \pi$ shift in the definition of $\theta^{(1)}_\JJ$ zeroes).
Note that ${\Asf}_3(\lambda)$ only depends on $\lambda^2$, since
$N$ is assumed to be even and the products in \eqref{A13-def} contain
even numbers of factors.  
Eqs.\,\eqref{baelat1},\,\eqref{bae2} can now be equivalently rewritten as
\begin{subequations}
\label{bag1}
\begin{align}
-1&=\displaystyle\re^{2\pi \ri (p_2-p_1)}\,\frac{{\mathsf f}(-\ri\lambda_\JJ^{(1)} q^{-1})}
{{\mathsf f}(\ri \lambda_\JJ^{(1)} q)}\,\frac{{\Asf}_3(\lambda_\JJ^{(1)}
 q)}
{{\Asf}_3(\lambda_\JJ^{(1)}\, q^{-1})}\,,
&\lambda_\JJ^{(1)}\equiv\re^{\theta^{(1)}_\JJ}\label{e1}
\\[.5cm]
-1&=\displaystyle\re^{-4\pi \ri p_1}\,
\frac{{\Asf}_3(\lambda^{(3)}_\ell\,
  q^2)}
{{\Asf}_3(\lambda^{(3)}_\ell q^{-2})}
\frac{{\Asf}_1(\lambda^{(3)}_\ell
  q^{-1})}
{{\Asf}_1(\lambda^{(3)}_\ell \,q)}
\frac{{\Asf}_1(-\lambda^{(3)}_\ell
  q^{-1})}
{{\Asf}_1(-\lambda^{(3)}_\ell \,q)}\,,
&\lambda_\ell^{(3)}\equiv\re^{\theta^{(3)}_\ell}\,,\label{e2}
\end{align}
\end{subequations}
where
\beq
{\mathsf f}(\re^\theta)=\big({\mathrm s}(\theta)\big)^N\,, 
\qquad q=\re^{\hf\ri \pi
    \delta}\label{f-def}
\eeq
and ${\mathrm s}(\theta)$  is defined in \eqref{phi-def}. The zeroes of ${\Asf}_3(\lambda)$ 
are determined by the variables
$\{\lambda^{(3)}_\ell\}$. Assume that these variables are fixed.  
Then Eq.\,\eqref{e1} can be viewed as  
an algebraic equation for zeroes of some Laurent polynomial ${\mathsf
  P}(\lambda)$ 
\beq\label{P-def1}
{\mathsf P}(\lambda)=\re^{\ri\pi(p_1-p_2)}\,
{\mathsf f}(\ri\lambda q)\ 
{\Asf}_3(\lambda q^{-1})+\re^{-\ri\pi(p_1-p_2)}\,
{\mathsf f}(-\ri\lambda q^{-1})\  
{\Asf}_3(\lambda q)
\eeq
of the degree $4N$. By construction, one half of its zeroes coincides 
with $\{\lambda^{(1)}_\JJ\}$, 
but there are $2N$ additional zeroes,
therefore, ${\mathsf P}(\lambda)$ factorizes as
\beq\label{P-def2}
{\mathsf P}(\lambda)={\rm const}\  {\Asf}_1(\lambda)\,
{\Asf}_2(-\lambda)\,,
\eeq
where ${\Asf}_2(\lambda)$ is some Laurent polynomial of the degree $2N$,
\beq\label{A2-def}
{\Asf}_2(\re^\theta)=
\prod_\JJ \,
\sinh\big({\textstyle\hf}(\theta-\theta^{(2)}_\JJ)\big)\,,
\qquad \lambda_\JJ^{(2)}=\re^{\theta_\JJ^{(2)}}\,.
\eeq  
It is easy to show that this polynomial satisfies the following BAE
\begin{subequations}
\label{bag2}
\bea
-1&=&\re^{-2\pi \ri (p_1+p_2)}\,
\frac{{\mathsf f}(-\ri\lambda^{(3)}_\ell)}{{\mathsf
    f}(\ri\lambda^{(3)}_\ell)} 
\,\frac{{\Asf}_1(\lambda^{(3)}_\ell q^{-1})}
{{\Asf}_1(\lambda^{(3)}_\ell \,q)} 
\,\frac{{\Asf}_2(\lambda^{(3)}_\ell \,q)}
{{\Asf}_2(\lambda^{(3)}_\ell q^{-1})}
\label{ba-new}
\\[.5cm]
-1&=&\displaystyle\re^{2\pi \ri (p_1-p_2)}\,\frac{{\mathsf
    f}(-\ri\lambda_\JJ^{(2)} q)} 
{{\mathsf f}(\ri \lambda_\JJ^{(2)}
  \,q^{-1})}\,\frac{{\Asf}_3(\lambda_\JJ^{(2)}  
 \,q^{-1})}
{{\Asf}_3(\lambda_\JJ^{(2)} q)}\qquad
\label{e12}
\\[.5cm]
\label{e22}
-1&=&\displaystyle\re^{-4\pi \ri p_2}\ 
\frac{{\Asf}_3(\lambda^{(3)}_\ell\,
  q^{-2})}
{{\Asf}_3(\lambda^{(3)}_\ell\, q^{2})}
\frac{{\Asf}_2(\lambda^{(3)}_\ell
  \,q)}
{{\Asf}_2(\lambda^{(3)}_\ell q^{-1})}
\frac{{\Asf}_2(-\lambda^{(3)}_\ell
  \,q)}
{{\Asf}_2(-\lambda^{(3)}_\ell q^{-1})}\,.
\eea
First, note that \eqref{e12} trivially follows from \eqref{P-def1} and 
\eqref{P-def2}.\footnote{%
Eq.\eqref{e12} is similar \eqref{e1}. A minor
difference caused by the minus sign in the argument ${\Asf}_2$ in
\eqref{P-def2}.}
Next, equating the ratios 
\newline
 ${\mathsf P}(\lambda^{(3)}_\ell q)/{\mathsf P}(\lambda^{(3)}_\ell q^{-1})$ 
obtained from alternative representations of ${\mathsf
  P}(\lambda)$ from 
\eqref{P-def1} and \eqref{P-def2}, and then using the fact that
${\Asf}_3(\lambda)={\Asf}_3(-\lambda)$,
one obtains
\beq\label{dual}
+1=\re^{2\pi \ri (p_1-p_2)}\
\frac{{\mathsf f}(-\ri\lambda^{(3)}_\ell)}{{\mathsf f}(\ri\lambda^{(3)}_\ell)}
\frac{{\Asf}_3(\lambda^{(3)}_\ell q^{-2})}{{\Asf}_3(\lambda^{(3)}_\ell q^{+2})}
\,\frac{{\Asf}_1(-\lambda^{(3)}_\ell q)}{{\Asf}_1(-\lambda^{(3)}_\ell q^{-1})}
\,\frac{{\Asf}_2(\lambda^{(3)}_\ell q)}{{\Asf}_2(\lambda^{(3)}_\ell
  q^{-1})}\ . 
\eeq
\end{subequations}
Together with \eqref{e2} this leads to \eqref{ba-new}. 
Finally, combining \eqref{dual} and \eqref{ba-new} one obtains \eqref{e22}.
Thus, we have obtained a set of coupled BAE for the zeroes of the
three Laurent polynomials ${\Asf}_1(\lambda), {\Asf}_2(\lambda)$ and
${\Asf}_3(\lambda)$. It is useful to
isolate the following three closed subsets of these equations:
\begin{enumerate}[(I)]
\item 
the pair of original equations
\eqref{e1} and \eqref{e2} involving 
${\Asf}_1$ and ${\Asf}_3$ only;
\item 
the pair of equations \eqref{e12} and \eqref{e22} involving 
${\Asf}_2$ and  ${\Asf}_3$ only;
\item
the set of three Eqs. \eqref{e1},
\eqref{ba-new} and \eqref{e12} involving all three ${\Asf}_1$,
${\Asf}_2$, ${\Asf}_3$.  
\end{enumerate}
The arrangement of the vacuum
roots is illustrated in Fig.\,\ref{lattice_string}.  

Note that under the ``particle-hole'' symmetry transformation
\beq\label{trans}
{\Asf}_1(\lambda)\leftrightarrow{\Asf}_2(\lambda)\,,
\qquad {\Asf}_3(\lambda)\leftrightarrow{\Asf}_3(\lambda)\,,\qquad
p_1\leftrightarrow p_2\,,\qquad 
q\mapsto q^{-1}, \qquad\delta\mapsto-\delta
\eeq
the sets (I) and (II) transform into one another, 
whereas the set (III) remains invariant. Obviously the sets (I) and
(II)  describe the fillings of the vacuum state for $\delta>0$ and
$\delta<0$, respectively. The above symmetry proves that the roots
$\{u_\ell\}$, which determine zeroes of ${\Asf}_3(\lambda)$,
remain unchanged under the transformation \eqref{trans} (this fact has
already been mentioned before Eq.\eqref{einf2}).

Even though the lattice-type regularization is used here as a
technical tool,   
it would be  interesting to construct a lattice model, which  
actually leads to the BAE presented above. 
This could provide ideas about  
a lattice regularization for non-linear sigma models, which so far has 
not been properly understood. 
From the point of view of quantum groups such
a lattice model
could be related to the quantized affine
superalgebra $U_q(\widehat{D}(2,1|\alpha))$ or
$U_q(\widehat{sl}(2|2))$ (see \cite{Bazhanov:2013cua} for a hidden 
quantum group structure of the general Fateev model). As noted in
\cite{Saleur:1998wa} it could also be
related to $U_q(osp(2|2))$ algebra $R$-matrices found in \cite{
Deguchi:1990,Gould:1997,Martins:1997ex}. 

\subsection{Scaling limit\label{sec::BAscale}}

In the scaling limit \eqref{qft-lim} 
the number of BA roots in the central region $|{\Re} e(\theta)|<\Theta$
in Fig.\,\ref{lattice_string} becomes infinite. 
With the standard
Weierstrass regularization the formulae \eqref{A13-def} and
\eqref{A2-def}, 
with 
$|{\Re}e(\theta)|<\Theta$, 
can be
easily turned into convergent infinite products, defining analytic 
functions ${\Asf}_1^{\rm (BL)}(\lambda)$, ${\Asf}_2^{\rm
  (BL)}(\lambda)$ and ${\Asf}_3^{\rm (BL)}(\lambda)$ of the
variable $\lambda$, having two essential singularities at $\lambda=0$
and $\lambda=\infty$: 
\bea
\Abl_i(\lambda)&=&{N}_i\ 
\,\re^{\alpha_i\lambda+\beta_i\lambda^{-1}}\,\prod_{n=1}^\infty\,
\Big(1-{\lambda}/{\lambda^{(i)}_n}\Big)\,
\re^{{\lambda}/{\lambda^{(i)}_n}}\  
\Big(1-{\lambda^{(i)}_{-n+1}}/{\lambda}\Big)\, 
\re^{{\lambda^{(i)}_{-n+1}}/{\lambda}}\qquad (i=1,2)\nonumber
\\[.3cm]
\Abl_3(\lambda)&=&{N}_3
\,\prod_{n=1}^\infty\,
\Big(1-{\lambda^2}/{\big(\lambda^{(3)}_n\big)^2}\Big)\
\Big(1-{\big(\lambda^{(3)}_{-n+1}\big)^2}/{\lambda^2}\Big)\,,\label{ABL}
\eea
where the coefficients $\alpha_i$ and $\beta_i$ are partially constrained by
\eqref{P-def1} and \eqref{P-def2}, but otherwise arbitrary (see a
remark after \eqref{ident} below). 
Obviously, the functions \eqref{ABL} can be regarded as vacuum
eigenvalues of suitable analogs of Baxter ${\bf Q}$-operators for the
BL model.

The asymptotic distribution of zeroes 
$\big\{\lambda^{(i)}_n\big\}_{n=-\infty}^\infty$
of ${\Asf}_1^{\rm
  (BL)}$ and ${\Asf}_3^{\rm (BL)}$ 
immediately follows from \eqref{assz} and \eqref{z-qft}, while for 
zeroes of ${\Asf}_2^{\rm (BL)}(\lambda)$ (which form 2-strings)
one needs to follow the reasonings of \cite{deVega:1989pj}. 
For $|n|\gg1$ one obtains  
\bea
r\,\sinh\big(\theta^{(1)}_n\big)&\asymp& 
{ -\frac{\pi}{\cos{\frac{\pi\delta}{2}}}\ 
\Big(n-\frac{1}{2}+\frac{2 p_2}{a_2}\Big)}+o(1)\nonumber\\[.2cm]
r\,\sinh\big(\theta^{(2)}_n\big) &\asymp & -2\pi\,
\re^{\pm \frac{\ri\pi}{2} \delta}\,
  \bigg(n - \frac{1}{2} +
 \frac{p_1}{a_1} + \frac{|p_2|}{a_2} \pm \ri\  \frac{\log
   (2)}{2\pi}\ \bigg)+o(1)\label{ZBL}
\\[.2cm]
r\,\sinh\big(\theta^{(3)}_n\big)& \asymp& \pm\, 2\pi
\,  \left(n+\frac{p_1}{a_1}+\frac{
  p_2}{a_2}-\frac{1}{2}\right)+o(1)\,,\nonumber
\eea
where $\lambda^{(i)}_n=\exp\big({\theta^{(i)}_n}\big)$.
The zeroes  of
${\Asf}_1^{\rm (BL)}(\lambda)$ are located 
on the real negative axis of
$\lambda$, accumulating towards $\lambda=0$ and $\lambda=-\infty$. 
The zeroes ${\Asf}_2^{\rm 
  (BL)}$ form 2-strings located just 
outside the wedge with an acute angle $\pi\delta $ centered around the
negative real axis of $\lambda$,
where signs ``$\pm$'' correspond to lower and upper branches of the wedge.
The zeroes of ${\Asf}_3^{\rm 
  (BL)}$ are located at the real axis of $\lambda$ (note, that 
${\Asf}_3^{\rm 
  (BL)}$ is, actually, a function of $\lambda^2$).
For small values of $r$ the above formulae give a very good
approximation to the position of zeroes even for small values of
$|n|$ (see Appendix~\ref{app:B}). 

The functions \eqref{ABL} satisfy the QFT
version of BAE \eqref{bag1},\,\eqref{bag2}.
The later are obtained by rewriting the coefficients  
there in the form
\beq\label{scale1}
\frac{{\mathsf f}(-\ri\,\re^\theta\,q^{\pm1})}
{{\mathsf f}(+\ri\,\re^\theta\,
  q^{\mp1})}=\re^{\ri N \myp(\theta,\,\pm\delta)}\,,
\qquad 
\frac{{\mathsf f}(-\ri\,\re^\theta)}{{\mathsf f}(+\ri\,\re^\theta)}
=\re^{\ri N \myp(\theta\,,0)}
\eeq
and making the substitution 
\beq\label{scale2}
{\Asf}_i(\lambda)\to {\Asf}_i^{{\rm (BL)}}(\lambda)\,,\qquad
N \myp(\theta,\,\pm\delta)\to r \cos(\pi \delta/2)\,\sinh(\theta)\, ,
\qquad
N \myp(\theta\,,0)\to r \,\sinh(\theta)
\eeq
which is reverse to \eqref{p-def}.

Finally, following the arguments of
\cite{Destri:1994bv,Lukyanov:2011wd}, 
one can show that in the scaling limit \eqref{qft-lim} 
an appropriately scaled regularized vacuum energy 
\beq
\lim_{\scriptstyle{\begin{subarray}{c}
N,\Theta\to\infty\\
r-\mbox{\scriptsize{fixed}}
\end{subarray}}}\Big(N\,E^{(1)}_N(\Theta,\delta)
-N^2\,\varepsilon_\infty^{(1)}\Big)\label{ceff-lat} 
={\mathfrak F}(r,{\bf k})-{\mathfrak F}(0,{\bf
  k})\, 
\eeq
is expressed in terms of the scaling function  \eqref{F-def}
of the continuous QFT \eqref{Lagr1}, where the renormalized field
theory twist
parameters in \eqref{apssspps} and their bare  
counterparts in the BAE \eqref{bae2},\,\eqref{bae1} are related as
\beq
k_\pm=k_1\pm k_2\,,\qquad k_i=2 p_i/a_i\,\qquad (i=1,2)\,.\label{twist2}
\eeq
A proof of this statement is presented in Appendix\,\ref{app:D}. 
Note that another variant of the formula \eqref{ceff-lat} for a simple momentum
cutoff regularization (rather that the lattice-type regularization of
this paper) was presented in \cite{Bazhanov:2016glt}, see
Eq.\,(5.14) therein. 
Remarkably, Eq.\,\eqref{ceff-lat} 
gives reasonably accurate results even for finite values of
$N$ and $\Theta$, since the main subleading term in the LHS is of the
order of $O(\re^{-2\Theta})$.

\section{Connection to classical sinh-Gordon equation \label{MShG}}
The Bethe ansatz equations derived in the previous section allow one to
make a connection of the BL model to the {\em classical} inverse scattering
problem method for the {\em modified} sinh-Gordon equation 
(see \cite{Lukyanov:2010rn,Lukyanov:2013wra,Bazhanov:2013cua,Bazhanov:2016glt})
\begin{equation}
\label{sinh-Gordon}
	\partial_z\partial_{\bar{z}}\eta-e^{2\eta}+
\rho^4\,|\mathcal{P}(z)|^2\, e^{-2\eta}=0
\end{equation}
for a complex-valued function $\eta(z)$ defined on the Riemann sphere 
with punctures. Here $\rho$ 
is an arbitrary constant, while $\mathcal{P}(z)$ is a  function    
with three singular points located at $z_1,z_2$ and $z_3$,
\begin{equation}\label{P-def}
	\mathcal{P}(z)=\frac{(z_3-z_2)^{a_1}(z_3-z_1)^{a_2} (z_2-z_1)^{a_3}
}{(z-z_1)^{2-a_1}(z-z_2)^{2-a_2}(z-z_3)^{2-a_3}}\,,
\end{equation}
where
\beq
a_1=1-\delta,\qquad a_2=1+\delta,\qquad a_3=0\,.\label{sregime}
\eeq
Eq.\eqref{sinh-eq}, given in  Introduction, is connected
to \eqref{sinh-Gordon} by a simple change of variables 
\beq\label{varchange}
\eta(z,\bar z)\to\hat\eta(w,\bar w)=\eta(z,\bar z)
-{\textstyle\frac{1}{2}}\log\big|\rho^2 \,{\mathcal P}(z)\big|\,,
\qquad \rd w=\rho\,\sqrt{{\mathcal P}(z)}\ \rd z
\eeq
and the domain ${\mathbb D}_{\rm BL}$, shown in Fig.\,\ref{fig3sgsg}, is
the image of the Riemann sphere with three punctures in
the coordinates $(w,\bar w)$. However, in this section we prefer to work
with Eq.\,\eqref{sinh-Gordon}.

For further references note also that Eq.\,\eqref{sinh-Gordon} with the choice
\eqref{sregime} is a particular
$a_3\to0$  case 
of a more general equation
\cite{Lukyanov:2013wra}, where 
\beq
a_1+a_2+a_3=2\,,\qquad 0<a_i<2\,,
\eeq
which is connected with the symmetric regime of the Fateev model
\cite{Fateev:1996ea}.


\subsection{Symmetries of the auxiliary linear problem}
The equation \eqref{sinh-Gordon}
is the zero-curvature condition for an $sl(2)$-valued connection 
\beq\label{A-def}
\begin{array}{rcl}
{\boldsymbol
  A}_z&=&-{\textstyle\frac{1}{2}}\ 
\partial_z\eta\,\sigma_3+\big(\,\re^\eta\,\sigma_+\,+\, 
\rho^2\,\lambda^{2\phantom{-}}
\,{\cal P}(z)\, \re^{-\eta}\,\sigma_-\, \big)\\[.3cm]
{ {\boldsymbol A}}_{\bar z}&=&\ \ \,
{\textstyle\frac{1}{2}}\ \partial_{\bar
  z}\eta\,\sigma_3+\big(\, 
\re^{\eta}\,\sigma_-\,+\,\rho^2{\lambda}^{-2}\, \,
{\bar {\cal P}}({\bar z})\,\re^{-\eta}\,\sigma_+\,\big)\ ,
\end{array}
\eeq
where $\bar{\mathcal P}(\bar{z})$ denotes the complex conjugate of ${\mathcal P}(z)$,
and $\lambda=\re^\theta$ is the multiplicative spectral parameter.

\begin{figure}
\centering
\includegraphics[width=6.  cm]{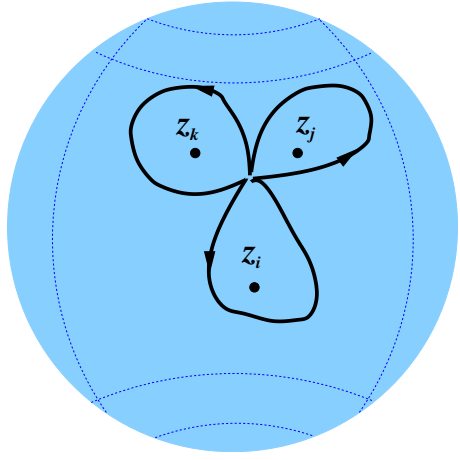}
\caption{The contractible loop $
\gamma_k\circ\gamma_j\circ\gamma_i=\gamma_i\circ\gamma_k\circ\gamma_j=
\gamma_j\circ\gamma_i\circ\gamma_k$ on
the sphere with three punctures.}
\label{fig1ay}
\end{figure}

Recall that the equations \eqref{bae2} and \eqref{bae1} are the main
ingredients 
of the coordinate Bethe ansatz solution for the Bukhvostov-Lipatov QFT \eqref{Lagr1}.
Remarkably, as we shell see below, exactly the same equations are 
also satisfied by
connection coefficients of the linear matrix
differential equations 
\bea
\label{auxlin}
(\partial_z-{\boldsymbol A}_z)\,{\boldsymbol \Psi}=0\,,\qquad
(\partial_{\bar z}-{ {\boldsymbol A}}_{\bar z})\,{\boldsymbol \Psi}=0\,
\eea
associated with modified sinh-Gordon equation \eqref{sinh-Gordon}.
The required calculations are only slightly different from those
related to the Fateev model, considered in
details in \cite{Bazhanov:2013cua}. Therefore, here we only briefly sketch
the main steps of working for the case of the vacuum state of the BL model. 
In this case the function $\re^{-\eta(z)}$ is a smooth, single-valued complex function without zeroes on a Riemann sphere 
with three punctures at $z=z_1, z_2, z_3$. The point $z=\infty$ 
is assumed to be a regular point on the  sphere, where
\bea
\label{asym-inf}
\re^{-\eta(z)}\sim |z|^{2}\ \ \ \ \ \ \ \  {\rm as}\ \ \ \ \ \  \ \   |z|\to  \infty\ .
\eea
\begin{subequations}\label{asym123}
The asymptotic conditions at the punctures are determined by  
\bea\label{asym12}
\eta(z)=-(1-a_i |k_i|)\log|z-z_i|+O(1)\ \ \ \ \  {\rm
  as}\ \ \ \ \ \  \ \ |z-z_i|\to  0\ \ \ \ \ \ (i=1,2)
\eea
and 
\bea\label{asym3}
\eta(z)=-\log|z-z_3|+\log\rho +o(1)\ \ \ \ \ \ \ \ \  \ {\rm
  as}\ \ \ \ \ \  \ \ |z-z_3|\to  0\, . \phantom{\ \ \ \ \ (i=1,2)}
\eea
Note that the above asymptotic conditions are equivalent to
\eqref{osasail}, \eqref{asym-z3} under the substitution
\eqref{varchange}. 
Below it will be convenient to use parameters ${\newp}_1,{\newp}_2$
related to $k_\pm$ in \eqref{apssspps} as,\footnote{%
This is, of course, the same parameters $p_1$ and $p_2$ that already
appeared in \eqref{twist2}, however, in this section they are
assumed to be positive and satisfying the relation \eqref{p-small}.}   
\end{subequations}
\beq\label{r-def}
k_\pm=k_1\pm k_2\,,\qquad
{\newp}_i=a_i |k_i| /2\,\qquad (i=1,2)\,.
\eeq
From now on we assume that $(i,j,k)$ is a cyclic permutation of
$(1,2,3)$. 
Consider the auxiliary linear  problem \eqref{auxlin}.
Introduce three matrix solutions 
\bea\label{matsol}
{\boldsymbol  \Psi}^{(i)}=\big({\boldsymbol  \Psi}^{(i)}_{-  }, {\boldsymbol  \Psi}^{(i)}_{+  }\,\big)\in
\mathbb{SL}(2,{\mathbb C})\ \ \ \ \ \ (i=1,2,3)
\eea
normalized by the
following asymptotic conditions 
\bea\label{psi12}
{\boldsymbol  \Psi}^{(i)}
&\to&
\big(2 {\newp}_i\big)^{-\hf\sigma_3}\,\re^{\ri \beta_i\sigma_3} \ \bigg(\,
\frac{{ z}-{ z}_i} {{\bar z}-{\bar z}_i}\,\bigg)^{\frac{1}{4}\,
  (1-2{\newp}_i)\,\sigma_3}\quad
{\rm as}\quad z\to z_i \ \ \ \quad (i=1,2)
\eea
and
\beq\label{psi3}
\massPsi^{(3)}_\pm \to 
\frac{1}{\sqrt{2\lambda}}
\left(\frac{z_{13}z_{32}}{z_{12}}\right)^{\mp
  \rho\lambda}
\left(\frac{\z_{13}\z_{32}}{\z_{12}}\right)^{\mp
  \rho\lambda^{-1}}
\left(
\begin{array}{c} 
\phantom{\pm\,\lambda}\,{(z-z_3)^{+\frac{1}{4}\pm\rho\lambda}}\ 
{(\bar{z}-\bar{z}_3)^{-\frac{1}{4}\pm\rho\lambda^{-1}}}\\[.3cm]
\pm\,\lambda\,{(z-z_3)^{-\frac{1}{4}\pm\rho\lambda}}\ 
{(\bar{z}-\bar{z}_3)^{+\frac{1}{4}\pm\rho\lambda^{-1}}}
\end{array}\right)
\eeq
for $z\to z_3$.
The constants $\beta_i$ are defined by
\beq
\label{betadef}
	e^{i\beta_i}=\left(\frac{z_{ji}z_{ik}}{z_{jk}}\frac{\z_{jk}}{\z_{ji}\z_{ik}}\right)^{\frac{{\newp}_i}{2}}\,\ \ \ \ \qquad (z_{ij}=z_i-z_j)\,.
   \eeq
The above conditions uniquely determine the solutions provided 
\beq
\big|{\Re e}\big(\rho(\lambda-\lambda^{-1})\big)\big|<\hf\,,\qquad 0<{\newp}_i<
\frac{a_i}{4}
\qquad\ \ ( i=1,2)\,. \label{p-small}
\eeq
The connection  matrices are defined  as
\bea\label{S-def}
{\boldsymbol  \Psi}^{(i)}={\boldsymbol  \Psi}^{(j)}\ 
{\boldsymbol S}^{(j, i)}(\lambda)\ .
\eea
They satisfy the obvious relations 
\bea\label{NS-rel2}
\det\big({\boldsymbol S}^{(j,i)}(\lambda)\big)=1\ ,\ \ \
{\boldsymbol S}^{(i,j)}(\lambda)\,{\boldsymbol S}^{(j,i)}(\lambda)={\boldsymbol I}\ ,\ \ \
{\boldsymbol S}^{(i,k)}(\lambda)\
{\boldsymbol S}^{(k,j)}(\lambda)\
{\boldsymbol S}^{(j,i)}(\lambda)={\boldsymbol I}
\eea
and
\beq
{\boldsymbol S}^{(i,j)}_{\sigma,\sigma'}(\lambda)=
\det\Big({\boldsymbol  \Psi}^{(j)}_{\sigma'}\,,\,
{\boldsymbol  \Psi}^{(i)}_{-\sigma}\Big)\,.
\eeq  
Further analysis is based on symmetries of the linear differential
equations \eqref{auxlin}. Let 
\beq 
\label{O-def}
\wh\Omega_i:\qquad z\mapsto \gamma_i\circ z\,,
\quad \qquad \bar{z}\mapsto \bar{\gamma_i}\circ \bar{z}\,,
\qquad \lambda\mapsto 
\lambda\,q_i^{-1}\, \qquad (i=1,2,3)
\eeq
be a transformation involving a translation of the independent 
variable $z$ along the contour $\gamma_i$, going around the point
$z_i$ anticlockwise (the variable ${\bar z}$ is translated along 
complex conjugate contour ${\bar \gamma}_i$),
accompanied by the
substitution $\lambda\mapsto \lambda\,q_i^{-1}$, where 
\bea\label{di-def}
q_1=\re^{-\ri \pi \delta}\,,\qquad \ \ \ 
q_2=\re^{+\ri \pi \delta}\,,\qquad\ \ \  
q_3=1,\quad \ \ \ q_1\,q_2\,q_3=1\,.
\eea
Using \eqref{P-def} and \eqref{A-def} it is easy to
check that the  substitutions \eqref{O-def} leave the system
\eqref{auxlin} unchanged.  Therefore
they act as linear transformations in the space of solutions. 
Namely, in the basis ${\boldsymbol\Psi}^{(i)}$ they read 
\begin{subequations}\label{N123}
\bea\label{N123a}
\wh\Omega_i\big({\boldsymbol\Psi}^{(i)}\big)&=&-{\boldsymbol
  \Psi}^{(i)}\,\re^{-2\pi \ri {\newp}_i(\lambda)\sigma_3}\\[.4cm] 
\wh\Omega_j\big({\boldsymbol\Psi}^{(i)}\big)
&=&-{\boldsymbol\Psi}^{(i)}\,{\boldsymbol
  S}^{(i,j)}(\lambda)\, 
\re^{-2\pi \ri {\newp}_j(\lambda) \sigma_3}\, {\boldsymbol S}^{(j,i)}(\lambda\,q_j^{-1}) \label{N123b}\\[.4cm] 
\wh\Omega_k\big({\boldsymbol\Psi}^{(i)}\big)
&=&-{\boldsymbol\Psi}^{(i)}\,{\boldsymbol
  S}^{(i,k)}(\lambda)\,  
\re^{-2\pi \ri {\newp}_k(\lambda) \sigma_3} \,{\boldsymbol S}^{(k,i)}(\lambda\,q_k^{-1})\, ,\label{N123c} 
\eea
\end{subequations}
where 
\beq
\label{pth-def}
{\newp}_1(\lambda)\equiv {\newp}_1,\qquad 
{\newp}_2(\lambda)\equiv {\newp}_2,\qquad 
{\newp}_3(\lambda)= \rho\,(\lambda-\lambda^{-1})\, .
\eeq
An important property of the linear system 
\eqref{auxlin} is that   a combined transformation 
$\wh\Omega_k\circ\wh\Omega_j\circ\wh\Omega_i$,
where $(i,j,k)$ is a cyclic permutation of $(1,2,3)$, is equivalent to
the identity transformation 
\beq\label{prop1}
\wh\Omega_k\circ\wh\Omega_j\circ\wh\Omega_i\,\big({\boldsymbol\Psi}^{(i)}\big)={\boldsymbol\Psi}^{(i)}\ .
\eeq
The proof follows from \eqref{di-def} and the fact that 
 $\gamma_k\circ\gamma_j\circ\gamma_i$ is a contractible contour which
loops around a regular point $z=\infty$ on the Riemann sphere, see Fig.\,\ref{fig1ay}.
Combining \eqref{N123} and \eqref{prop1} with the definition
\eqref{S-def} one easily obtains
\beq\label{S-rel2}
{\boldsymbol S}^{(i,k)}(\lambda)\ \re^{-2\pi \ri {\newp}_k(\lambda) \sigma_3}\ 
{\boldsymbol S}^{(k,j)}(\lambda\,q_k^{-1})\ \re^{-2\pi \ri {\newp}_j(\lambda\,q_k^{-1}) \sigma_3}\ 
{\boldsymbol S}^{(j,i)}(\lambda\,q_i)\ \re^{-2\pi \ri {\newp}_i(\lambda\,q_i) \sigma_3}=
-{\boldsymbol I}\ .
\eeq

\subsection{Functional relations for connection coefficients}
The non-trivial functional relation \eqref{S-rel2}, complemented with
the asymptotic (WKB) analysis of the differential equations
\eqref{auxlin}, allows one to completely determine all connection matrices.
To within some simple $\sqrt{\lambda}$ factors, 
the connection
coefficients are meromorphic function of $\lambda$ with two essential
singularities at $\lambda=0$ and $\lambda=\infty$.
For the normalization \eqref{psi12}, \eqref{psi3} they could 
have poles for real negative $\lambda$'s 
solving the equations   
\beq\label{poles}
2\rho\, \big(\lambda-\lambda^{-1}\big)=4 \rho\,\sinh(\theta)= n\,,\qquad
{\Im m}(\theta)=\pi\ 
\qquad
\ \ (n\in {\mathbb Z})\,.
\eeq  
To take this into account it is convenient to introduce a 
function 
\bea
{\Zbs(\lambda)}\Big\vert_{\lambda={\rm e}^\theta}
&=&\frac{\big(2\rho\  \re^{\gamma_E-1}\big)^{-2\rho\cosh(\theta)}}{4\,\sqrt{\pi
    \rho}\,\cosh(\frac{\theta}{2})}\,
\prod_{n=1}^\infty\left(\sqrt{1+\Big(\frac{4\rho}{n}\Big)^2}
+\frac{4\rho\cosh(\theta)}{n}\right)^{-1}\,
\displaystyle\re^{\frac{4\rho}{n}\cosh(\theta)}\nonumber\\[.4cm]
&=&\exp\left(4\rho\,\theta\sinh\theta-\int_{-\infty}^{\infty}
\frac{\rd \tau}{2\pi}\ \frac{\log(1-\re^{-8\pi
    \rho\cosh(\tau)})}{\cosh(\theta-\tau)}\right)\,.
\eea
Note that $\sqrt{\lambda}\,\Zbs(\lambda)$ is a meromorphic function of $\lambda$.
It satisfies the functional relation 
\beq
\Zbs(\lambda)\,\Zbs(\re^{\ri\pi}\lambda)=
\frac{1}{2\ri\,\sin\big(2\pi\rho(\lambda-\lambda^{-1})\big)}
\eeq
and has the following asymptotics at large real $\theta$
\beq\label{Zass}
\log\big(\Zbs(\re^\theta)\big)%
\raisebox{-2.pt}{$\Big\vert$}_{\theta\to\pm\infty}= 
4\rho\,\theta\sinh(\theta)-
\frac{\re^{\mp\theta}}{4\rho}\ 
{\mathfrak f}_{\rm B}(8\pi\rho)+O\big(\re^{\mp 3\theta}\big)\,,
\eeq
where ${\mathfrak f}_{\rm B}(\beta)$ is the free-boson free energy,
which has already appeared in \eqref{free-boson}. The 
function $\Zbs(\lambda)$ was originally introduced in \cite{Lukyanov:2000jp}.
To parameterize the connection coefficients introduce a set of functions 
${\Qbs}^{(1)}_{\sigma'}(\lambda)$, ${\Qbs}^{(2)}_{\sigma}(\lambda)$,  
${\Qbs}^{(3)}_{\sigma\sigma'}(\lambda)$ of the variable $\lambda$,
which are analytic everywhere, except the points 
$\lambda=0$ and $\lambda=\infty$. They are defined as
\bea
\boldsymbol{S}^{(2,3)}_{-\sigma',\sigma''}(\lambda)&=&-\frac{\ri\sigma'}
{\sqrt{2}}
\ 
\Qbs^{(1)}_{\sigma'}(\sigma''\,\lambda)\ \Zbs(\sigma''\,\lambda)\,
\re^{\ri\pi\sigma''\rho(\lambda-\lambda^{-1})}
\nonumber \\[.3cm]
\boldsymbol{S}^{(3,1)}_{-\sigma'',\sigma}(\lambda)&=&-\frac{\ri\sigma''}
{\sqrt{2}} 
\ 
\Qbs^{(2)}_{\sigma}(\sigma''\,\lambda)\ 
\Zbs(\sigma''\,\lambda)\,
\re^{\ri\pi\sigma {\newp}_1}
\label{Ws-def}\\[.3cm]
\boldsymbol{S}^{(1,2)}_{-\sigma,\sigma'}(\lambda)&=&\frac{\ri\sigma}
{2}
\ 
\Qbs^{(3)}_{\sigma',\sigma}(\lambda\,\re^{\frac{\ri
    \pi\delta}{2}})\ \re^{\ri\pi\sigma' {\newp}_2}\ ,\nonumber
\eea
where 
$\sigma,\sigma',\sigma''=\pm1$. Note that
$\Qbs^{(3)}_{\sigma'\sigma}(\lambda)$ is, actually, a function of $\lambda^2$.

Specializing Eq.\eqref{S-rel2} for different choices of the
indices $(i,j,k)$ and using \eqref{NS-rel2} one can 
obtains a number of important 
functional relations between $\Qbs$'s (see Appendix \ref{app:A} for details)
\begin{subequations}\label{fr-all}
\begin{align}
\label{fr11}
2\,
\Qbs^{(1)}_{\sigma}(\lambda)\,
\Qbs^{(1)}_{\sigma}(-\lambda)&=
	\re^{2\pi\ri {\newp}_1}\,\Qbs^{(3)}_{\sigma-}(\lambda q)\,
\Qbs^{(3)}_{\sigma+}(\lambda/ q)- 
\re^{-2\pi \ri {\newp}_1}\,\Qbs^{(3)}_{\sigma+}(\lambda q)
\,\Qbs^{(3)}_{\sigma-}(\lambda/ q)\\[.5cm]
\label{fr22}
2\,
\Qbs^{(2)}_{\sigma}(\lambda)\,
\Qbs^{(2)}_{\sigma}(-\lambda)&=
\re^{2\pi\ri {\newp}_2}\,\Qbs^{(3)}_{-\sigma}(\lambda q)\,
\Qbs^{(3)}_{+\sigma}(\lambda /q)- 
\re^{-2\pi \ri {\newp}_2}\,\Qbs^{(3)}_{+\sigma}(\lambda q) \,
\Qbs^{(3)}_{-\sigma}(\lambda /q)\\[.5cm]
2\,
\Qbs^{(1)}_{\sigma'}(\lambda)\,
\Qbs^{(2)}_{\sigma}(-\lambda)&=
\re^{-\ri\pi ({\newp}_1 \sigma-{\newp}_2 \sigma'
-\rho(\lambda-\lambda^{-1}))}\,\Qbs^{(3)}_{\sigma'\sigma}(\lambda
q)\, 
+\re^{\ri \pi ({\newp}_1 \sigma-{\newp}_2 \sigma' -\rho(\lambda-\lambda^{-1}))}
\,\Qbs^{(3)}_{\sigma'\sigma}(\lambda/q)
\label{fr12}\\[.5cm]
\Qbs^{(3)}_{\sigma'\sigma}(
\lambda)\,{{\Tbs}^{(3)}
(\lambda)}&=
\re^{\ri\pi({\newp}_1\sigma+{\newp}_2\sigma'-
\rho(\lambda-\lambda^{-1})\cos({\frac{\pi\delta}{2}}))}\ 
\Qbs^{(1)}_{\sigma'}(\lambda q)
\Qbs^{(2)}_{\sigma}(\lambda/q)\nonumber\\[.5cm]
&\qquad \qquad \qquad+
\re^{-\ri\pi({\newp}_1\sigma+{\newp}_2\sigma'-\rho
(\lambda-\lambda^{-1})\cos({\frac{\pi\delta}{2}}))}\ 
\Qbs^{(1)}_{\sigma'}(\lambda/q)
\Qbs^{(2)}_{\sigma}(\lambda q))\,,\label{fr33}
\end{align}
where $q=\re^{\ri\frac{\pi\delta}{2}}$ 
and 
\begin{align}
\Tbs^{(3)}(\lambda)&=
\frac{\re^{+\rho(\lambda+\lambda^{-1})\sin(\frac{\pi\delta}{2})}}{2\ri}\ \Big(
\re^{-2\pi \ri {\newp}_2}\  \Qbs_{+}^{(1)}(\lambda/ q)\,
\Qbs_{-}^{(1)}(\lambda q)-
\re^{2\pi \ri {\newp}_2}\  \Qbs_{+}^{(1)}(\lambda q)\,
\Qbs_{-}^{(1)}(\lambda /q)\Big)\label{t3-def1}\\[.5cm]
&=
\frac{\re^{-\rho(\lambda+\lambda^{-1})\sin(\frac{\pi\delta}{2})}}
{2\ri}\Big(
\re^{-2\pi \ri {\newp}_1}\  \Qbs_{+}^{(2)}(\lambda q)\,
\Qbs_{-}^{(2)}(\lambda/q)-
\re^{2\pi \ri {\newp}_1} \ \Qbs_{+}^{(2)}(\lambda/ q)\,
\Qbs_{-}^{(2)}(\lambda q)\Big)\,.\label{t3-def2}
\end{align}
\end{subequations}

\subsection{Asymptotic expansions\label{as-exp}}
The asymptotic
behavior of the $\Qbs$-functions  
can be found from the WKB analysis of the linear differential equations
\eqref{auxlin}. The required calculations are similar to those of the
Fateev model (see Sect.\,9 of Ref.\!\cite{Bazhanov:2013cua}). 
For ${\Re}e(\theta)\to\pm\infty$ and $\big|{\Im m}
(\theta)\big|<\frac{\pi}{2}$, one obtains 
\bea\nonumber
\log\Big(\Zbs\big(\re^\theta\big)\,
\Qbs_\s^{(i)}\big(\re^\theta\big)\Big)\raisebox{-2.pt}{$\Big\vert$}_{{\Re e}(\theta)\to\pm\infty}
&=&
2\rho\,C^{(i)}_{-1}\,\cosh(\theta)
-\s |k_{3-i}|\,\theta-\tshf\log\sin(\pi |k_{3-i}|)\\[.4cm]
&\pm& {\textstyle\frac{1}{4}}\,\s\,
\log\big({\mathfrak
  S}_{3-i}\big)+C_{1}\,\re^{\mp\theta}+O\big(\re^{\mp2\theta/(1+\delta)}\big)\, ,
\label{ass1}
\eea
where $k_1$ and $k_2$ are the same as in \eqref{r-def}.
The leading coefficient is given by 
\beq\label{cm1}
C^{(i)}_{-1}=-\gamma_E-\psi(a_j/2)\,,
\eeq
where 
$\gamma_E$ is the Euler constant and $\psi(x)=\partial_x\log\Gamma(x)$. 
The constant terms (see Eq.\,(9.19) of Ref.\!\cite{Bazhanov:2013cua})
\beq
	\sqrt{{\mathfrak S}_i}=
\left(\frac{\rho}{a_i}\right)^{-2 |k_i|}\ 
\frac{\Gamma(1+|k_i|)}{\Gamma(1-|k_i|)}\ 
\frac{\exp(\eta^{(i)}_{\rm reg}) }{a_i |k_i|}\  
\left|\frac{z_{jk}}{z_{ji}z_{ik}}\right|^{-a_i |k_i|}\,\qquad\ \ 
(z_{ij}=z_i-z_j)\,,
\eeq
are expressed through the regular parts $\eta^{(i)}_{\rm reg}$ of the
expansions of the solution
\bea\label{asym12a}
\eta(z)=-(1-a_i|k_i|)\log|z-z_i|+\eta^{(i)}_{\rm reg}+o(1)\ \ \ \ \  {\rm
  as}\ \ \ \ \ \  \ \ |z-z_i|\to  0\ \ \ \ \ \ (i=1,2)
\eea
near the punctures (see \eqref{asym12} and footnote\,\ref{f:sigma} on page 
\pageref{f:sigma}). 

Further, taking the limit $a_3\to0$ in the results
\cite{Lukyanov:2013wra} and   
\cite{Bazhanov:2013cua}, devoted to the symmetric regime
\eqref{sregime} of the Fateev model,\footnote{Namely, one needs to
  combine Eqs.\,(5.27), (6.1) and (9.26) of Ref.\cite{Bazhanov:2013cua}
with (4.27) of Ref.\cite{Bazhanov:2016glt} and then take the limit
$a_3\to0$.} 
one can express 
the coefficient $C_1$ through the solution $\hat\eta(w)$ of \eqref{sinh-eq},
satisfying the
asymptotic conditions \eqref{osasail},\,\eqref{asym-z3},
\beq\label{C1-pde}
C_1=\frac{1}{4\rho}
\left(-\frac{8}{\pi}\,\int_{{\mathbb D}_{\rm BL}}\rd^2 w\sinh^2(\hat\eta)+
2\,\sum_{i=1}\,a_i\, (|k_i|-\tshf)^2\right)\,,
\eeq
where the integral is taken over domain ${\mathbb D}_{\rm BL}$ shown
in Fig.\ref{fig3sgsg}. 
The asymptotic expansion of  $\Qbs_{\sigma'\sigma}^{(3)}$ is simply
determined by \eqref{ass1} and the relation \eqref{fr12}.

\subsection{Connection to the Bethe ansatz}
Remarkably, it turns out that the zeroes of the functions
$\Qbs^{(i)}(\lambda)$ (introduced above as 
connection coefficients for the differential operators \eqref{auxlin})
satisfy exactly the same equations \eqref{bag1},\,\eqref{bag2}, as the
zeroes of the functions $\Asf^{\rm (BL)}_i(\lambda)$, arising in the
Bethe ansatz description of the vacuum state 
of the BL model. 
To make a precise correspondence let us first 
relate the dimensionless parameter 
$r=MR$, used in Sects.\,1-3, to the parameter $\rho$, used 
in \eqref{sinh-Gordon} and further down in this section,
\beq\label{rho-r}
\rho=\frac{r}{4\pi}\  \cos\Big(\frac{\pi\delta}{2}\Big)\ .
\eeq
Next, recall that the coefficients $\Qbs^{(i)}(\lambda)$
depend on the parameters ${\newp}_1,{\newp}_2$ defined in 
\eqref{r-def}. To indicate this dependence explicitly, we will write these
coefficients as
$\Qbs^{(i)}(\lambda\,|\,{\newp}_1,
{\newp}_2)$.
Similarly, the functions $\Asf^{\rm  (BL)}_i(\lambda)$, introduced 
in Sect.\,\ref{sec::BAscale}, depend on the twist parameters $p_1$ and
$p_2$, so we will write them as $\Asf^{\rm  (BL)}_i(\lambda\,|\,p_1,p_2)$.

Below we are going to establish that   
\bea\nonumber
\Qbs_{\s}^{(1)}(\lambda\,|\,{\newp}_1,{\newp}_2)&=&
\Asf^{\rm (BL)}_i(\lambda\,|\,{\newp}_1,\s {\newp}_2)\\[.3cm]
\Qbs_{\s}^{(2)}(\lambda\,|\,{\newp}_1,{\newp}_2)&=&
 \Asf^{\rm (BL)}_i(\lambda\,|\,\s {\newp}_1,{\newp}_2)\label{ident}\\[.3cm]
\Qbs_{\s'\s}^{(3)}(\lambda\,|\,{\newp}_1,{\newp}_2)&=&
\Asf^{\rm (BL)}_3(\lambda\,|\,\s {\newp}_1,\s' {\newp}_2)\,,\nonumber
\eea
where $\s,\s'=\pm$. Recall that here we assume that $p_1$ and $p_2$
are positive, whereas in Sect.\,\ref{CBA} and Sect.\,\ref{lat-type} they were
taking both signs. The correspondence is achieved by using the sign
variables $\s,\s'$ in RHS of \eqref{ident}.  
Moreover, in writing \eqref{ident} 
we have assumed a particular choice of normalization
factors and
the coefficients $\alpha_i,\beta_i$ in \eqref{ABL}, which so far were
at our disposal since they do not affect
the position of zeroes of $\Asf^{\rm (BL)}(\lambda)$. %
With this identification it is easy to check the 
that the functional equations \eqref{fr-all} 
imply all the BAE \eqref{bag1} and \eqref{bag2} in their scaling limit 
form (recall that the latter is obtained with the help of 
substitutions \eqref{scale1} and
\eqref{scale2}). For instance, \eqref{fr12} immediately leads the
scaling limit of \eqref{e1} and \eqref{e12}. Similarly, \eqref{fr33}
leads to the scaling limit of \eqref{ba-new}. Finally, setting
$\lambda=q\,\lambda^{(3)}_n$ and $\lambda=q^{-1}\,\lambda^{(3)}_n$ in
\eqref{fr12}  and combining the resulting relations one arrives 
to \eqref{e2}. Eqs.\eqref{e22} and \eqref{dual} are derived in a
similar way.

To complete the identification
\eqref{ident} one also needs to check that the
zeroes of the connection coefficients $\Qbs^{(i)}(\lambda)$, 
determined by the differential equations \eqref{auxlin},  
have precisely the same phase
assignments \eqref{phases} as the vacuum state zeroes of the BAE. 
In particular, that the phases are given by consecutive integers
containing no ``holes'' in their distribution. 
Alternatively, it is enough to prove that $\Qbs^{(i)}(\lambda)$ and 
$\Asf^{(i)}(\lambda)$ have the same loci of zeroes. We have verified
this statement in the small-$r$ limit, when the differential equations
\eqref{auxlin} simplify considerably and become ODE's with explicity
known algebraic potentials (see Appendix\,\ref{app:B}).  
Using asymptotic (WKB) analysis of these ODE's we show that the
corresponding distributions of zeroes precisely match \eqref{ZBL} in
the small-$r$ limit. Moreover we have confirmed this coincidence 
by extensive numerical checks. On this basis we assume \eqref{ident} to hold.

\section{\label{secnew4}Non-linear integral equations}
As is well known \cite{Klumper:1991,Destri:1992qk}, the BAE can be
transformed into 
NLIE
which allow one to accurately calculate vacuum energies as well as
eigenvalues of all higher integrals of motion. There are several
ways to proceed starting with different sets
of BAE discussed after Eq.\,\eqref{bag2}. 
It turns out that 
the set (III) appears to be most convenient for our purposes. 
Indeed, in this case the resulting equations admit a regular expansion for
small values $\delta$, which is very convenient for the comparison with
perturbation theory calculations of our previous paper \cite{Bazhanov:2016glt}.
The working is presented in  Appendix~\ref{app:C}.  
The approach is somewhat 
similar to that of the regime $a_i>0$ of the Fateev model, 
considered in details in Ref.\!\cite{Bazhanov:2013cua}.
However, the presence of 2-strings among the vacuum roots makes the
considerations rather tedious. As a final result, one obtains
the following system of two 
NLIE:
\bea\label{DDV}
\varepsilon_\sigma(\theta)=r\,\sinh(\theta-\ri\chi_\sigma)-2\pi k_\sigma+
\sum_{\sigma'=\pm}\,
\int_{-\infty}^{\infty}\frac{\rd\theta'}{\pi}\,
{G}_{\sigma\sigma'}(\theta-\theta')\
\Im m\Big[\log\big(1+e^{-\ri\varepsilon_{\sigma'}(\theta'-\ri 0)}\big)\Big]\, .
\eea
Here $\sigma=\pm$,  $(\chi_+,\chi_-)=(0,\,\pi a_1/2)$
and the kernels are given by the relations
\bea
G_{\pm\pm}(\theta)=G_{a_1}(\theta)+G_{a_2}(\theta)\ ,\ \ \ \ \ \ 
{ G}_{\pm\mp}(\theta)={\hat G}_{a_1}(\theta)-{\hat G}_{a_2}(\theta)
\eea
with
\bea
G_a(\theta)&=&
\int_{-\infty}^\infty\rd\nu \ \frac{\re^{\ri\nu\theta}\, \sinh(\frac{\pi\nu}{2}(1-a))}
{2\cosh(\frac{\pi \nu}{2})\sinh(\frac{\pi \nu a}{2})}\\
{\hat G}_a(\theta)&=&
\int_{-\infty}^\infty \rd\nu\ \frac{\re^{\ri\nu\theta}\, \sinh(\frac{\pi\nu }{2})}
{2\cosh(\frac{\pi \nu}{2})\sinh(\frac{\pi \nu a}{2})}\ .\nonumber
\eea
Once the numerical data for $\varepsilon_\pm(\theta)$ are available, 
any (regularized) sum over the Bethe zeroes can be calculated via
fastly converging integral representations (see
Appendix~\ref{app:D}). 
All information about the vacuum eigenvalues of
local and nonlocal integrals of motion
is contained in the
coefficients of the asymptotic expansions, 
\begin{subequations}\label{Qasym}
\bea
&&\log\big({\Asf}_i^{\rm (BL)}(\re^\theta)\big)\Big|_{\theta\to+\infty}
\asymp-4\,\rho\, \theta
  \sinh(\theta)+2\rho\, C_{-1}^{(i)}\,\cosh(\theta) 
-k_j\theta -\tshf\log\big(\sin(\pi k_j)\big)\nonumber
\\[.4cm]
&&\ \ \ \ \ \ \ \ \ \ \ \ \ \ \ \ \ \ 
-\sum_{n=1}^\infty\,\frac{\re^{-(2n-1)\theta}\,I_{2n-1}\,}
{\big(M\cos(\frac{\pi\delta}{2})\big)^{2n-1}}
+\sum_{n=0}^\infty\,\frac{(-1)^n\re^{-{2n\theta}/{a_j}}\,H_n^{(j)}}
{2\cos(\frac{n \pi}{a_j})\,\big(M\cos(\frac{\pi\delta}{2})\big)^{2n/a_j}}
\label{Qasymp}
\eea
and 
\bea
&&\log\big({\Asf}_i^{\rm (BL)}(\re^\theta)\big)\Big|_{\theta\to-\infty}
\asymp-4\,\rho\, \theta
  \sinh(\theta)+2\rho\, C_{-1}^{(i)}\,\cosh(\theta) 
-k_j\theta -\tshf\log\big(\sin(\pi k_j)\big)\nonumber
\\[.4cm]
&&\ \ \ \ \ \ \ \ \ \ \ \ \ \ \ \ \ \ 
-\sum_{n=1}^\infty\,\frac{\re^{+(2n-1)\theta}\,\bar I_{2n-1}\,}
{\big(M\cos(\frac{\pi\delta}{2})\big)^{2n-1}}
+\sum_{n=0}^\infty\,\frac{(-1)^n\re^{+{2n\theta}/{a_j}}\,\bar H_n^{(j)}}
{2\cos(\frac{n \pi}{a_j})\,\big(M\cos(\frac{\pi\delta}{2})\big)^{2n/a_j}}\ ,
\label{Qasymm}
\eea
\end{subequations}
where $(ij)=(12)$ or $(21)$, the parameter $\rho$ is defined in \eqref{rho-r}
and the coefficients $C_{-1}^{(i)}$ are
given by \eqref{cm1}.
The quantities $\{I_{2n-1},{\bar I}_{2n-1}\}$ and $\{H_n, {\bar H}_n\}$ are 
vacuum eigenvalues of the local and nonlocal integrals of motion (for
the vacuum state $I_{2n-1}={\bar I}_{2n-1}$). 
The above expansions for  ${\Asf}_1^{\rm (BL)}(\re^\theta)$ are
valid for $|{\Im} m(\theta)|<\pi$, while for 
${\Asf}_2^{\rm (BL)}(\re^\theta)$ they are valid only for 
$|{\Im}m(\theta)|<\pi/2$.

The asymptotic expansion of ${\Asf}_3^{\rm 
  (BL)}$ does not contain new coefficients, since it can
be obtained by combining \eqref{Qasym} and \eqref{fr12} (with an account of 
\eqref{ident}), 
\bea
\label{ass3}
&&\log\big(\Asf_3^{\rm (BL)}(\pm\,\ri\,\re^\theta)\big)\Big|_{\theta\to+\infty}
\asymp\frac{2\pi\rho\,\cosh(\theta)}{\cos({\frac{\pi\delta}{2})}} 
-\big(k_1+k_2\big)\theta-\tshf\log\big(\sin(\pi k_1)\big)
-\tshf\log\big(\sin(\pi k_2)\big)\nonumber\\[.3cm]
&&\ \ \ \ \ \ \ \ \ \ \ \ \ \ \ \ \ \ 
+\sum_{n=0}^\infty\,\frac{\re^{-{2n\theta}/{a_1}}\,H_n^{(1)}}
{2\cos(\frac{n \pi}{a_1})\,(M\cos\big(\frac{\pi\delta}{2})\big)^{2n/a_1}}
+\sum_{n=0}^\infty\,\frac{\re^{-{2n\theta}/{a_2}}\,H_n^{(2)}}
{2\cos(\frac{n \pi}{a_2})\,(M\cos(\frac{\pi\delta}{2})\big)^{2n/a_2}}
\,.
\eea
This expansion is valid for $|{\Im}m(\theta)|<\pi/2$. A similar expansion 
where $H_n^{(i)}$ are replaced by $\bar H_n^{(i)}$ holds for 
$\theta\to-\infty$.
 
The scaling function \eqref{F-def} 
can be computed by means of the relation 
\bea\label{F-ddv}
{\mathfrak F}(r,{\bf k})=\frac{R}{2\pi}\big(I_1+\bar I_1)=
\pm \frac{ r}{\pi}\,\Im m\Big(L_+(\pm \ri)+
\re^{\mp\frac{\ri\pi}{2}a_1}\,L_-(\pm\ri)\,
\Big)\ ,
\eea
which is valid for  both choices  of
the signs $\pm$, where
\beq\label{kajsjas}
L_\sigma(\nu)=\int_{-\infty}^\infty\frac{\rd \theta}{\pi}\  \re^{-\ri \nu \theta}\,
\log\Big(1+\re^{-\ri\varepsilon_\sigma(\theta-\ri 0)}\Big)\ .
\eeq
Moreover, it is worth noting that the lattice-type
regularization formula \eqref{ceff-lat}, written via the integral
\eqref{EN12}, leads 
precisely to the same expression \eqref{F-ddv}.
On the other hand, in view of the identification \eqref{ident}, 
one can compare $\re^{\mp\theta}$ terms of the expansions \eqref{ass1}
and \eqref{Qasym}. Then, taking into account \eqref{Zass}, one arrives
to the alternative expression \eqref{F-sinh} for the scaling function
in terms of solutions of the classical sinh-Gordon equation
\eqref{sinh-eq}, stated in  Introduction. 

As demonstrated in \cite{Bazhanov:2016glt} the exact formula
\eqref{F-ddv} is in a perfect agreement with the results of renormalized
perturbation theory and conformal perturbation theory.
It is possible to show that
\eqref{F-ddv} implies the following large-$r$ asymptotics,
\bea\label{hauasyu}
{\mathfrak F}(r,{\bf k})&=&{\mathfrak F}_{0}(r,{\bf k})+
{\mathfrak f}_{ \rm B}(2r)-{\mathfrak f}_{\rm B}\big(2r c\big({\textstyle\frac{\delta}{2}}\big)\big)
+
\frac{16 r}{\pi^2}\ \sum_{i=1}^2\int_{-\infty}^\infty \frac{\rd\nu}{2\pi}\
\\
&\times&
 \Big(c^2(k_1)\,c^2(k_2)-c^2(k_i)\,\cosh^2\big({\textstyle \frac{\pi\nu}{2}}\big) \Big)\ 
K_{\ri\nu}(r)K_{1-\ri\nu}(r)
\ \frac{\sinh(\frac{\pi\nu}{2}(1-a_i))}
{\cosh(\frac{\pi\nu}{2})\sinh(\frac{\pi\nu}{2} a_i)}+o\big(\re^{-2r}\big)\, ,\nonumber
\eea
where $K_\alpha(x)$ is the modified Bessel function, 
$k_1=k_++k_-,\,k_2=k_+-k_-$ and $c(x)\equiv\cos(\pi x)$.

Besides the vacuum energies the non-linear integral
equations allows one to determine the vacuum eigenvalues of all
higher Integral of Motions (IM). In particular, the vacuum eigenvalues
of the local IM, \ $\big\{{ I}_{2 n-1},\, {\bar  I}_{2 n-1}\big\}$ for $n>1$,
are given by the formula generalizing \eqref{F-ddv},
\beq\label{In-def}
I_{2n-1}={\bar I}_{2n-1}=\pm\frac{(M \cos(\frac{\pi\delta}{2})\big)^{2n-1}}
{\cos\big(\frac{\pi\delta}{2}(2n-1)\big)}\,\Im m\Big(L_+\big(\pm\ri (2n-1)\big)+
\re^{\mp\frac{\ri\pi}{2}(2n-1) a_1}\,L_-\big(\pm\ri (2n-1)\big)
\Big)\, . 
\eeq
Some details concerning small- and large-$R$ behavior of $I_{2n-1}$,  along with   numerical data for $I_{3}$ and $I_5$
can be found  in Appendix\,\ref{app:E}.

\begin{figure}
[ht!]
\centering
\includegraphics[width=15cm]{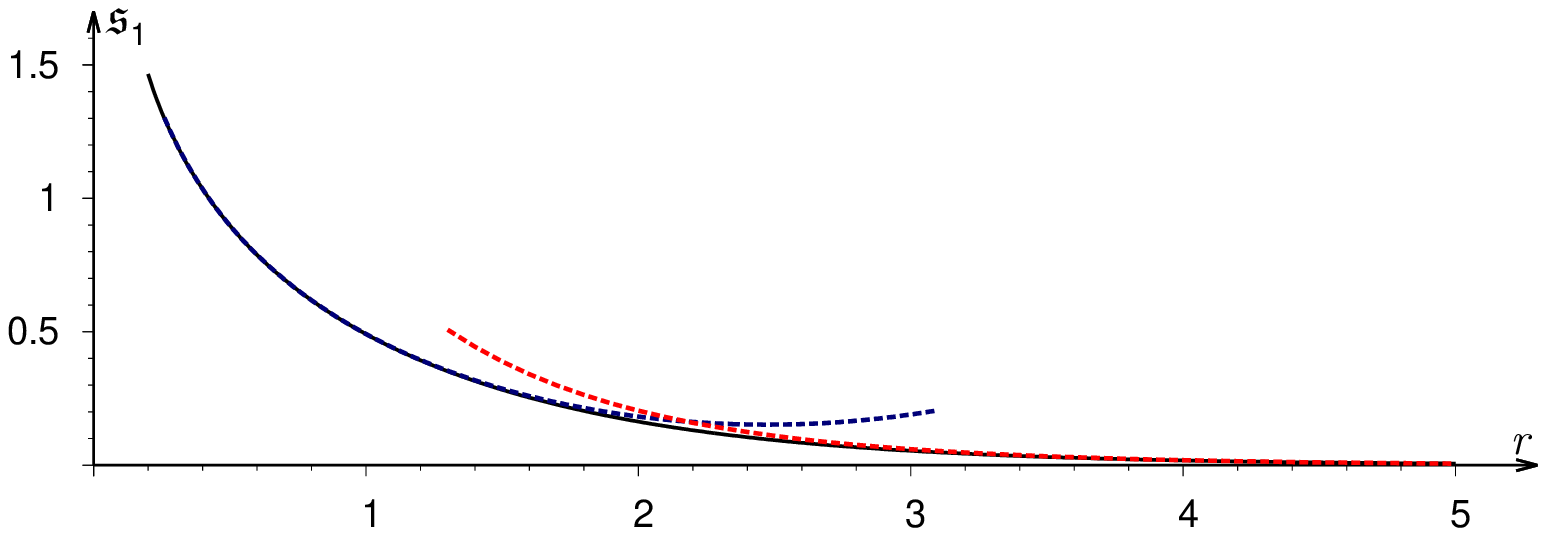}
\includegraphics[width=15cm]{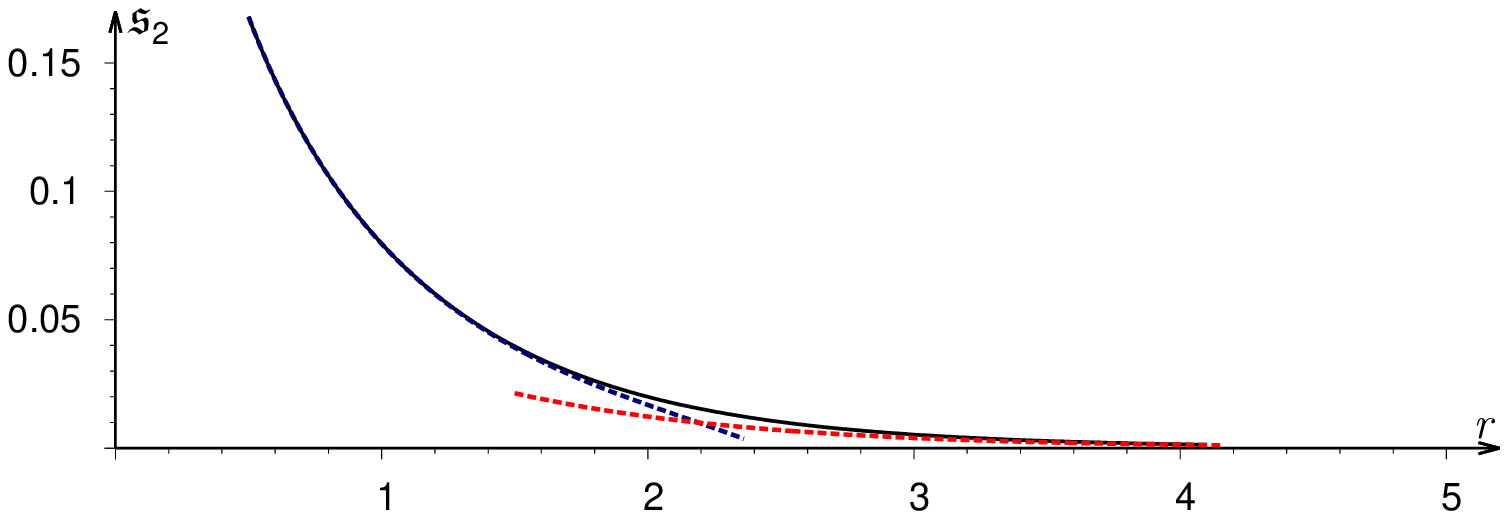}
\caption{
${\mathfrak s}_i=\frac{1}{2}\ \log ({\mathfrak S}_i)\ (i=1,2)$ vs $r$ for 
$\delta=\frac{17}{47}=0.36\ldots,\ k_1=\frac{47}{150},\ k_2=\frac{47}{640}$.
The solid lines were obtained from numerical
integration of eqs.\,\eqref{DDV},\,\eqref{jashssjsa} (for the
 numerical data, see Tabs.\,\ref{tab3},\,\ref{tab4} in Appendix\,\ref{app:E}).
The dashed lines represent the large-$r$  approximation\   \eqref{large}
and the  small-$r$ expansions
${\mathfrak s}_1=  -0.6266666667\, \log(r)+0.4555032510 + 0.03493323398\, r^2 + 0.001340210\, r^4+O(r^6)$
and
${\mathfrak s}_2=-0.1468750000\, \log(r) +0.0613720808 + 0.01949676545\, r^2 - 0.001294871\, r^4+O(r^6)$.
}
\label{fig5aa}
\end{figure}

Among the nonlocal IM the special r${\hat{\rm o}}$le is played by the 
so-called reflection operators \cite{Lukyanov:2013wra}.
Their vacuum  eigenvalues ${\mathfrak S}_{i}$ $(i=1,\,2)$ are related 
to the scaling function   ${\mathfrak F}(r,{\bf k})$
\cite{Lukyanov:2013wra},
\bea\label{aassspsa}
\log\big( {\mathfrak S}_{i}\big)
=\frac{2}{a_i}\ \int_{r}^\infty \frac{\rd r}{r}\ \frac{\partial }{\partial k_i}\ {\mathfrak F}(r,{\bf k})\ ,
\eea
and can be calculated through the solution of the system\,\eqref{DDV}:
\bea\label{jashssjsa}
\log\big( {\mathfrak S}_{i}\big)=2H^{i}_0=-2\bar H^{(i)}_0=
\frac{2}{ a_i}\, \Im m\Big(L_+(0)+(-1)^{i-1}\, L_-(0)\Big)\ .
\eea
On the other hand, taking into account \eqref{ident} and 
comparing the constant terms in the expansions \eqref{ass1} and
\eqref{Qasym}, one arrives 
to an alternative expression \eqref{C1-pde}
of ${\mathfrak S}_i$ in terms of the 
regular part of the solution of the modified sinh-Gordon equation 
\eqref{sinh-Gordon},\,\eqref{asym123}.\footnote{%
\label{f:sigma}
The expression \eqref{C1-pde} can be rewritten in terms of the
function $\hat\eta(w,\bar w)$, satisfying
\eqref{sinh-eq} and \eqref{osasail}, 
\beq\label{sigma-alt}
{\mathfrak s}_i=\tshf\log({\mathfrak S}_i)
=\lim_{|w-w_i|\to 0}\big({\hat \eta}(w,{\bar
  w})-(2|k_i|-1)\,\log|w-w_i|\big)+\log\big(\gamma(|k_i|)\big)-(1-2|k_i|)\,\log(2)\,,
\eeq
where $i=1,2$ and $\gamma(x)=\Gamma(x)/\Gamma(1-x)$.}

The large-$r$ behavior of ${\mathfrak S}_{i}$   
follows immediately from  Eqs.\eqref{hauasyu},\,\eqref {aassspsa}:
\bea\label{large}
\log\big({\mathfrak S}_{i}\big)&=&\frac{1}{a_i}\ \log\big({\mathfrak S}^{(0)}_i\big)+
\frac{16}{a_i\pi}\  \sin(2\pi k_i) \, \int_{-\infty}^\infty
\frac{\rd\nu}{2\pi} \ 
K^2_{\ri\nu}(r)\ \bigg(
\ \frac{\cosh(\frac{\pi\nu}{2})\sinh(\frac{\pi\nu}{2}(1-a_i))}
{\sinh(\frac{\pi\nu}{2} a_i)} \nonumber\\[.3cm]
&&\qquad\qquad\qquad-\cos^2(\pi k_{3-i})\ \sum_{k=1}^2
\ \frac{\sinh(\frac{\pi\nu}{2}(1-a_k))}
{\cosh(\frac{\pi\nu}{2})\sinh(\frac{\pi\nu}{2} a_k)}\,\bigg)+o\big(\re^{-2r}\big)\  ,
\eea
where 
\beq
\log\big({\mathfrak S}^{(0)}_i\big)=2 \int_{-\infty}^\infty\frac{\rd\theta}{\pi}\,
 \Im m\bigg(\log\Big(1+\re^{2\ri \pi k_+-r\cosh(\theta)}\Big)
+(-1)^{i-1}\ \log\Big(1+\re^{2\ri \pi k_--r\cosh(\theta)}\Big)\bigg)\,.
\eeq
Using the  small-$r$ asymptotic expansion 
\bea\label{apspassp}
{\mathfrak F}(r,{\bf k})\asymp -\frac{1}{3}+2k_+^2+2k_-^2-4\delta\,k_+k_--
16\,\rho^2\, \log(\rho)-
\sum_{n=1}^\infty e_n(\delta)\ (2\rho)^{2n}\ ,
\eea
where a few first coefficients was calculated in
\cite{Bazhanov:2016glt} (see Eq.(2.19) therein) one also   finds
\bea\label{small}
\log\big({\mathfrak S}_{i}\big)\asymp 2 \log\big(S(p_i|p_{3-i})\big)+\sum_{n=1}^{\infty}\frac{\partial  e_n(\delta)}{\partial p_i}\ 
\frac{(2\rho)^{2n}}{2n}\ .
\eea
The first term in the RHS  reads explicitly \cite{Lukyanov:2013wra}
\bea
S(p_i|q)=\bigg(\frac{\rho}{a_i}\bigg)^{-\frac{4p_i}{a_i}}\ 
\frac{\Gamma(\frac{1}{2}+p_i+q)\Gamma(\frac{1}{2}+p_i-q)}
{\Gamma(\frac{1}{2}-p_i+q)\Gamma(\frac{1}{2}-p_i-q)}
\, \frac{\Gamma(1-2p_i)}{\Gamma(1+2p_i)}\,
 \frac{\Gamma(1+\frac{2p_i}{a_i})}{\Gamma(1-\frac{2p_i}{a_i})}\ .
\eea
In Fig.\,\ref{fig5aa} the numerical results for $\frac{1}{2} \log({\mathfrak S}_i)$  are compared
against the large- and small-$r$ asymptotic formulae \eqref{large} and
\eqref{small}.

\noindent
{}\ \ Finally,  there is an  infinite set of non-local IM,
whose vacuum eigenvalues\ $\big\{{ H}^{(i)}_{2 n-1},\!{\bar {
    H}}^{(i)}_{2 n-1}\big\}_{n=1}^\infty$ $(i=1,2),$
are given by the relations:
\bea\label{jassa}
H^{(i)}_n&=&+\frac{2}{a_i}\ \big(M \cos({\textstyle\frac{\pi\delta}{2}})\big)^{\frac{2n}{a_i}}\
\Im m\Big(\,L_+\big(+{\textstyle\frac{2\ri n}{a_i}}\big)+(-1)^{i-1}\,L_-\big(+{\textstyle\frac{2\ri n}{a_i}}\big)\,\Big)\\
{\bar H}^{(i)}_n&=&-\frac{2}{a_i}\ \big(M \cos({\textstyle\frac{\pi\delta}{2}})\big)^{\frac{2n}{a_i}}\
\Im m\Big(\,L_+\big(-{\textstyle\frac{2\ri n}{a_i}}\big)+
(-1)^{i-1}\,L_-\big(-{\textstyle\frac{2\ri n}{a_i}}\big)\,\Big)\ .\nonumber
\eea
Some further details can be found in Appendix\,\ref{app:E}.

\section{Summary}
This work completes our study of the Bukhvostov-Lipatov (BL) model,
started in \cite{Bazhanov:2016glt}. 
Besides its applications to the instanton calculus in the $O(3)$ non-linear
sigma model, the BL model provides an ideal testground to developing
new and refining existing methods for integrable QFT's. 
Our attention was mostly devoted to calculation of 
the scaling function \eqref{F-def} (which is simply related to the
so-called effective central charge) of the BL model in a finite volume
and the
quasiperiodic boundary conditions.  
The scaling function 
was computed in a variety of different ways. For the readers'
convenience let us briefly summarize our main results below: 
\begin{enumerate}[(i)]
\item
{\em the conformal perturbation theory\/} for 
a few first terms in the small-$r$ expansion of the scaling function;
\item
{\em the renormalized Matsubara perturbation theory\/} 
(recall that we consider the finite-volume theory) for two
non-trivial orders of the expansion of the scaling function  
in powers of the coupling constant;
\item {\em an exact formula for the scaling function\/} in terms
  of different sets of non-linear integral equations (NLIE) derived
  from the Bethe ansatz and functional relations;
\item
{\em an exact formula for the
scaling function\/} in terms of a special
solution of the classical  sinh-Gordon  equation, based
on the ODE/IQFT correspondence and the classical inverse scattering transform;
\item
{\em renormalization of the coordinate Bethe ansatz} results for the
  scaling function both with the lattice-type regularization and the
  simple momentum cut-off. 
\end{enumerate} 
Remarkably, all the above approaches perfectly agree to each
other. This allowed us to establish the ODE/IQFT correspondence
between the quantum Bukhvostov-Lipatov model and the classical
sinh-Gordon equation.   
Note that this correspondence 
has recently been extended to 
the strong coupling
regime of the BL model with $\delta>1$ \cite{Bazhanov:2017nzh}. 
In this case it provides a dual description of
the 2D sausage \cite{Fateev:1992tk}, which includes 
the $O(3)$-sigma model when $\delta\to\infty$.

It would be interesting to further develop the
``quantum-classical''  
duality, in particular, 
to explore its possible connections to the duality between
quantum and classical systems with a finite number of degrees of
freedom, studied in \cite{Gorsky:2013xba}.

\section*{Acknowledgment}
The authors thank R.\,Baxter, M.\,Batchelor, 
G.\,Dunne, A.\,Gorsky, M.\,Jimbo, I.\,Krichever,
J-M.\,Maillet, A.\,Marshakov, A.\,Pogrebkov and F.\,Smirnov for fruitful
discussions. 

\bigskip
\noindent
Research of S.L. was
supported by the NSF under grant number
NSF-PHY-1404056.

\newpage
\app{\label{app:A}}
Here we present details of derivation of the individual functional
relations \eqref{fr-all} for
the connection coefficients, starting from the main
functional relation (\ref{S-rel2}) for the connection matrices, 
\begin{equation}
\label{FRloop}
	\Ss{i,k}{}(\gla)\ \re^{-2\pi \ri {\newp}_k(\gla)\s_3}\ \Ss{k,j}{}(\gla q_k^{-1})\ \re^{-2\pi \ri {\newp}_j(\gla q_k^{-1})\,\s_3}\ \Ss{j,i}{}(\gla q_i) \ \re^{-2\pi \ri  {\newp}_i (\gla q_i)\,\s_3}=-{\boldsymbol I}\ .
\end{equation}
Take $(i,j,k)=(1,2,3)$ and express $\Ss{1,3}{}$ 
through $\Ss{1,2}{}$ and $\Ss{2,3}{}$ in two different ways: from the 
main relation (\ref{FRloop}) 
\begin{equation}
	\Ss{1,3}{\s,\spp}(\gla)=-\sum\limits_{s=\pm} \re^{-2\pi\ri {\newp}_1 \s-2\pi \ri {\newp}_2 s-2\pi \ri \rho\,(\gla-\gla^{-1})}\ \Ss{1,2}{\s,s}(\gla \re^{-\ri\pi\delta})\,\Ss{2,3}{s,\spp}(\gla)
\nonumber
\end{equation}
and from the simple  properties (\ref{NS-rel2}) of the connection
matrices, 
\begin{equation}
\label{trivial}
	\Ss{1,3}{\s,\spp}(\gla \re^{-\ri\pi\delta})=\sum_{s=\pm}\Ss{1,2}{\s,s}(\gla \re^{-\ri\pi\delta})\ \Ss{2,3}{s,\spp}(\gla \re^{-\ri\pi\delta})\ .
	\end{equation}
This system contains both $\Ss{1,2}{\s,+}(\gla \re^{-\ri\pi\delta})$ and  $\Ss{1,2}{\s,-}(\gla \re^{-\ri\pi\delta})$.
Choose $\spp=+$ in both equations, fix a particular sign $\sp=\pm$, 
and then exclude $\Ss{1,2}{\s,-\sp}(\gla \re^{-\ri\pi\delta})$ 
among the above two equations. In this way one obtains
\begin{align}
&\Ss{3,1}{-,-\s}(\gla \re^{-\ri\pi\delta})\ \Ss{2,3}{-\sp,+}(\gla)
	+
	\re^{2\pi\ri (\s {\newp}_1-\s' {\newp}_2)+2\ri\pi \rho\,(\gla-\gla^{-1})} 
\ \Ss{3,1}{-,-\s}(\gla)\ \Ss{2,3}{-\sp,+}(\gla \re^{-\ri\pi\delta})
\nonumber\\[.4cm]
&=
\s\Ss{1,2}{\s,\sp}(\gla \re^{-\ri\pi\delta})\ 
\left(
	\Ss{2,3}{\sp,+}(\gla \re^{-\ri\pi\delta})\ 
        \Ss{2,3}{-\sp,+}(\gla)\,
	-\Ss{2,3}{\sp,+}(\gla )\ \Ss{2,3}{-\sp,+}(\gla
        \re^{-\ri\pi\delta})
\ \re^{-4\pi \ri \sp {\newp}_2}\right)
\nonumber
\end{align}
which is equivalent to (\ref{fr33}) upon the substitution (\ref{Ws-def}).

Next, take main relation (\ref{FRloop}) with $(i,j,k)=(1,2,3)$ and express $\Ss{1,3}{}$ therein in terms of product of $\Ss{1,2}{}$ and $\Ss{2,3}{}$ using relation similar to (\ref{trivial}) with shifted spectral parameter $\lambda \mapsto \lambda \re^{\frac{\ri\pi\delta}{2}}$.
It follows then
\begin{equation}
	\sum_{\spp=\pm}
	 \re^{2\pi \ri \spp \rho (\lambda-\lambda^{-1})}\ \Ss{2,3}{\s,\spp}(\gla)\ \Ss{3,2}{\spp,\sp}(\gla)=
	-\sum_{\spp=\pm} \re^{-2\pi \ri {\newp}_1\spp-2\pi \ri {\newp}_2\s}\ \Ss{2,1}{\s,\spp}(\gla)\ \Ss{1,2}{\spp,\sp} (\gla 
	\re^{-\ri\pi\delta})\,.
	\nonumber
\end{equation}
Setting $\sigma=\sigma'$ and combining like terms  one obtains the relation
\begin{align}
	2\ri\ \sin\left(2\pi\rho\,(\gla-\gla^{-1})\right)&\ \re^{2\pi \ri \sp {\newp}_2}\ 
	\Ss{2,3}{\sp,+}(\gla)\ \Ss{2,3}{\sp,-}(\gla)\nonumber \\[.4cm]
	=&\ \Ss{1,2}{-,-\sp}(\gla)\ \Ss{1,2}{+,-\sp}(\gla \re^{-\ri\pi\delta})\ \re^{-2\pi \ri {\newp}_1}
 -\Ss{1,2}{+,-\sp}(\gla)\ \Ss{1,2}{-,-\sp}(\gla \re^{-\ri\pi\delta})\ \re^{2\pi \ri {\newp}_1}\nonumber
\end{align}
which is equivalent to (\ref{fr11}) upon the substitution
(\ref{Ws-def}). 
One can repeat the above steps excluding $\Ss{2,3}{}$ from
(\ref{FRloop}). This leads to the relation (\ref{fr22}).

Finally, 
express $\Ss{1,2}{\s,\sp} (\gla \re^{-\ri \pi \delta})$ from (\ref{FRloop}) as
\begin{equation}
	\Ss{1,2}{\s,\sp}(\gla \re^{-\ri\pi\delta})=-\sum_{\spp=\pm}\re^{2\pi \ri 
	(\sigma {\newp}_1+\sp {\newp}_2)+2\pi\ri\spp \rho\,(\gla-\gla^{-1})}\ \Ss{1,3}{\s,\spp}(\gla)\ \Ss{3,2}{\spp,\sp}(\gla)
	\nonumber
\end{equation}
and $\Ss{1,2}{\s,\sp}(\gla)$  from (\ref{trivial}) (with shifted spectral parameter)
\begin{equation}
	\Ss{1,2}{\s,\sp} (\gla) =\sum_{\spp=\pm}\Ss{1,3}{\s,\spp}(\gla)\ \Ss{3,2}{\spp,\sp}(\gla)\ .
	\nonumber
\end{equation}
Then the product $\Ss{1,3}{\s,-}\Ss{3,2}{-,\sp}$ 
can be excluded from the above equations, leading to 
\begin{align}
	 2\ri\, \sin \left(2\pi \rho\,(\gla-\gla^{-1})\right) \ \Ss{1,3}{\s,\spp}(\gla )&\ \Ss{3,2}{\spp,\sp}(\gla)\nonumber
	 \\[0.4cm]
 =&\ \re^{-2\pi \ri (\s {\newp}_1+\sp {\newp}_2)}\ \Ss{1,2}{\s,\sp}(\gla \re^{-\ri\pi\delta})
 +\re^{-2\pi \ri \spp \rho\,(\gla-\gla^{-1})}\ \Ss{1,2}{\s,\sp}(\gla)\nonumber
\end{align}
which is equivalent to (\ref{fr12}) upon the substitution (\ref{Ws-def}).


\app{Conformal field theory limit\label{app:B}}
To understand properties of the connection coefficients it is
instructive to first consider the small-distance limit $\rho\to0$. 
The linear equations \eqref{auxlin} can be reformulated as second
order ordinary differential equations. For example, making the substitution 
\beq
\label{confz}
	\massPsi=\left(
	\begin{array}{c}
	e^{\frac{\eta}{2}}\ \confpsi\\
	e^{-\frac{\eta}{2}}\ (\partial_z+\partial_z\eta)\  \confpsi
	\end{array}
	\right)
\eeq
in the first equation of \eqref{auxlin}, one 
obtains
\beq\label{confeq}
(\partial_z^2-u(z,\z)-\rho^2\,\lambda^2\, {\cal P}(z))\,\confpsi=0\,,\qquad
	u(z,\z)=\partial_z^2\, \eta- (\partial_z\,\eta)^2\ .
\eeq
Let $\confpsi^{(i)}_\pm$\ $(i=1,2,3)$ denote the bases of solutions of
this second order equation, which under the substitution \eqref{confz}
precisely correspond to the solutions $\massPsi^{(i)}_\pm$, defined
by \eqref{psi12},\,\eqref{psi3}.
Then the connection problem \eqref{S-def} can be rewritten as
\beq
\boldsymbol{\confpsi}^{(i)} =
\boldsymbol{\confpsi}^{(j)}\,\boldsymbol{S}^{(j,i)}(\lambda)  \ ,
\nonumber
\eeq
where $\boldsymbol{\confpsi}^{(i)}=(\confpsi^{(i)}_-
  ,\confpsi^{(i)}_+)$. Note also, that \eqref{confz} implies 
\beq
\label{detwr}
\det\big(\massPsi^{(i)}_\sigma,\massPsi^{(j)}_{\sigma'}\big)=
{\rm Wr}[\,\confpsi^{(i)}_\sigma,\,\confpsi^{(j)}_{\sigma'}\,]\, ,
\eeq
where ${\rm Wr}[f,g]=f\partial_z g-g\partial_z f$ denotes the usual
Wronskian.

Consider now the limit when $\rho\rightarrow
0$, but the product $\mu=\rho \lambda$ is kept fixed. 
In this limit the variable $\z$ completely decouples from \eqref{confeq}
and the potential term there simplifies
\beq
u(z,\z){\Big\vert}_{\rho\to 0}
=-\frac{z_{12}\,z_{23}\,z_{31}}{(z-z_1)(z-z_2)(z-z_3)}\left(
	\frac{{\newp}_1^2-\frac{1}{4}}{(z-z_1)\,z_{23}}
	+\frac{{\newp}_2^2-\frac{1}{4}}{(z-z_2)\,z_{31}}
	-\frac{1}{4(z-z_3)\,z_{12}}
	\right)\,,
	\label{T0}
\eeq
where $z_{ij}=z_i-z_j$.
The basis solutions, corresponding to \eqref{psi12} and \eqref{psi3}, become
\bea
	\confpsi^{(i)}_{\pm}(z)&\sim& 
\frac{1}
{\sqrt{2{\newp}_i}}\,
\re^{\mp\,\hf\eta^{(i)}_{\rm reg}\,\,\mp\, \ri \beta_i}\,
(z-z_i)^{\frac{1}{2}\pm {\newp}_i}\,,\quad \quad z\to z_i\, \quad (i=1,2)
\nonumber
\\[.3cm]
\label{psi3c}
\confpsi^{(3)}_{\pm}(z)&\sim& 
\frac{1}{\sqrt{2\mu}}\,
\left(\frac{z_{13}z_{32}}{z_{12}}\right)^{\mp \mu}\,
(z-z_3)^{\frac{1}{2}\pm \mu}\,,\quad\quad z\to z_3\,,\nonumber
\eea
where the constants $\eta^{(i)}_{\rm reg}$ should be computed at $\rho=0$,
while $\mu=\rho\lambda$ is kept finite.

Further, by a change of variables 
\beq
\label{confodemapx}
	\re^x=\frac{(z-z_3)(z_2-z_1)}{(z-z_1)(z_3-z_2)}\,,
\ \ \ \ \ \ \  \confPsi (x)= \confpsi (z)\,\left (\frac{\rd z}{\rd x}\right)^{-\frac{1}{2}}\ ,
\nonumber
\eeq
the differential equation \eqref{confeq},\,\eqref{T0} 
can be cast into a form more suitable for analytical and numerical analysis,
\beq
\label{diffa} \bigg[ -{\frac{\rd^2}{\rd x^2}}   +
p^2_1\   { \frac{\re^x}{1+\re^x}} +
\big({\textstyle\frac{1}{4}}-{\newp}_2^2\,\big) \ {\frac{\re^x}{(1+\re^x)^2}} +
\mu^2\  \big(1+ \re^x\big)^{\delta-1} \bigg]\, \confPsi(x) = 0\, .
\eeq
The branch cuts of the power function $\big(1+ \re^x\big)^{\delta-1}$ are
chosen to be on the lines, where $(1+ \re^x)$ is real and
negative. Therefore, there are no branch cuts in the $x$-plane, when
${\Re}e (x)<0$. The symmetry transformations $\wh\Omega_1$ and $\wh\Omega_3$
defined in \eqref{O-def} above act as follows: 
\bea
\wh\Omega_1:\qquad x&\to& x-2\pi\ri\,,\quad
\mu\to\re^{\ri\pi\delta}\mu\,\quad\ \mbox{for}\quad x>0\nonumber
\\[.1cm] 
\wh\Omega_3:\qquad x&\to& x+2\pi\ri\,,
\quad \mu\to\mu\quad\qquad \mbox{for}\quad x<0\,.
\nonumber
\eea
Introduce two bases of solutions $\confPsi_\pm^{(1)}$ and 
$\confPsi_\pm^{(3)}$, which are uniquely defined by their asymptotic
behavior 
\beq
\confPsi_{\pm}^{(1)}(x)
\sim\re^{\mp {\newp}_1 x}\quad\text{as}\quad x\rightarrow+\infty\,,\qquad\quad  
\confPsi_\pm^{(3)}(x)\sim \re^{\pm \mu x}
\quad\text{as}\quad x\rightarrow-\infty
\nonumber
\eeq
and symmetry transformation properties
\beq
\wh\Omega_1\big(\confPsi_{\pm}^{(1)}\big)
=\re^{\pm 2\pi \ri {\newp}_1}\,\confPsi_{\pm}^{(1)}\,,
\qquad\qquad
\wh\Omega_3\big(\confPsi_{\pm}^{(3)}\big)
=\re^{\pm 2\pi \ri \mu}\,\confPsi_{\pm}^{(3)}\,.
\nonumber
\eeq 
Note  that $\confPsi_-(x|\mu)=\confPsi_+(x|-\mu)$. 
With these definitions introduce the functions 
\beq
{\Wsf}^{(2)}_{\pm}(\mu)={\rm Wr}[\,\confPsi^{(1)}_\pm,
\confPsi^{(3)}_{+}\,]\,,\label{W2-def}
\eeq
defined as Wronskian determinants. 
Further, interchanging $z_1$ and $z_2$ in \eqref{confodemapx} and
proceeding as above one defines another pair of connection
coefficients 
\beq
{\Wsf}^{(1)}_\pm(\mu)\equiv{\Wsf}^{(1)}_\pm(\mu\,|\,{\newp}_1,{\newp}_2,\delta)=
{\Wsf}^{(2)}_\pm(\mu\,|\,{\newp}_2,{\newp}_1,-\delta)
\label{W1-def}
\eeq
which are obtained from ${\Wsf}^{(2)}$ by interchanging
${\newp}_1$ and ${\newp}_2$ and the negation of $\delta$.

\begin{figure}[!ht]
\centering
\includegraphics[height=8cm]{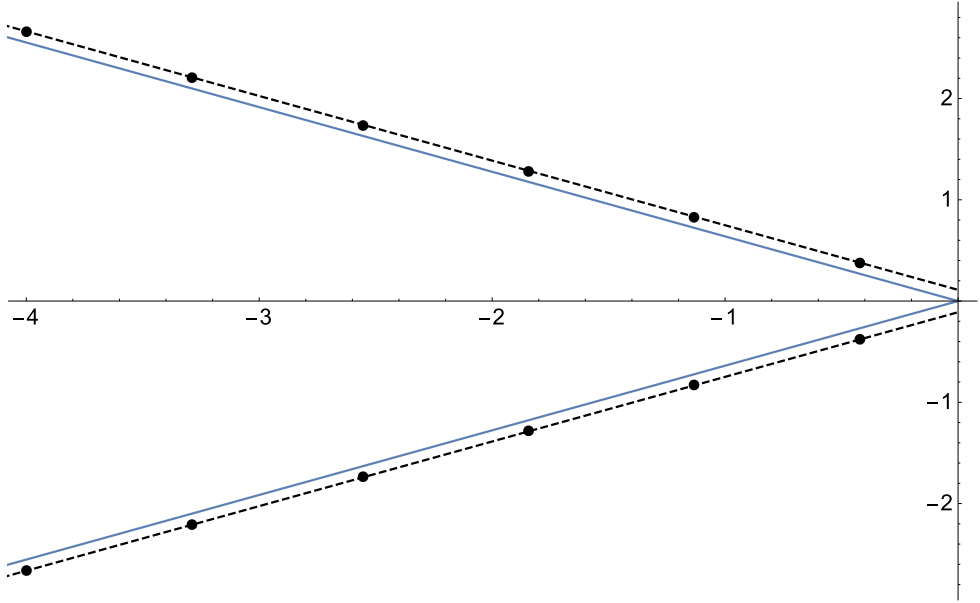}
\caption{Zeroes of the function $\Wsf^{(2)}_+(\mu)$ (shown by dots) 
for $\delta=17/47$, ${\newp}_1 = 1/10$, ${\newp}_2 = 1/20$ calculated
from \eqref{W2-def}. 
The dashed lines are shifted by $\pm \ri (\log2)/(2\pi)$ from the solid
lines where $\arg \mu=\pm \pi (1+\delta)/2$. The regularity of
the distribution of the ``2-string'' zeroes w.r.t. these lines is rather remarkable.}
\label{W2-zer}
\end{figure}
Yet another form of \eqref{diffa} 
\beq
\label{diffd} \bigg[ -{\frac{\rd^2}{\rd y^2}}   +
p^2_2\  { \frac{\re^y}{1+\re^y}} +
 {\frac{{\newp}_1^2}{(1+\re^y)}} +
{\frac{\re^y}{4(1+\re^y)^2}} +
\bar\mu^2\ \re^{(1-\delta)y}\big(1+ \re^y\big)^{-2} \bigg]\, 
\Theta(y) = 0
\eeq
is obtained by the substitution 
\beq
	\re^{y}=-1-\re^{-x},\quad\quad \Theta (y) = \confPsi (x)\, 
\left(\frac{\rd x}{\rd y}\right)^{-\frac{1}{2}}\,,\quad\quad
\bar\mu=\-\ri \,\mu  \,
\re^{\hf\ri\pi\delta} 
\nonumber
\eeq
which includes a phase rotation of the spectral parameter $\mu$. 
Similarly to the above, introduce the solutions
$\Theta^{(1)}_{\pm}(y)$ and $\Theta^{(2)}_{\pm}(y)$ 
defined by their asymptotics
\beq
\Theta^{(1)}_{\pm}(y)\sim\re^{\pm {\newp}_1 y}\quad {\rm as}\quad 
y\to -\infty\,,\quad\quad
\Theta^{(2)}_{\pm}(y)\sim \re^{\mp {\newp}_2 y}\quad {\rm as}\quad 
y\to+\infty
\nonumber
\eeq
and symmetry properties
\beq
\wh\Omega_1\big(\Theta_{\pm}^{(1)}\big)
=\re^{\pm 2\pi \ri {\newp}_1}\,\Theta_{\pm}^{(1)}\,,
\qquad\qquad
\wh\Omega_2\big(\Theta_{\pm}^{(2)}\big)
=\re^{\pm 2\pi \ri {\newp}_2}\,\Theta_{\pm}^{(2)}\,.
\nonumber
\eeq 
Now define the functions  
\beq
\Wsf^{(3)}_{\sigma',\sigma}(\bar{\mu})={\rm Wr}[\Theta^{(2)}_{\sigma'},\Theta^{(1)}_{\sigma}]\,\ \
\qquad
(\sigma,\sigma'=\pm)\ .\label{W3-def}
\eeq

From the standard analysis of the differential equation \eqref{diffa}
it is easy to see that the functions 
${\Wsf}^{(1,2)}_\pm(\mu)/\Gamma(1+2\mu)$ are
entire  functions of the complex variable $\mu$, while the
${\Wsf}^{(3)}_{\sigma',\sigma}(\mu)$ are entire
functions of the variable $\mu^2$.  
The
asymptotic expansion of ${\Wsf}^{(1,2)}_{\pm}(\mu)$ 
at large $\mu$ is given by
\beq
{\Wsf}^{(i)}_{\pm}(\mu)=-\Gamma(1\pm
k_j)\,\sqrt{\frac{a_j\mu}{\pi}}\,\Big(\frac{\mu}{a_j}\Big)^{\mp k_j}
\,D^{(i)}_\pm(\mu)\ \ \ \ \qquad (i=1,2)\, ,
\eeq
where $(i,j)=(1,2)$ or $(2,1)$,
\beq
\log D_\pm^{(i)}(\mu)\asymp\, C^{(i)}_{-1} \mu
+\sum_{n=1}^\infty\, C_{2n-1}\,\mu^{1-2n}-
\sum_{n=1}^\infty x_n(\pm {\newp}_j|{\newp}_i;a_j)\,\mu^{-\frac{2n}{a_j}}
\nonumber
\eeq
as $\mu\to\infty$ and $|\!\arg(\mu)|\leq \frac{\pi}{2}$. 
A few first expansion coefficients is known explicitly, in particular, 
\beq
C^{(i)}_{-1}=-\gamma_E-\psi(a_j/2)\,,\qquad
C_{+1}=\frac{1}{8}-\frac{{\newp}_1^2}{a_1}-\frac{{\newp}_2^2}{a_2}\ ,
\nonumber
\eeq
where $\gamma_E$ stands
for Euler constant, $\psi(x)\equiv\partial_x \log\Gamma(x)$ and
$a_1=2-a_2=1-\delta$. Moreover,
\beq
x_1(p|q;a)=\Big(\frac{2}{a}\Big)^{-\frac{2}{a}}\ \frac{\Gamma\big(\frac{1}{a}\big)
\Gamma\big(\hf-\frac{1}{a}\big)}{4\sqrt{\pi}}\ 
\frac{\Gamma\big(1+\frac{1}{a}+\frac{2p}{a}\big)}{
\Gamma\big(-\frac{1}{a}+\frac{2p}{a}\big)}\ \Big(\frac{q^2}{p^2-\frac{1}{4}}+
\frac{2-a}{2+a}\Big)\, .
\nonumber
\eeq

The coefficients ${\tt
  W}^{(3)}_{\sigma'\sigma}(\mu)\ \ (\sigma,\sigma'=\pm)$ are
entire functions of the complex variable  
$\mu^2$. Their asymptotic expansion at large $\mu$ is given by 
\bea
{\tt W}^{(3)}_{\sigma'\sigma}(\mu)=
\Gamma\big(1+ {\textstyle \frac{2\sigma {\newp}_1}{a_1}}\big)\ \Gamma\big(1+ {\textstyle \frac{2\sigma' {\newp}_2}{a_2}}\big)
\ \frac{\sqrt{a_1a_2}}{2\pi}\
\Big(\frac{\mu}{a_1}\Big)^{-\frac{2\sigma {\newp}_1}{a_1}}\
\Big(\frac{\mu}{a_2}\Big)^{-\frac{2\sigma'{\newp}_2}{a_2}}
{ D}^{(3)}_{\sigma'\sigma}(\mu)
\eea
and as $\mu\to\infty$ and $|\!\arg(\mu)|< \frac{\pi}{2}$
\bea
\log\big({ D}^{(3)}_{\sigma'\sigma}(\mu)\big)\asymp \frac{\pi\mu }{\sin(\frac{\pi a_1}{2})}+
\sum_{n=1}^\infty \Big(x_n(\sigma {\newp}_1|{\newp}_2; a_1)\,\mu^{-\frac{2n}{a_1}}+
x_n(\sigma' {\newp}_2|{\newp}_1; a_2)\,\mu^{-\frac{2n}{a_2}}\Big)\ .
\nonumber
\eea

\setlength{\arrayrulewidth}{.15mm}

\begin{table}[h]
\centering
\begin{tabular}{|c|c|c|}
\hline
\multicolumn{3}{|c|}{\bf{Zeroes of $W^{(1)}_+(\mu)$}}\\\hline
$n$& {$\mu_n$ (numerical)}&$\mu_n$ (asymptotic formula \eqref{jashsaysa}) \\ \hline
 1 & -0.32118644 & -0.28671875 \\
 2 & -0.78656554 & -0.78671875 \\
 3 & -1.28902812 & -1.28671875 \\
 4 & -1.78780916 & -1.78671875 \\
 5 & -2.28754568 & -2.28671875 \\
\hline
\multicolumn{3}{|c|}{\bf{Zeroes of $W^{(1)}_-(\mu)$}}\\\hline
$n$& \rm{$\mu_n$ (numerical)}&$\mu_n$ (asymptotic formula \eqref{jashsaysa}) \\ \hline
 1 & -0.25404467 & -0.21328125 \\
 2 & -0.71063637 & -0.71328125 \\
 3 & -1.21516013 & -1.21328125 \\
 4 & -1.71396329 & -1.71328125 \\
 5 & -2.21383761 & -2.21328125 \\
\hline
\end{tabular}
\caption{Zeroes of the functions $W^{(1)}_\pm$
for $\delta=17/47$, ${\newp}_1 = 1/10$, ${\newp}_2 = 1/20$ calculated from \eqref{W1-def}.}
\label{t:zerW1}
\end{table}

\begin{table}[ht]
\centering
\begin{tabular}{|c|c|c|}
\hline
\multicolumn{3}{|c|}{\bf{Zeroes of $W^{(2)}_+(\mu)$}}\\\hline
$n$& $\mu_n$ (numerical)&$\mu_n$ (asymptotic formula \eqref{assyaasys}) \\ \hline
1 & -0.41729701-0.37695011 i & -0.44259358-0.39285900 i \\
 2 & -1.12935605-0.82780500 i & -1.15306097-0.84640446 i \\
 3 & -1.83960684-1.28084953 i & -1.86352835-1.29994992 i \\
 4 & -2.54998314-1.73418285 i & -2.57399573-1.75349537 i \\
 5 & -3.28446342-2.20704109 i & -3.28446311-2.20704083 i \\
 6 & -3.99491662-2.66056721 i & -3.99493049-2.66058629 i \\
 7 & -4.70539788-3.11413172 i & -4.70539787-3.11413175 i \\
 8 & -5.41586525-3.56767720 i & -5.41586525-3.56767720 i \\
 9 & -6.12633265-4.02122266 i & -6.12633263-4.02122266 i \\
\hline
\end{tabular}
\caption{Zeroes of 
of the function $W^{(2)}_{+}$
for $\delta=17/47$, ${\newp}_1 = 1/10$, ${\newp}_2 = 1/20$ calculated from \eqref{W2-def}.}
\label{t:zerW2}
\end{table}

\begin{table}
\centering
\begin{tabular}{|c|c|c|}
\hline
\multicolumn{3}{|c|}{\bf{Zeroes of $W^{(3)}_{++}(\mu)$}}\\\hline
$n$& $\mu_n$ (numerical)&$\mu_n$ (asymptotic formula \eqref{gassatast}) \\ \hline
 1 & 0.58940086 i& 0.58444921 i \\
 2 & 1.42941557 i & 1.42734148 i  \\
 3 & 2.27132015 i & 2.27023375 i \\
 4 & 3.11381245 i & 3.11312602 i \\
 5 & 3.95650092 i & 3.95601829 i \\
 6 & 4.79927351 i & 4.79891057 i \\
 7 & 5.64208861 i & 5.64180284 i \\
 8 & 6.48492773 i & 6.48469511 i \\
 9 & 7.32778156 i & 7.32758738 i \\
 10 & 8.17064498 i & 8.17047965 i \\
\hline
\end{tabular}
\caption{Zeroes 
of the function $W^{(3)}_{++}$ 
for $\delta=17/47$, ${\newp}_1 = 1/10$, ${\newp}_2 = 1/20$ calculated from \eqref{W3-def}.}
\label{t:zerW3}
\end{table}

Let $\big\{\mu_n^{(i)}\big\}|_{n=1}^\infty$ $(i=1,2,3)$ denote the zeroes of 
$\Wsf^{(i)}(\mu)$. For sufficiently small ${\newp}_{1,2}\geq 0$ and large
$n\gg1$ they are distributed as follows:
\begin{itemize}
\item the zeroes of $\Wsf^{(1)}_{\sigma'}(\mu)$ are located on the
  real negative axis of variable $\mu$ 
\begin{equation}
\label{jashsaysa}
\mu^{(1)}_{n}\asymp-\frac{1}{2}\ 
\Big(n-\frac{1}{2}\pm\frac{2 {\newp}_2 \sigma'}{a_2}\Big)+o(1)\,\ \ \ \ \quad (n\gg1)\,,
\end{equation}
\item
the zeroes of 
${\Wsf}^{(2)}_\s(\mu)$ combine into complex conjugate pairs
$(\mu_n,\mu_n^*)$\  $(n=1,2,\ldots,\infty)$ in the left
half-plane ${\Re}e(\mu)<0$, located just outside the
wedge with an acute angle $\pi\delta$ centered around the negative
real axis of $\mu$ (see Fig.\,\ref{W2-zer})
\bea\label{assyaasys}
\mu_{n}\asymp
\re^{\frac{\ri\pi }{2}(1+a_1)}\ 
\cos\big(\frac{\pi\delta}{2}\big)\ \Big(n+ \frac{\sigma {\newp}_1}{a_1}
+\frac{{\newp}_2}{a_2}-\frac{1}{2}-\frac{\ri}{2\pi}\,\log (2)\,\Big)+o(1)\, \  \quad (n\gg1)\,,
\eea
\item
the zeroes $\big\{\mu_n^2\big\}_{n=1}^\infty$ of 
${\tt W}^{(3)}_{\sigma'\sigma}$ 
are simple, real and negative,
\bea\label{gassatast}
\mu_n\asymp \pm \ri\,  \cos\Big(\frac{\pi\delta}{2}\Big)\ \Big(n+\frac{\sigma {\newp}_1}{a_1}+\frac{\sigma' {\newp}_2}{a_2}-\frac{1}{2}\Big)+
o(1)\, \ \ \ \quad (n\gg1)\,.
\eea
\end{itemize}
The above asymptotic formulae are in a very good agreement with
the numerical values of the zeroes obtained from a direct solution 
of the differential equation \eqref{diffa},\,\eqref{diffd} even for
small values of $n$ (see Tables\,\ref{t:zerW1},\ref{t:zerW2},\ref{t:zerW3}).


\app{\label{app:C} Derivation of non-linear integral equations}

\begin{figure}[ht]
\centering
\includegraphics[width=17cm]{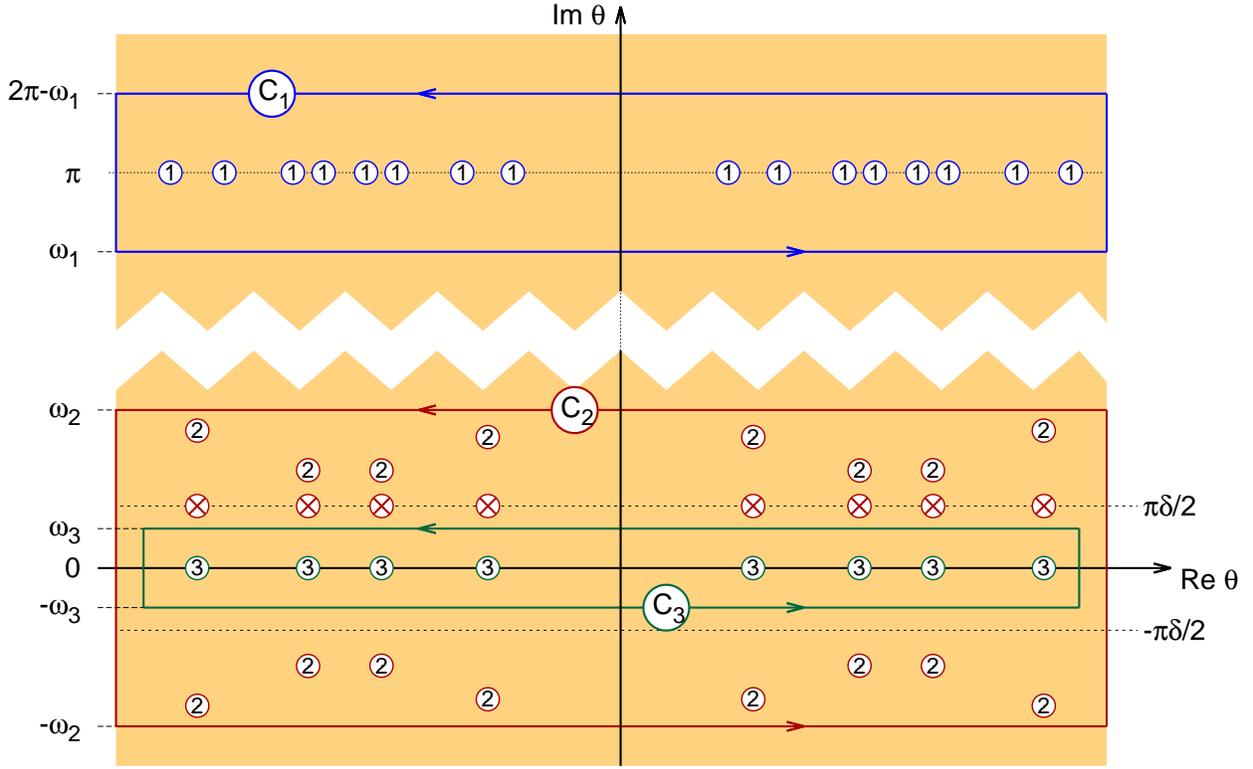}
\caption{Contours of integration involved in the derivation of the non-linear
  integral equations. The encircled numbers 1, 2 and 3 represent,
  respectively, the zeroes
  of ${\Asf}_1\big(\re^\theta\big)$, ${\Asf}_2\big(\!-\re^\theta\big)$ and 
${\Asf}_3\big(\re^\theta\big)$ in the complex plane of the variable $\theta$. 
The encircled crosses correspond to the positions of poles of
$f_1(\theta)$ 
  contributing to the integral over contour $C_2$ in \eqref{c2poles}.}
\label{plot_contours}
\end{figure}

As noted above one can derive several different, but equivalent
variants of the non-linear integral equations, starting with different BAE.
Below we will use the three equations \eqref{e1},
\eqref{ba-new} and \eqref{e12}.
Introduce two functions 
\bea
f_1(\theta)&=&
\displaystyle\re^{2\pi \ri (p_2-p_1)}\,
\left.\frac{{\mathsf f}(-\ri\lambda q^{-1})}
{{\mathsf f}(\ri \lambda q)}\,\frac{{\Asf}_3(\lambda
 q)}
{{\Asf}_3(\lambda\, q^{-1})}\right\vert_{\textstyle\lambda=\re^\theta}
\\[.5cm]
f_3(\theta)&=&\re^{-2\pi \ri (p_1+p_2)}\,
\left.\frac{{\mathsf f}(-\ri\lambda)}{{\mathsf
    f}(\ri\lambda)} 
\,\frac{{\Asf}_1(\lambda q^{-1})}{{\Asf}_1(\lambda \,q)} 
\,\frac{{\Asf}_2(\lambda \,q)}{{\Asf}_2(\lambda q^{-1})}
\right\vert_{\textstyle\lambda=\re^\theta}\nonumber
\eea
which appear in 
 the RHS's of \eqref{e1} and \eqref{ba-new}. The other relevant
 notations are given in \eqref{phi-def},\,\eqref{f-def},\,\eqref{A13-def} and \eqref{A2-def}. 
These functions are periodic with the
 period $2 \pi\ri$ 
\beq
f_1(\theta+2\pi\ri)=f_1(\theta)\,,\qquad 
f_3(\theta+2\pi\ri)=f_3(\theta)\nonumber
\eeq
and possess the following symmetry 
\beq
{f_1(\theta^*)}\, \big(f_1(\theta))^*=1\,,\qquad
{f_3(\theta^*)}\, \big(f_3(\theta))^*=1\,,\nonumber
\eeq
under the complex conjugation. Note that in the strip $-\pi<\Im m( \theta)\le\pi
$ the function $f_1(\theta)$ has following poles and zeroes,
\begin{subequations}\label{poles0}
\bea
\mbox{poles:\ } \ \theta&=&\theta_\ell^{(3)}+\tshf\,\ri \pi\,\delta\,,\qquad \theta=
\pm\Theta-\tshf\,\ri\pi\,(1+\delta)\label{poles1}\\[.2cm]
\mbox{zeroes:\ }\ \theta&=&\theta_\ell^{(3)}-\tshf\,\ri \pi\,\delta\,,\qquad \theta=
\pm\Theta+\tshf\,\ri\pi\,(1+\delta)\,.\label{poles2}
\eea
\end{subequations}
Taking the logarithms, one obtains
\begin{subequations}\label{lba0}
\begin{align}
\frac{1}{2\pi \ri}\log \big(f_1(\theta)\big)&=
N\,{\myp}(\theta,-\delta)-p_1+p_2-
\sum_{\ell} {\mye}_{2\delta}(\theta-\theta^{(3)}_\ell)\label{lba1}
\\[.2cm]
\frac{1}{2\pi \ri}\log\big( f_3(\theta)\big)&=
N\,{\myp}(\theta,0)-p_1-p_2+
\sum_{\JJ} {\mye}_{\delta}\big(\tshf(\theta-\theta^{(1)}_\JJ)\big)
-
\sum_{\JJ} {\mye}_{\delta}\big(\tshf(\theta-\theta^{(2)}_\JJ)\big)\,,\label{lba2} 
\end{align}
\end{subequations}
where $p(\theta,\delta)$ and $\mye_\alpha(\theta)$ are defined in \eqref{p-def}
and \eqref{e-def}. The indices $\JJ$ and $\ell$ run over the values 
$\JJ\in\{-N+1,-N+2,\ldots,N\}$, $\ell\in\{-\hf N+1,-\hf
N+2,\ldots,\hf N\}$ with $N\ge2$ being an even integer.
The (complex) numbers $\{\theta^{(1)}_\JJ\}$,\,$\{\theta^{(2)}_\JJ\}$ and 
$\{\theta^{(3)}_\ell\}$ stand for the zeroes of the functions
${\Asf}_1(\re^\theta)$, ${\Asf}_2(\re^\theta)$ and
${\Asf}_3(\re^\theta)$, 
as defined by \eqref{A13-def},\,\eqref{A2-def}.
With these notations BAE \eqref{e1},
\eqref{ba-new} and \eqref{e12} take the form
\beq
f_1(\theta^{(1)}_\JJ)=-1\,,\qquad f_1(\theta^{(2)}_\JJ+\ri \pi)=-1\,,\qquad 
f_3(\theta^{(3)}_\ell)=-1\, . \nonumber
\eeq
Fig.\,\ref{plot_contours} shows 
the contours $C_1$, $C_2$ and $C_3$ enclosing zeroes of 
${\Asf}_1(\re^\theta)$, ${\Asf}_2(\re^{\theta+\ri\pi})$ and ${\Asf}_3(\re^\theta)$, respectively.
The horizontal segments of these contours go
along the lines $\Im m(\theta)=\pm\omega_i \pmod{2\pi}$, $i=1,2,3$, with
the positive constants $\omega_1$, $\omega_2$ and $\omega_3$,
satisfying the inequalities 
\beq
0<\pi-\omega_1<\tshf\, \pi \delta-\omega_3\,,\qquad
\pi-\tshf\,\pi \delta-\omega_3>\omega_2>\omega_2^{\rm (min)}\,,
\qquad \omega_2 > \frac{\pi\delta}{2}+\omega_3\,,
\label{devi}
\eeq
where 
\beq
\omega_2^{\rm(min)}=\max_{\{\theta^{(2)}_\JJ\}}\Big|\Im m \,(\ri\pi-\theta^{(2)}_\JJ)\Big|\nonumber
\eeq
is the maximal deviation of the roots $\theta^{(2)}_\JJ$ from the line
$\Im m( \theta)=\pi$. The second inequality in \eqref{devi} ensures
that the contour $C_2$ in Fig.\,\ref{plot_contours} encloses all zeroes
of ${\Asf}_2(-\re^\theta)$.
Next, the sum in \eqref{lba1} and the first sum in \eqref{lba2} can be 
easily written as contour integrals, for instance  
\begin{subequations}\label{contours}
\beq
\sum_{\JJ} {\mye}_{\delta}\big(\tshf(\theta-\theta^{(1)}_\JJ)\big)=\frac{1}{2\pi \ri}
\oint_{{C_1}} \rd\theta'\, {\mye}_{\delta}\big(\tshf(\theta-\theta')\big)
\, \partial_{\theta'}\log\big(1+f_1(\theta')\big)
\eeq
and similarly 
\beq
\sum_{\ell} {\mye}_{2\delta}(\theta-\theta^{(3)}_\ell)=\frac{1}{2\pi \ri}
\oint_{{C_3}}\rd\theta'\,  {\mye}_{2\delta}(\theta-\theta')
 \,\partial_{\theta'}\log\big(1+f_3(\theta')\big)\,.
\eeq
\end{subequations}
The handling of the second sum in \eqref{lba2} 
is slightly more complicated,
since the contour $C_2$  (which is enclosing the zeroes of ${\Asf}_2(\re^{\theta+\ri \pi})$), also surrounds (unwanted) poles of $f_1(\theta)$ 
located on the line $\Im m(\theta)=\pi\delta/2$. Taking this into account one
obtains
\begin{align}\nonumber
\sum_{\JJ} {\mye}_{\delta}\big(\tshf(\theta-\theta^{(2)}_\JJ)\big)=&
\frac{1}{2\pi \ri}\oint_{{C_2}} \rd\theta'\, {\mye}_{\delta}\big(\tshf(\theta-\theta'+\ri\pi)\big)
\, \partial_{\theta'}\log(1+f_1(\theta'))\\
&+\frac{1}{2\pi \ri}\oint_{{C_3}}\rd\theta'\,  {\mye}_{\delta}\big(\tshf(\theta-\theta'-\tshf\ri\pi\delta)\big)
\, \partial_{\theta'}\log(1+f_3(\theta'))\, .
\label{c2poles}
\end{align}
Further considerations follow the standard way
\cite{Klumper:1991,Destri:1992qk} of deriving the non-linear integral
equations. Substitute \eqref{contours} back into \eqref{lba0}.
Integrating by parts, stretching the contours towards both infinities
along the real axis of $\theta$ and introducing the pseudo-energies
\beq
f_1(\theta-\ri \omega_1)=\re^{\ri\varepsilon_1(\theta)}\,,\qquad
f_1(\theta-\ri \omega_2)=\re^{\ri\varepsilon_2(\theta)}\,,\qquad
f_3(\theta-\ri \omega_3)=\re^{\ri\varepsilon_3(\theta)}\,,
\label{eps-def}
\eeq
one obtains 
\begin{align}
&\varepsilon_i(\theta)-\sum_j\, \int_{-\infty}^{\infty}\rd\theta'\,
R^{(+)}_{ij}(\theta-\theta')\,\varepsilon_j(\theta')=2\pi\,
{\myp}_i(\theta)-2\pi\,{\mathsf C}_i \label{preddv}\\[.6cm]
&\ \ \ \ \ \ \, +\,\frac{1}{\ri}\,\sum_j\,
\int_{-\infty}^{\infty}\rd\theta'\,\Big(R^{(+)}_{ij}(\theta-\theta')\,\log\big(1+
\re^{-\ri\varepsilon_j(\theta')}\big)\,\rd\theta'
-R^{(-)}_{ij}(\theta-\theta')\,\log\big(1+\re^{-\ri\varepsilon^*_j(\theta')}\big)\Big),\nonumber
\end{align}
where $i,j=1,2,3$,
\beq
{\mathsf C}_1={\mathsf C}_2=p_1-p_2\,,\qquad
{\mathsf C}_3=p_1+p_2
\nonumber
\eeq
and 
\beq
{\myp}_1(\theta)={\myp}(\theta-\ri \omega_1,-\delta)\,,\qquad
{\myp}_2(\theta)={\myp}(\theta-\ri \omega_2,-\delta)\,,\qquad
{\myp}_3(\theta)={\myp}(\theta-\ri \omega_3,0)\,.
\nonumber
\eeq
The integral operators $R^{(\pm)}_{ij}(\theta)$
are conveniently defined by their Fourier transforms. We use the
following general convention 
\beq\label{Fdef}
F(\theta)=
\frac{1}{2\pi}\int_{-\infty}^\infty \,d\nu\, \hat F(\nu)\,\re^{\ri \nu
  \theta}\,.
\eeq
The matrix elements $\hat R^{(\pm)}_{ij}(\nu)$ then take the form 
\beq
\hat R^{(\pm)}_{ij}(\nu)=d_i(\nu)\,\hat R_{ij}(\nu)\,d_j(\mp\nu)\,,
\nonumber
\eeq
where
\beq\label{Rdef2}
d_1(\nu)=\re^{(\omega_1-\pi)\nu}\,,\quad
d_2(\nu)=\re^{\omega_2\nu}
\,,\quad
d_3(\nu)=\re^{\omega_3\nu}\nonumber
\eeq
and 
\beq
\begin{array}{rclrcl}
\hat R_{11}(\nu)&=&\hat R_{12}(\nu)=0\,,\qquad
&\ds\hat R_{13}(\nu)&=&-\ds\frac{\sinh(\frac{\pi\nu(1-\delta)}{2})}
{\sinh(\frac{\pi\nu}{2})}\nonumber\\[.6cm]
\hat R_{21}(\nu)&=&\hat R_{22}(\nu)=0\,,
&\ds\hat R_{23}(\nu)&=&\phantom{+}\ds\frac{\sinh(\frac{\pi\nu\delta}{2})}
{\sinh(\frac{\pi\nu}{2})}\re^{-\frac{\pi\nu}{2}}\\[.6cm]
\hat R_{31}(\nu)&=&\ds\hat R_{32}(\nu)
=\ds\frac{\sinh(\frac{\pi\nu\delta}{2})}
{\sinh({\pi\nu})}\qquad
&\hat R_{33}(\nu)&=&\phantom{+}\ds\frac{\sinh(\frac{\pi\nu\delta}{2})}
{\sinh{(\pi\nu)}}\,
\re^{\frac{\pi\nu\delta}{2}}\,.
\end{array}
\eeq
For completeness we present also the Fourier transform of the source term
in \eqref{preddv}
\bea
\hat{\myp}_1(\nu)&=&+2\ri\, d_1(\nu)\ \frac{\cos(\Theta \nu)  
\sinh\big(\frac{\pi \nu
    (1+\delta)}{2}\big)}{\nu\sinh(\pi\nu)}\quad\nonumber\\[.4cm]
\hat{\myp}_2(\nu)&=&-2\ri\, d_2(\nu)\ \frac{ \cos(\Theta \nu) 
\sinh(\frac{\pi \nu
    (1-\delta)}{2})}{\nu\,\sinh(\pi\nu)}\nonumber
    \\[.4cm]
\hat{\myp}_3(\nu)&=&-2\ri\, d_3(\nu)\ 
\frac{\cos (\Theta\nu) \sinh(\frac{\pi \nu
   }{2})} {\nu\sinh({\pi \nu})}\nonumber
\eea
for which one should use the principal value integral in \eqref{Fdef}. 
Multiplying both sides of \eqref{preddv} by the inverse of the
matrix integral operator
$\big(\delta(\theta)\,\delta_{ij}-R^{(+)}_{ij}(\theta)\big)$, 
one obtains
\bea\nonumber
\varepsilon_i(\theta)&=&2\pi N\,{\mathsf z}_i(\theta)-2\pi{\mathsf c}_i+
\frac{1}{\ri}\,\sum_{j=1}^3\,
\int_{-\infty}^{\infty}\rd\theta'\,
G^{(+)}_{ij}(\theta-\theta')\  
\log\big(1+\re^{-\ri\varepsilon_j(\theta')}\big)\\[.5cm]
&&-\frac{1}{\ri}\,\sum_{j=1}^3\,
\int_{-\infty}^{\infty}\rd\theta'\,
G^{(-)}_{ij}(\theta-\theta')\ 
\log\big(1+\re^{-\ri\varepsilon^*_j(\theta')}\big)\ ,\label{set3ddv}
\eea
where the Fourier transform of the kernel $G^{(\pm)}(\theta)$ is given
by 
\beq
\hat G^{(\pm)}_{ij}(\nu)=d_i(\nu)\,\hat G_{ij}(\nu)\,d_j(\mp\nu)
\nonumber
\eeq
with
\beq
\begin{array}{lll}
&\hat G_{11}(\nu)=\hat G_{12}(\nu)=\ds-\frac{\sinh(\frac{\pi \nu \delta}{2})}{2
\cosh(\frac{\pi \nu}{2}) \sinh(\frac{\pi \nu a_2}{2})}\, ,
&\hat G_{13}(\nu)
=\ds-\frac{\sinh(\frac{\pi \nu}{2})}{\sinh(\frac{\pi \nu a_2}{2})}
\label{Gf-def}\\[.8cm]
&\hat G_{21}(\nu)=\hat G_{22}(\nu)=\ds\frac{\sinh^2(\frac{\pi \nu
    \delta}{2})\, \re^{-\frac{\pi \nu}{2}}}{2\, \cosh(\frac{\pi
    \nu}{2}) \sinh(\frac{\pi \nu a_1}{2} )
\sinh(\frac{\pi \nu a_2}{2})}\, ,
&\hat G_{23}(\nu)=\ds\frac{\sinh(\frac{\pi \nu}{2})\,\sinh(\frac{\pi \nu \delta}{2})\, \re^{-\frac{\pi \nu}{2}}}
{\sinh(\frac{\pi \nu a_1}{2})
\sinh(\frac{\pi \nu a_2}{2})}\\[.8cm]
&\hat G_{31}(\nu)=\hat G_{32}(\nu)=\ds\frac{\sinh(\frac{\pi \nu}{2})\,
\sinh(\frac{\pi \nu \delta}{2})}{2\, \cosh(\frac{\pi \nu}{2}) \sinh(\frac{\pi \nu a_1}{2})
\sinh(\frac{\pi \nu a_2}{2})}\, ,
&\hat G_{33}(\nu)=\ds\frac{\sinh^2(\frac{\pi \nu \delta}{2})}{\sinh(\frac{\pi \nu a_1}{2})
\sinh(\frac{\pi \nu a_2}{2})}\,.\nonumber
\end{array}
\eeq
For the source terms in \eqref{set3ddv} one obtains   
\beq
{\mathsf c}_1=-\frac{2p_2}{a_2}\,,\qquad 
{\mathsf c}_2=\frac{p_1}{a_1}-\frac{p_2}{a_2}\,,\qquad
{\mathsf c}_3=\frac{p_1}{a_1}+\frac{p_2}{a_2}
\nonumber
\eeq
and 
\beq
{\mathsf z}_1(\theta)=-{\mathsf z}_\theta(\theta-\ri
\omega_1+\ri\pi)\,,\quad 
{\mathsf z}_2(\theta)={\mathsf z}_u(\theta-\ri
\omega_2+\tshf\ri\pi\delta)\,,\qquad
{\mathsf z}_3(\theta)={\mathsf z}_u(\theta-\ri
\omega_3)\,,
\nonumber
\eeq
where the functions ${\mathsf z}_\theta$ and ${\mathsf z}_u$ were already
defined in \eqref{zcf}. 

Equation \eqref{set3ddv} can be considerably simplified thanks to 
some additional symmetries. Indeed, using the Fourier transform 
of the integral operator $G^{(\pm)}_{ij}(\theta)$ one can see 
that all the dependence on the functions 
$\varepsilon_i(\theta)$ in RHS of \eqref{set3ddv} reduces  
to the integrals 
\beq
g_k(\nu)=\re^{-\nu\omega_k}\,\int_{-\infty}^\infty\,\rd \theta \,\re^{-\ri \nu \theta}\,
\Big(\log\big(1+\re^{-\ri\varepsilon_k(\theta)}\big)-
\log\big(1+\re^{-\ri\varepsilon_k(-\infty)}\big)\Big)\,,\label{gnu-def}
\eeq
where $\re^{-\ri\varepsilon_k(+\infty)}=\re^{-\ri\varepsilon_k(-\infty)}$ 
for any finite $N$
and $\Theta$. Then using the definitions \eqref{eps-def} and the fact
that the function $1+a_1^{-1}(\theta)$ 
does not have poles or zeroes \eqref{poles0} in the   
strip $-\pi<\Im m (\theta)<-\omega_2^{\rm(min)}$ one can shift the
integration contours in \eqref{gnu-def} to prove that 
\beq
g_1(\nu)= g_2(\nu)\,.
\label{g12rel}
\eeq
Thus, one of the functions $\varepsilon_1(\theta)$ or
$\varepsilon_2(\theta)$ can be excluded from \eqref{set3ddv}.

Consider now the scaling limit \eqref{qft-lim} of \eqref{set3ddv}.
For the central region $|\theta|\ll\Theta$ one needs to make the substitution 
\beq
2\pi N {\mathsf z}_\theta(\theta)=2r \cos\big(\tshf\pi\delta\big)\,
\sinh(\theta)+O\big(\re^{-\Theta})\,, \quad 
2\pi N {\mathsf z}_u(\theta)=r 
\sinh(\theta)+O(\re^{-\Theta}) 
\label{z-qft}
\nonumber
\eeq 
for $\Theta\to+\infty$, 
while the rest of \eqref{set3ddv} remains intact. Next, using the
symmetry \eqref{g12rel} to exclude $\varepsilon_1(\theta)$, redenoting
$\varepsilon_3(\theta)\to \varepsilon_+(\theta)$ and  
$\varepsilon_2(\theta)\to \varepsilon_-(\theta)$, and then choosing
$\omega_2=\pi/2$ and $\omega_3$ to be infinitesimally small, one
arrives to Eq.\,\eqref{DDV}, given in the main text.

\app{\label{app:D} Integral representations for sums over the Bethe
  roots} 
Once the non-linear integral equations \eqref{set3ddv} are solved,
every sum over the vacuum Bethe roots can be effectively computed via
exact integral representations, considered below. The technique works
equally well both for finite values of $N$ (when the number of Bethe
roots is finite) as well as in the scaling limit (when $N\to \infty$
and \eqref{set3ddv} is replaced by its field theory analog
\eqref{DDV}).

Let us start with the case of finite $N$. Consider, for instance,
the product formulae \eqref{A13-def} and \eqref{A2-def}. 
With an account of \eqref{set3ddv}  they can be converted into integral 
representations
\bea\label{Aintrep}
\log \bigg(\frac{{\Asf}_i(\kappa_i \,\re^\theta)}
{{\Asf}_i(\kappa_i)}\bigg)
& =&\log\big({\Asf}_i^{(as)}(\kappa_i \,\re^\theta) \big)
+\sum_{k=1,3}\int_{-\infty}^\infty \rd\theta'\,F_{ik}^{(+)}(\theta-\theta'+\ri \omega_k) 
\log\big(1+\re^{-\ri\varepsilon_k(\theta')}\big)\nonumber\\[.3cm]
&&-\sum_{k=1,3}\int_{-\infty}^\infty\rd\theta'\, F_{ik}^{(-)}(\theta-\theta'-\ri \omega_k) 
\log\big(1+\re^{-\ri\varepsilon_k^*(\theta')}\big)\, ,
\eea
where $\kappa_1=\kappa_2=1$, $\kappa_3=\ri$,
\bea
\log\big({\Asf}_i^{(as)}(\re^\theta)\big)&=&
-\frac{2\,p_{3-i}}{a_{3-i}}\,\theta\,+
2N\,\int_{-\infty}^{\infty}\frac{\rd\nu}{\nu}\ \frac{\sin^2(\hf \theta\nu)\cos(\Theta\nu)
  \cosh(\hf \pi \delta)}{\sinh(\pi
  \nu)\cosh(\hf\pi\nu)}\,\quad (i=1,2)
\nonumber\\[.4cm]
\log\big({\Asf}_3^{(as)}(\ri\,\re^\theta)\big)&=&
-\Big(\frac{2\,p_1}{a_1}+\frac{2\,p_2}{a_2}\Big)
\,\theta+2N\,\int_{-\infty}^{\infty}\frac{\rd\nu}{\nu}\ 
\frac{\sin^2(\hf  \theta\nu)\cos(\Theta\nu)} {\sinh(\pi
  \nu)}\, .\label{Aasrep}
\eea
The Fourier
transform of the kernel is defined by \eqref{Fdef}
(but with the principal value integral) where   
\bea
\hat{F}_{11}^{(\pm)}(\nu)
&=&+\frac{1}{2\sinh(\pi\nu)}\ \Big[\big(\hat{G}_{11}(\nu)+1\big)\,\re^{\pm\pi\nu}
+\hat{G}_{12}(\nu)\Big]\nonumber\\[.3cm]
\hat{F}_{13}^{(\pm)}(\nu)&=&+\frac{1}{2\sinh(\pi\nu)}\ \hat{G}_{33}(\nu)
\nonumber
\label{FfromG}
\\[.3cm]
\hat{F}_{21}^{(\pm)}(\nu)&=&-\frac{1}{2\sinh(\pi\nu)}\ \Big[\big(\hat{G}_{21}(\nu)
+\re^{\frac{\pi\delta\nu}{2}}\,\hat{G}_{31}(\nu)\big)\,\re^{\pm\pi\nu}
+\hat{G}_{22}(\nu)+1+\re^{\frac{\pi\delta\nu}{2}}\,\hat{G}_{32}(\nu)\Big]
\nonumber
\\[.3cm]
\hat{F}_{23}^{(\pm)}(\nu)&=&-\frac{1}{2\sinh(\pi\nu)}\ \Big[\hat{G}_{23}(\nu)
+\re^{\frac{\pi\delta\nu}{2}}\big(\hat{G}_{33}(\nu)+1\big)\Big]
\\[.3cm] 
\hat{F}_{31}^{(\pm)}(\nu)&=&-\frac{\cosh(\frac{\pi\nu}{2})}{\sinh(\pi\nu)}\ 
\Big[\hat{G}_{31}(\nu)     
\,\re^{\pm\pi\nu}+\hat{G}_{32}(\nu)\Big]\,,\nonumber\\[.3cm]
\hat{F}_{33}^{(\pm)}(\nu)&=&-\frac{\cosh(\frac{\pi\nu}{2})}{\sinh(\pi\nu)}\ \Big[\hat{G}_{33}(\nu)
+1\Big]\,.\nonumber
\eea
The derivation is standard and very much similar to that of the equation
\eqref{set3ddv} itself. The only complication is that one needs to
systematically take into account the fact that summing over the zeroes of
${\Asf}_2(\re^\theta)$ with integrals over 
the contour $C_2$ in Fig.\,\ref{plot_contours} (similarly to
\eqref{c2poles}) contains extra
contributions determined by the zeroes of ${\Asf}_3(\re^\theta
q^{-1})$. Moreover, the function $\varepsilon_2(\theta)$ was excluded
from the RHS of \eqref{Aintrep} in favor of $\varepsilon_1(\theta)$
of by virtue of \eqref{g12rel}.
We have presented the intermediate formulae \eqref{FfromG}
to guide the reader through this tedious derivation. From \eqref{Gf-def} one
obtains 
\beq
\begin{array}{rclrcl}
\hat{F}_{11}^{(\pm)}(\nu)&=&\ds+\frac{\re^{\pm\frac{1}{2} \pi  (1+a_2)\nu} }
{4\cosh(\frac{\pi \nu}{2})
\sinh(\frac{\pi a_2\nu}{2}  )}\, ,\qquad
&\hat{F}_{13}^{(\pm)}(\nu)&=&\ds-\frac{1}
{4\cosh(\frac{\pi \nu}{2})
\sinh(\frac{\pi a_2\nu}{2}  )}\\[.7cm]
\hat{F}_{21}^{(\pm)}(\nu)&=&\ds-\frac{\re^{\pm\frac{1}{2} \pi  (1-a_1)\nu} }
{4\cosh(\frac{\pi \nu}{2})
\sinh(\frac{\pi a_1\nu}{2}  )}\,,
&\hat{F}_{23}^{(\pm)}(\nu)&=&\ds-\frac{1}
{4\cosh(\frac{\pi \nu}{2})
\sinh(\frac{\pi a_1\nu}{2}  )}\nonumber\\[.7cm]
\hat{F}_{31}^{(\pm)}(\nu)&=&\ds-\frac{\re^{\pm\frac{\pi
      \nu}{2}}\,\sinh(\frac{\pi \delta \nu}{2})}{
2\sinh(\frac{\pi a_1\nu}{2})\,\sinh(\frac{\pi a_2\nu}{2})}\,,
&\hat{F}_{33}^{(\pm)}(\nu)&=&\ds-\frac{\sinh(\frac{\pi\nu}{2})}{
2\sinh(\frac{\pi a_1\nu}{2})\,\sinh(\frac{\pi a_2\nu}{2})}\,.
\end{array}
\eeq

Similar representations can be obtained for the two 
expressions for the energy \eqref{energy}, when the vacuum state is
filled with real roots (1-strings) and complex roots (2-strings), respectively,
\beq
E^{(1)}_N=\sum_{\JJ} \myenergy(\theta^{(1)}_\JJ,\delta)\,,\qquad  
E^{(2)}_N=\sum_{\JJ} \myenergy(\theta^{(2)}_\JJ,\delta)\,.
\nonumber
\eeq
Proceeding as above, one obtains for $i=1,2$
\bea
E^{(i)}_N&=&N\,\varepsilon_\infty^{(i)}+
\frac{1}{2\pi}\sum_{k=1,3}\,
\int_{-\infty}^{\infty}\rd\theta\,
H^{(+)}_{ik}(-\theta+\ri\omega_k)\  
\log\big(1+\re^{-\ri\varepsilon_k(\theta)}\big)\nonumber\\[.3cm]
&&-\frac{1}{2\pi}\sum_{k=1,3}\,
\int_{-\infty}^{\infty}\rd\theta\,
H^{(-)}_{ik}(-\theta-\ri\omega_k)\ 
\log\big(1+\re^{-\ri\varepsilon^*_k(\theta)}\big)\, ,\label{EN12}
\eea
where 
\beq
\begin{array}{rclrcl}
\hat{H}_{11}^{(\pm)}(\nu)&=&\ds+\re^{\pm\frac{1}{2} \pi  (1+a_2)\nu}
 \hat{H}_{13}^{(\pm)}(\nu)\,,\qquad
&\hat{H}_{13}^{(\pm)}(\nu)&=&\ds-\frac{2 \sin (\Theta\nu)}
{\cosh(\frac{\pi \nu}{2})}\nonumber\\[.7cm]
\hat{H}_{21}^{(\pm)}(\nu)&=&\ds-\re^{\pm\frac{1}{2} \pi  (1-a_1)\nu}
\hat{H}_{13}^{(\pm)}(\nu)\,,\qquad
&\hat{H}_{23}^{(\pm)}(\nu)&=&\hat{H}_{13}^{(\pm)}(\nu)\, .
\end{array}
\eeq
In particular for the difference between LHS and RHS of
\eqref{identity} one gets
\beq
8\pi\, \Im m\bigg[ \log\Big(\big(1+f_1(\Theta+\tshf\ri\pi(1+\delta)\big)
\big(1+f_1(-\Theta+\tshf\ri\pi(1+\delta)\big)\Big)\bigg]=0
\nonumber
\eeq
which is zero, since $f_1(\pm\Theta+\tshf\ri\pi(1+\delta))$ 
vanishes, as stated in \eqref{poles0}.

\bigskip
{\bf Scaling limit}. Consider now the scaling limit \eqref{qft-lim}. Then \eqref{Aintrep}
can be used to obtain integral representations for the
$\Asf^{\rm (BL)}$-functions \eqref{ABL} for the BL model. The only modification
is that for $i=1,2$ one needs to add to the RHS of \eqref{Aintrep} the entire
functions 
$\alpha_i\re^\theta+\beta_i\re^{-\theta}$, that appear in \eqref{ABL}.   
Assume that $|\theta|\ll\Theta$. Then for the leading terms in \eqref{Aintrep}
one obtains in the limit \eqref{qft-lim},
\bea\nonumber
\!\!\!\!\!\!\!\!&&\log\big({\Asf}_i^{(as)}(\re^\theta)\big)=-\frac{r \cos(\hf \pi \delta)}
{\pi}\,\theta\, \sinh(\theta)+\frac{r\,{\mathsf
    B(\Theta)}}{2\pi}\,
    \sinh^2\big(\tshf\,\theta\big) -\frac{2\, p_{3-i}}{a_{3-i}}\,
\theta +O\big(\Theta\re^{-\Theta}\big) \quad ( i=1,2)\\[.4cm]
\!\!\!\!\!\!\!\!&&\log\big({\Asf}_3^{(as)}(\ri\,\re^\theta)\big)=
r\sinh^2\big(\tshf\,\theta\big)
-\Big(\frac{2\,p_1}{a_1}+\frac{2\,p_2}{a_2}\Big)\,\theta+O\big(\re^{-\Theta}\big)\,,
\nonumber
\eea
where 
\beq
{\mathsf B}(\Theta)=4\,(1+\Theta)\cos\big(\tshf\pi\delta\big)+2\pi\delta
\sin\big(\tshf\pi\delta\big)\,.
\nonumber
\eeq
The undetermined coefficients $\alpha_i,\beta_i$ are fixed by the
requirement that the functions \eqref{ABL} satisfy the functional
relation \eqref{fr12} (upon the identification \eqref{ident}). The
integral representations for $\Asf_i^{\rm (BL)}$ obtained in this way,
lead to the
asymptotic expansions \eqref{Qasym}, with the coefficients given by
\eqref{In-def} and \eqref{jassa}. 

Finally, note that in the scaling limit \eqref{qft-lim} 
formula \eqref{EN12} leads precisely
to \eqref{ceff-lat}. First, let us split the integration in \eqref{EN12}
into two regions, where $|\theta'|<\Theta$ 
and $|\theta'|>\Theta$.
Then, it is not difficult to see that the contribution of 
the first integral to the LHS of \eqref{ceff-lat} exactly equals to 
${\mathfrak
  F}(r,{\bf k})$ (in the form \eqref{F-ddv}), while the second
integral gives the term ``$-{\mathfrak
  F}(0,{\bf k})$'' in the RHS of \eqref{ceff-lat}.

\app{\label{app:E} Small  and large-$R$ asymptotics of IM}

Here we collect some facts concerning the vacuum eigenvalues of integral of motions in
the BL model.  
The local IM  are normalized in such a way that 
\bea\label{kaaaajssa}
\lim_{R\to 0}\big({\textstyle \frac{R}{2\pi}}\big)^{2n-1} \, I_{2n-1}=
 F_{n}(k_1^2,k_2^2)\ .
\eea
Here $F_n(x,y)$ are certain polynomials
of two variables   of degree $n$, such that
\bea
F_{n}(x,y)= 
\frac{(2n-2)!}{(-4)^n}\sum_{i+j=n}
\frac{{\textstyle\big ((1-\delta)(\frac{1}{2}-n)\big)_{j}\big ((1+\delta)(\frac{1}{2}-n)\big)_{i}}}
{{\textstyle\big ((1-\delta)(\frac{1}{2}-n)\big)_{n}\big ((1+\delta)(\frac{1}{2}-n)\big)_{n}}}\,
(1-\delta)^{2i} (1+\delta)^{2j}\  \frac{x^{i}y^{j}}{i!j!}+\ldots
\nonumber
\eea
where $(a)_n$ is the Pochhammer symbol and dots
stand for  monomials with degree  lower than $n$.
The polynomial $F_2$ and $F_3$  read   explicitly 
\cite{Lukyanov:2003nj}:\footnote{The polynomials $I_{2k-1}(P,Q)$ from Ref.\!\cite{Lukyanov:2003nj} are
related to $F_{k}(x,y)$ as $$I_{2k-1}(P,Q)=2^{1-2 k}\,F_{k}\Big(\!-\frac{P^2}{n}, \frac{Q^2}{n+2}\Big)\Big|_{n=\delta-1}\, .$$}
\bea\label{asosisasaisaoi}
F_2(x,y)&=&\frac{1}{12(1-9\delta^2)}\ \Big(\, (1+3 \delta) (1-\delta)^3 \  x^2+6\, (1-\delta^2)^2\ x y+
(1- 3 \delta)(1+\delta)^3\  y^2\nonumber \\
&-&
2\,(1 + 2 \delta)(1 - \delta)^2 \   x-
2\,(1 - 2 \delta)(1 + \delta)^2 \   y+
{\textstyle \frac{1}{15}}\, (7-18 \delta^2)\,\Big)
\eea
and
\bea\label{ahsahasj}
\!\!\!&&F_3(x,y)=\frac{1}{10(1-25\delta^2)(9-25 \delta^2)}\Big(
(1 - \delta)^5 (1 + 5 \delta) (3+5 \delta)\, x^3+
(1 + \delta)^5 (1 - 5 \delta) (3-5 \delta)\, y^3\nonumber \\[0.2cm]
\!\!\!&&\ \ \ \ +\,
15\,  (1-\delta)^4 (1+\delta)^2 (3+5 \delta)\, x^2 y
+\,
15\, (1+\delta)^4 (1-\delta)^2 (3-5 \delta)\, y^2 x\nonumber\\[0.2cm]
\!\!\!&&
\ \ \ \ -\,5\, (1 - \delta)^4 (1 + 3 \delta) (3 + 5 \delta)\, x^2 -
5\, (1 + \delta)^4 (1 - 3 \delta) (3 - 5 \delta)\, y^2-
30\, (1 - \delta^2)^2 (3 - 5 \delta^2)\,  x y\nonumber\\[0.2cm]
\!\!\!&&\ \ \ \ -\,
(1-\delta)^2\, \big(76\, \delta^3+ 22\, \delta^2- 47\, \delta  -21  )\, x+(1+\delta)^2 (76\, \delta^3- 22\, \delta^2
- 47\, \delta  +21  \big)\, y\nonumber\\[0.2cm]
\!\!\!&&\ \ \ \ -\,
{ \frac{1}{63}}\,  (1420 \delta^4 - 1135\, \delta^2+279)\,\Big)\, .
\eea

The vacuum eigenvalues of local IM \eqref{In-def} can be written in the form
\bea
\label{i2nm1}
I_{2n-1}= \big(M \cos({\textstyle\frac{\pi\delta}{2}})\big)^{2 n-1}\ \Bigg( 
\, (-1)^n\ f_n\big(2r\,c\big({\textstyle\frac{\delta}{2}}\big)\big)
\pm \frac{\zeta\big(\pm  \ri (2n-1)\, |\, k_i,k_{3-i}\big)}{\pi \cos(\frac{\pi\delta}{2}(2n-1)\big)}\,\Bigg)\ ,
\eea
\begin{figure}
[!ht]
\centering
\includegraphics[width=15cm]{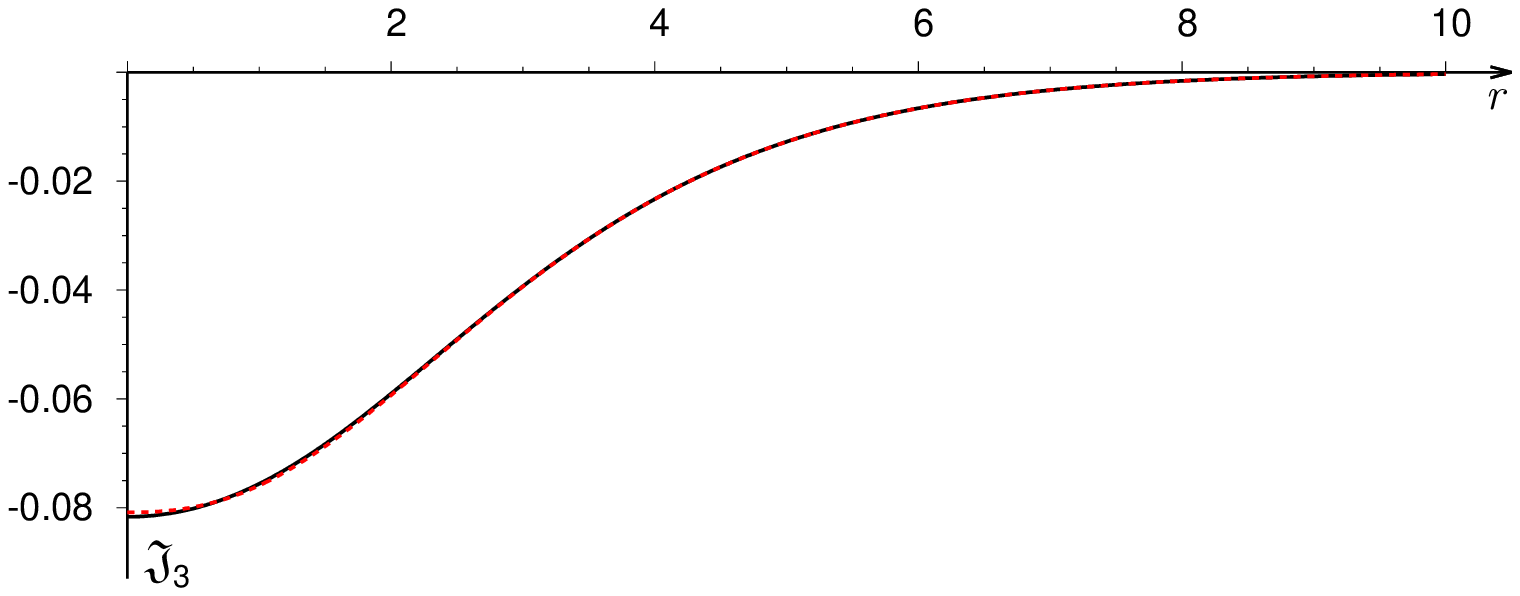}

\vspace{.5cm}
\includegraphics[width=15cm]{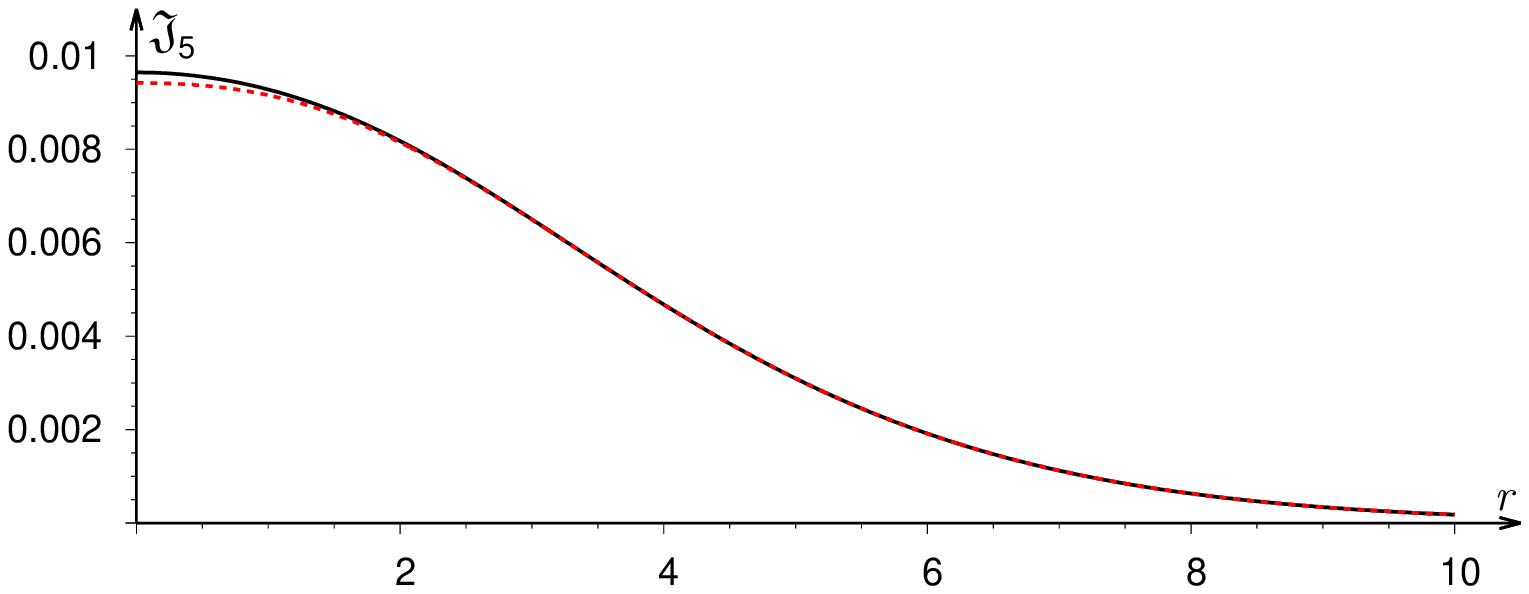}
\caption{
Plots of $\big({\textstyle \frac{R}{2\pi}}\big)^{3}I_3=
{\mathfrak I}_3$ (top) and  
$\big({\textstyle \frac{R}{2\pi}}\big)^{5}I_5= {\mathfrak I}_5$
(bottom)  versus $r$ for 
$\delta=\frac{17}{47}=0.36\ldots,\ k_1=\frac{47}{150},\ k_2=\frac{47}{640}$.
The solid lines were obtained from numerical
integration of eqs.\,\eqref{DDV},\,\eqref{In-def}.
The dashed lines  represent the large-$r$
approximation\ \eqref{i2nm1},\,\eqref{manssaj}  
 (for the
 numerical data, see Tab.\,\ref{tab2}).}
\label{fig8aa}
\end{figure}

where 
\bea
f_n(\beta)=
\int_{-\infty}^\infty\frac{\rd\theta}{\pi}\ \re^{(2 n-1)\theta}\ \log\big(1-\re^{-\beta\cosh(\theta)}\big)
\nonumber
\eea
and,  as $r\to \infty$,
\bea\label{manssaj}
\zeta(\nu|k_i,k_j)=\zeta^{(0)}(\nu|k_i,k_j)+\zeta^{(1)}(\nu|k_i,k_j)+o\big(\re^{-2 r}\big)
\eea
with 
\bea\label{aoosioasasos}
\zeta^{(0)}(\nu|k_i,k_j)=
4\, c(k_j)\,s\big(k_i+{\textstyle\frac{\ri\nu}{2}}\big)\, K_{\ri\nu}(r) + 4\, \Big(\, c^2(k_j)\,s\big({\textstyle\frac{\ri\nu}{2}}\big)-
 c(k_i)\,s\big(k_i+{\textstyle \frac{\ri\nu}{2}}\big)c(2k_j) \,\Big)
\ K_{\ri\nu}(2r)
\nonumber
\eea
and  
\bea\label{asiuussaui}
&&\zeta^{(1)}(\nu|k_i,k_j)=
16
\int_{-\infty}^\infty \frac{\rd\mu}{2\pi}\
K_{\ri\mu}(r) K_{\ri\nu-\ri\mu}(r)\nonumber\\
&&\times
\Bigg[
\frac{c^2(k_j)c(k_i)s\big(k_i+{\textstyle\frac{\ri\nu}{2}}\big)}                                                                                
{\cosh(\frac{\pi\mu}{2})}                                                                                                                       
\bigg(
\frac{\sinh(\frac{\pi\mu }{2} (a_2-1))}{\sinh(\frac{\pi\mu}{2} a_2)}+                                                                           
\frac{\sinh(\frac{\pi \mu}{2} (a_1-1))}{\sinh(\frac{\pi\mu}{2} a_1)}\bigg)\nonumber\\                                                           
&&-
c(k_i)s\big(k_i+ {\textstyle\frac{\ri\nu}{2}}\big)\ \cosh\big({\textstyle\frac{\pi\mu}{2}}\big)\                                                
\frac{\sinh(\frac{\pi \mu}{2} (a_j-1))}{\sinh(\frac{\pi \mu}{2} a_j)}                                                                           
+\ri\,  c^2(k_j)\ \sinh\big({\textstyle\frac{\pi(\mu-\nu)}{2}}\big)\                                                                            
\frac{\sinh(\frac{\pi \mu}{2} (a_i-1))}{\sinh(\frac{\pi \mu}{2} a_i)} \,\Bigg]\, .\nonumber                                                      
\eea
The relations \eqref{i2nm1} 
are valid for any choice of the sign $\pm$  and $i=1,\,2$.
Also notice that for the non interacting  case
\bea
\!\!\!&&\zeta(\nu|k_i,k_j)|_{\delta=0}=
\frac{1}{2\ri}\ 
\int_{-\infty}^\infty\rd\theta\, \re^{-\ri\nu\theta}\
 \Bigg(
\re^{-\frac{\pi\nu}{2}}\,
\log\Bigg[
\frac{\big(1+\re^{\ri\pi (k_i+k_j)-r\cosh(\theta)}\big)\big(1+\re^{\ri\pi (k_i-k_j)-r\cosh(\theta)}\big)}
{1-\re^{-2r\cosh(\theta)}}\Bigg]\nonumber\\
\!\!\!&&\ \ \ \ \ \ \ \ \ \ \ \ \ \ \ \ \ \  -
\re^{+\frac{\pi\nu}{2}}\,
\log\Bigg[
\frac{\big(1+\re^{-\ri\pi (k_i+k_j)-r\cosh(\theta)}\big)\big(1+\re^{-\ri\pi (k_i-k_j)-r\cosh(\theta)}\big)}
{1-\re^{-2r\cosh(\theta)}}\Bigg]\,\Bigg)
\nonumber
\eea
and $\zeta(\nu|k_i,k_j)|_{\delta=0}=\zeta^{(0)}(\nu|k_i,k_j)+o\big(\re^{-2 r}\big)$.
\begin{figure}
[!ht]
\centering
\includegraphics[width=7.5cm]{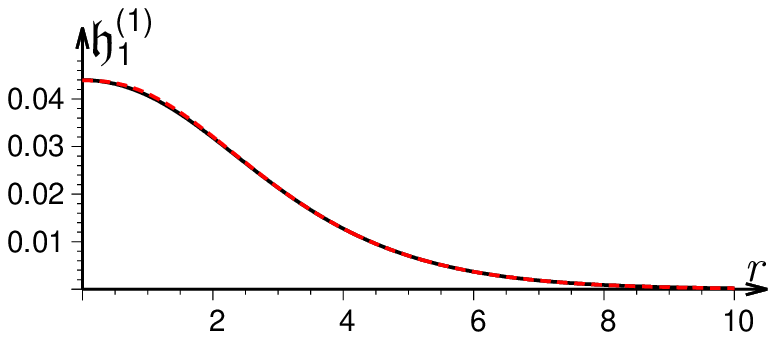}
\includegraphics[width=7.5cm]{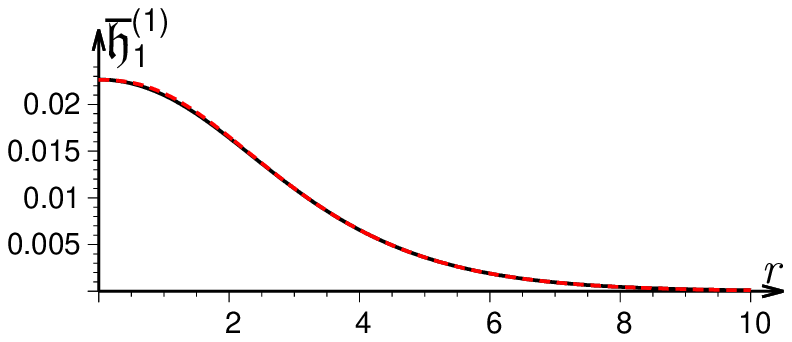}\\
\includegraphics[width=7.5cm]{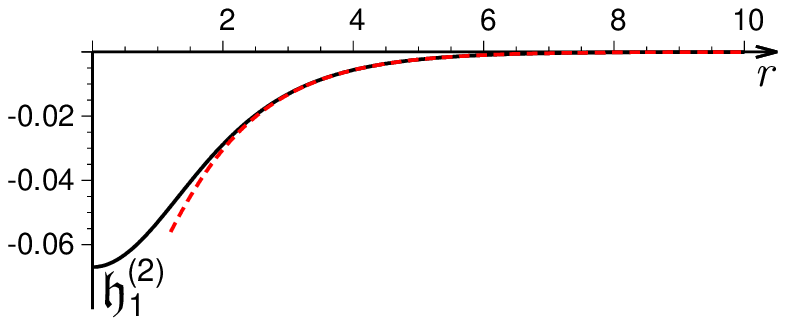}
\includegraphics[width=7.5cm]{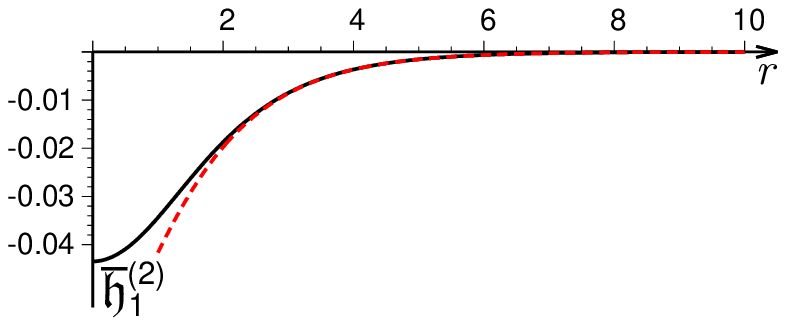}
\caption{
${\mathfrak h}_1^{(1)}=\big({\textstyle \frac{R}{2\pi}}\big)^{\frac{2}{a_1}}H_1^{(1)}$
(upper left panel),   ${\bar {\mathfrak h}}_1^{(1)}=
\big({\textstyle \frac{R}{2\pi}}\big)^{\frac{2}{a_1}}{\bar H}_1^{(1)}$ 
(upper right panel),
 ${\mathfrak h}_1^{(2)}=
\big({\textstyle \frac{R}{2\pi}}\big)^{\frac{2}{a_2}}H_1^{(2)}$ 
(lower  left panel) and  
${\bar {\mathfrak h}}_1^{(2)}=
{\big({\textstyle \frac{R}{2\pi}}\big)^{\frac{2}{a_2}}{\bar
    H}_1^{(2)}}$ 
(lower right panel) 
 versus $r$ for 
$\delta=\frac{17}{47}=0.36\ldots,\ k_1=\frac{47}{150},\ k_2=\frac{47}{640}$.
The solid lines were obtained from numerical
integration of eqs.\,\eqref{DDV},\,\eqref{jassa} (for the
 numerical data, see Tabs.\,\ref{tab5},\,\ref{tab6}).
The dashed lines  represent the large-$r$
approximation\  \eqref{manssaj},\,\eqref{jsaisaiu}. Note that for ${\mathfrak
  h}_1$ and $\bar{\mathfrak h}_1$ the two lines are practically
indistinguishable. }
\label{fig7aa}
\end{figure}

In Fig.\,\ref{fig8aa},
the numerical data for    vacuum eigenvalues of the  first local IM
are  compared against the  large-$r$ approximation \eqref{i2nm1},\,\eqref{manssaj}.

Similarly to the formula \eqref{i2nm1}, the  vacuum eigenvalues of  nonlocal IM
\eqref{jassa} are expressed in terms of the $\zeta$-function \eqref{manssaj}:
 \bea\label{jsaisaiu}
H^{(i)}_{n}
&=&\frac{2}{\pi a_i}\  \big(M \cos({\textstyle\frac{\pi\delta}{2}})\big)^{\frac{2n}{a_i}}\ 
\zeta\big( {\textstyle\frac{2\ri n}{a_i}}\big|\! +k_i,k_{3-i}\big)\nonumber\\
{\bar H}^{(i)}_{n}
&=&\frac{2}{\pi a_i}\  \big(M \cos({\textstyle\frac{\pi\delta}{2}})\big)^{\frac{2n}{a_i}}\ 
\zeta\big({\textstyle\frac{2\ri n}{a_i}}\big| \!- k_i,k_{3-i}\big)\,.
\eea
This gives the corresponding  large-$r$ asymptotics (see Fig.\,\ref{fig7aa}).
In the small-$r$ limit one has
\bea\label{kajsska}
\lim_{R\to 0}\big({\textstyle \frac{R}{2\pi}}\big)^{\frac{2n}{a_i}} \, H^{(i)}_{n}=h_{n}(k_i,k_{3-i}\,)\ ,\ \ \  \ \
\lim_{R\to 0}\big({\textstyle \frac{R}{2\pi}}\big)^{\frac{2n}{a_i}} \, {\bar H}^{(i)}_{n}=h_{n}(-k_i,k_{3-i})\ .
\eea
An integral representation for these limiting values where discussed in Ref.\!\!\cite{Bazhanov:2013cua}. 
Using the results of this work,  it is possible to
show that
\bea\label{lsakslsakl}
h_1(k_i,k_j)=(a_i)^{\frac{2}{a_i}}\  \frac{\sqrt{\pi}\ \Gamma(\frac{1}{a_i})\, \Gamma(1+\frac{1}{a_i}+k_i)}
{2\, \Gamma(\frac{1}{2}+\frac{1}{a_i}) \Gamma(-\frac{1}{a_i}+k_i)}\ 
        \bigg(\frac{ (2-a_i)^2  k_j^2}{a_i^2 k^2_i-1}+\frac{2-a_i}{2+a_i}\bigg)\ .
\eea

\begin{table}
[!ht]
\begin{center}
\begin{tabular}{| c || l| l | l|}
\hline \rule{0mm}{3.6mm}
$r=MR$&$\ \ \ \ \ \ \ \ \ \ {\mathfrak s}_1$&  $\ \ \ \ \ \  \ ({\mathfrak s}_1)_{\rm UV}$&
$\ \ \ \ \ \ \ ({\mathfrak s}_1)_{\rm IR}$\\
\hline
$0.1$  & $1.8988060413448$   &$1.8988060423805$               & $\ \ \ \ \ \ \ \ \star\star\star$\\
$0.2$  & $1.4654837862197$   &$1.4654838165411$               & $\ \ \ \ \ \ \ \ \star\star\star$\\
$0.3$  & $1.2131473840105$   &$1.2131477218447$               & $\ \ \ \ \ \ \ \ \star\star\star$\\
$0.4$  & $1.0353338553850$   &$1.0353357364881$               & $\ \ \ \ \ \ \ \ \star\star\star$\\
$0.5$  & $0.8986854527823$   &$0.8986925558030$               & $\ \ \ \ \ \ \ \ \star\star\star$\\
$0.6$  & $0.7883493477945$   &$0.7883702973779$               & $\ \ \ \ \ \ \ \ \star\star\star$\\
$0.7$  & $0.6964065576931$   &$0.6964586183195$               & $0.6481367015033$\\
$0.8$  & $0.6181320446112$   &$0.6182460963202$               & $ 0.5892787485706$\\
$0.9$  & $0.5504776094433$   &$0.5507044055490$               & $0.5341533278400$\\
$1.0$  & $0.4913590816186$   &$0.4917766951250$               & $0.4829990934522$\\
$1.1$  & $0.4392846752853$   &$0.4400069531085$               & $0.4358646597841$\\
$1.2$  & $0.3931455547148$   &$0.3943313254242$               & $0.3926746860951$\\
$1.3$  & $0.3520905225173$   &$0.3539532515462$               & $0.3532745611255$\\
$1.4$  & $0.3154474446990$   &$0.3182650059333$               & $0.3174611565686$\\
$1.5$  & $0.2826721374302$   &$0.2867963735527$               & $0.2850041743179$\\
$1.6$  & $0.2533141494059$   &$0.2591799235187$               & $0.2556609735028$\\
$1.7$  & $0.2269933362366$   &$0.2351268290042$               & $0.2291867944274$\\
$1.8$  & $0.2033835365588$   &$0.2144096090884$               & $0.2053416941707$\\
$1.9$  & $0.1822010282640$   &$0.1968495429423$               & $0.1838951179057$\\
$2.0$  & $0.1631962542843$   &$ 0.1823073160659$              & $0.1646287674316$\\
$2.4$  & $0.1048626544268$   &$\ \ \ \ \ \ \ \ \ \star\star\star$     & $0.1054983760734$\\
$2.8$  & $0.0672585487613$   &$\ \ \ \ \ \ \ \ \ \star\star\star$     & $0.0675070860423$\\
$3.2$  & $0.0431232955868$   &$\ \ \ \ \ \ \ \ \ \star\star\star$     & $0.0432143523328$\\
$3.6$  & $0.0276705670714$   &$\ \ \ \ \ \ \ \ \ \star\star\star$     & $0.0277027234064$\\
$4.0$  & $0.0177821908482$   &$\ \ \ \ \ \ \ \ \ \star\star\star$     & $0.0177933005407$\\
$4.4$  & $0.0114493801119$   &$\ \ \ \ \ \ \ \ \ \star\star\star$     & $0.0114531669215$\\
$4.8$  & $0.0073871082019$   &$\ \ \ \ \ \ \ \ \ \star\star\star$     & $0.0073883880176$\\
$5.2$  & $0.0047760242018$   &$\ \ \ \ \ \ \ \ \ \star\star\star$     & $0.0047764543816$\\
$5.6$  & $0.0030940448336$   &$\ \ \ \ \ \ \ \ \ \star\star\star$     & $0.0030941889171$\\
$6.0$  & $0.0020081910916$   &$\ \ \ \ \ \ \ \ \ \star\star\star$     & $0.0020082392392$\\
$6.4$  & $0.0013057031049$   &$\ \ \ \ \ \ \ \ \ \star\star\star$     & $0.0013057191700$\\
$6.8$  & $0.0008503266370$   &$\ \ \ \ \ \ \ \ \ \star\star\star$     & $0.0008503319922$\\
$7.2$  & $0.0005545896959$   &$\ \ \ \ \ \ \ \ \ \star\star\star$     & $0.0005545914800$\\
$7.6$  & $0.0003622001925$   &$\ \ \ \ \ \ \ \ \ \star\star\star$     & $0.0003622007867$\\
$8.0$  & $0.0002368464526$   &$\ \ \ \ \ \ \ \ \ \star\star\star$     & $0.0002368466504$\\
$8.4$  & $0.0001550534176$  & $\ \ \ \ \ \ \ \ \ \star\star\star$     & $  0.000155053483$\\
$8.8$  & $0.0001016134826$  & $\ \ \ \ \ \ \ \ \ \star\star\star$     & $0.0001016135046$\\
$9.2$  & $0.0000666561518$  & $\ \ \ \ \ \ \ \ \ \star\star\star$     & $0.0000666561591$\\
$9.6$  & $0.0000437638138$  & $\ \ \ \ \ \ \ \ \ \star\star\star$     & $0.0000437638163$\\
$10.0$& $0.0000287572005$  & $\ \ \ \ \ \ \ \ \ \star\star\star$     & $0.0000287572013$\\
\hline
\end{tabular}
\end{center}
\caption{Numerical data for the left panel of Fig.\,\ref{fig5aa}.
The column ${\mathfrak s}_1=\frac{1}{2}\,\log({\mathfrak S}_1)$  was obtained from the numerical
solution of the system  \eqref{DDV},\,\eqref{jashssjsa}
for 
$\delta=\frac{17}{47}=0.36\ldots,\ k_1=\frac{47}{150},\ k_2=\frac{47}{640}$.
The column $({\mathfrak s}_1)_{\rm UV}$ represents the small-$r$  expansion\   \eqref{small}, which reads explicitly as
${\mathfrak s}_1=  -0.6266666667\, \log(r)+0.4555032510 + 0.03493323398\, r^2 + 0.001340210\, r^4+O(r^6)$.
The column $({\mathfrak s}_1)_{\rm IR}$ represents the large-$r$  approximation\   \eqref{large}.}
\label{tab3}
\end{table}

\begin{table}
[!ht]
\begin{center}
\begin{tabular}{| c || l| l | l|}
\hline \rule{0mm}{3.6mm}
$r=MR$&$\ \ \ \ \ \ \ \ \ \ {\mathfrak s}_2$&  $\ \ \ \ \ \  \ ({\mathfrak s}_2)_{\rm UV}$&$\ \ \ \ \ \ \ ({\mathfrak s}_2)_{\rm IR}$\\
\hline
$0.1$   & $0.3997591044995$   &$0.3997591045008$               & $\ \ \ \ \ \ \ \ \star\star\star$\\
$0.2$   & $0.2985360750023$   &$0.2985360730126$               & $\ \ \ \ \ \ \ \ \star\star\star$\\
$0.3$   & $0.2399498302262$   &$0.2399498068682$               & $\ \ \ \ \ \ \ \ \star\star\star$\\
$0.4$   & $0.1990387499302$   &$0.1990386158103$               & $\ \ \ \ \ \ \ \ \star\star\star$\\
$0.5$   & $0.1679718595306$   &$0.1679713348500$               & $\ \ \ \ \ \ \ \ \star\star\star$\\
$0.6$   & $0.1432522266963$   &$0.1432506145302$               & $\ \ \ \ \ \ \ \ \star\star\star$\\        
$0.7$   & $0.1230054237481$   &$0.1230012296587$               & $0.1683612865238$\\
$0.8$   & $0.1061034977430$   &$0.1060938404966$               & $0.1407371805351$\\
$0.9$   & $0.0918099684157$   &$0.0917897214819$               & $0.1182847381657$\\
$1.0$   & $0.0796133720389$   &$0.0795739749350$               & $0.0998514610923$\\
$1.1$   & $0.0691407797493$   &$0.0690686632434$               & $0.0846006892845$\\
$1.2$   & $0.0601093192474$   &$0.0599838992351$               & $0.0719057512702$\\
$1.3$   & $0.0522973702302$   &$0.0520885811040$               & $0.0612858142565$\\
$1.4$   & $0.0455266456450$   &$0.0451920036845$               & $0.0523649516275$\\
$1.5$   & $0.0396506206560$   &$0.0391318292769$               & $0.0448448812229$\\
$1.6$   & $0.0345468193747$   &$0.0337659486565$               & $0.0384861361738$\\
$1.7$   & $0.0301115232549$   &$0.0289668138657$               & $0.0330946521765$\\
$1.8$   & $0.0262560384939$   &$0.0246173933341$               & $0.0285119577425$\\
$1.9$   & $0.0229039873715$   &$0.0206082220787$               & $0.0246078368913$\\
$2.0$   & $0.0199892836248$   &$0.0168352094152$               & $0.0212747360195$\\
$2.4$   & $0.0116633336664$   &$\ \ \ \ \ \ \ \ \ \star\star\star$    & $0.0120763504827$\\ 
$2.8$   & $0.0068826994564$   &$\ \ \ \ \ \ \ \ \ \star\star\star$    & $0.0070144600406$\\
$3.2$   & $0.0041150495660$   &$\ \ \ \ \ \ \ \ \ \star\star\star$    & $0.0041570364532$\\
$3.6$   & $0.0024935093461$   &$\ \ \ \ \ \ \ \ \ \star\star\star$    & $0.0025069200083$\\
$4.0$   & $0.0015302172810$   &$\ \ \ \ \ \ \ \ \ \star\star\star$    & $0.0015345175108$\\
$4.4$   & $0.0009498222274$   &$\ \ \ \ \ \ \ \ \ \star\star\star$    & $0.0009512072298$\\
$4.8$   & $0.0005954275349$   &$\ \ \ \ \ \ \ \ \ \star\star\star$    & $0.0005958755398$\\
$5.2$   & $0.0003764155775$   &$\ \ \ \ \ \ \ \ \ \star\star\star$    & $0.0003765610654$\\
$5.6$   & $0.0002396445866$   &$\ \ \ \ \ \ \ \ \ \star\star\star$    & $0.0002396919980$\\
$6.0$   & $0.0001534665449$   &$\ \ \ \ \ \ \ \ \ \star\star\star$    & $0.0001534820419$\\
$6.4$   & $0.0000987574561$   &$\ \ \ \ \ \ \ \ \ \star\star\star$    & $0.0000987625346$\\
$6.8$   & $0.0000638079185$   &$\ \ \ \ \ \ \ \ \ \star\star\star$    & $0.0000638095865$\\
$7.2$   & $0.0000413648757$   &$\ \ \ \ \ \ \ \ \ \star\star\star$    & $0.0000413654246$\\
$7.6$   & $0.0000268906240$   &$\ \ \ \ \ \ \ \ \ \star\star\star$    & $0.0000268908049$\\
$8.0$   & $0.0000175221586$   &$\ \ \ \ \ \ \ \ \ \star\star\star$    & $0.0000175222183$\\
$8.4$   & $0.0000114402348$   &$\ \ \ \ \ \ \ \ \ \star\star\star$    & $0.0000114402545$\\
$8.8$   & $0.0000074819722$   &$\ \ \ \ \ \ \ \ \ \star\star\star$    & $0.0000074819787$\\
$9.2$   & $0.0000049003651$   &$\ \ \ \ \ \ \ \ \ \star\star\star$    & $0.0000049003672$\\
$9.6$   & $0.0000032135761$   &$\ \ \ \ \ \ \ \ \ \star\star\star$    & $0.0000032135769$\\
$10.0$ & $0.0000021097379$   &$\ \ \ \ \ \ \ \ \ \star\star\star$    & $0.0000021097381$\\
\hline
\end{tabular}
\end{center}
\caption{Numerical data for the right panel of Fig.\,\ref{fig5aa}.
The column ${\mathfrak s}_2=\frac{1}{2}\,\log({\mathfrak S}_2)$ 
 was obtained from the numerical
solution of the system  \eqref{DDV},\,\eqref{jashssjsa}
for 
$\delta=\frac{17}{47}=0.36\ldots,\ k_1=\frac{47}{150},\ k_2=\frac{47}{640}$.
The column $({\mathfrak s}_2)_{\rm UV}$ represents the small-$r$  expansion\   \eqref{small}, which reads explicitly as
${\mathfrak s}_2=-0.1468750000\, \log(r) +0.0613720808 + 0.01949676545\, r^2 - 0.001294871\, r^4+O(r^6)$.
The column $({\mathfrak s}_2)_{\rm IR}$ represents the large-$r$  approximation\   \eqref{large}.}
\label{tab4}
\end{table}

\begin{table}
[!ht]
\begin{center}
\begin{tabular}{| c || l| l| | l|l|}
\hline \rule{0mm}{3.6mm}
$r=MR$&$\ \ \ \ \ \ \ \ \  {\mathfrak I}_3$&  $\ \ \  \ \ ({\mathfrak I}_3)_{\rm IR}$&$\ \ \ \ \ \ \ \ {\mathfrak I}_5$&  
$\ \ \ \ \  ( {\mathfrak I}_5)_{\rm IR}  $  \\
\hline
$0.0$& $-0.0816581987$   &$-0.0808549145$               & $0.0096460439$&$0.0094192397$\\
\hline\hline
$0.1$& $-0.0815973708$   &$-0.0808281809$               & $0.0096424920$&  $0.0094176779$\\
$0.2$& $-0.0814148360$   &$-0.0807342479$               & $0.0096315680$&  $0.0094123784$\\
$0.3$& $-0.0811104634$   &$-0.0805545500$               & $0.0096134830$&  $0.0094025372$\\
$0.4$& $-0.0806841094$   &$-0.0802729358$               & $0.0095881340$&  $0.0093874663$\\
$0.5$& $-0.0801357223$   &$-0.0798759163$               & $0.0095554317$&  $0.0093665661$\\
$0.6$& $-0.0794654791$   &$-0.0793529541$               & $0.0095153239$&  $0.0093393111$\\
$0.7$& $-0.0786739415$   &$-0.0786966296$               & $0.0094678595$&  $0.0093052411$\\
$0.8$& $-0.0777622192$   &$-0.0779026603$               & $0.0094129858$&  $0.0092639570$\\
$0.9$& $-0.0767321288$   &$-0.0769697851$               & $0.0093506182$&  $0.0092151198$\\
$1.0$& $-0.0755863309$   &$-0.0758995406$               & $0.0092806976$&  $0.0091584511$\\
$1.1$& $-0.0743284379$   &$-0.0746959640$               & $0.0092032710$&  $0.0090937349$\\
$1.2$& $-0.0729630794$   &$-0.0733652478$               & $0.0091183243$&  $0.0090208194$\\
$1.3$& $-0.0714959216$   &$-0.0719153739$               & $0.0090258954$&  $0.0089396185$\\
$1.4$& $-0.0699336369$   &$-0.0703557460$               & $0.0089259636$&  $0.0088501127$\\
$1.5$& $-0.0682838279$   &$-0.0686968360$               & $0.0088186043$&  $0.0087523493$\\
$1.6$& $-0.0665549097$   &$-0.0669498539$               & $0.0087040703$&  $0.0086464410$\\
$1.7$& $-0.0647559610$   &$-0.0651264497$               & $0.0085824477$&  $0.0085325638$\\
$1.8$& $-0.0628965528$   &$-0.0632384496$               & $0.0084539580$&  $0.0084109544$\\
$1.9$& $-0.0609865667$   &$-0.0612976294$               & $0.0083187596$&  $0.0082819052$\\
$2.0$& $-0.0590360125$   &$-0.0593155233$               & $0.0081772410$&  $0.0081457607$\\
$2.4$& $-0.0510235231$   &$-0.0511886302$               & $0.0075547640$&  $0.0075386051$\\
$2.8$& $-0.0430938981$   &$-0.0431807976$               & $0.0068628911$&  $0.0068549471$\\
$3.2$& $-0.0356791140$   &$-0.0357213978$               & $0.0061327265$&  $0.0061289408$\\
$3.6$& $-0.0290413046$   &$-0.0290607646$               & $0.0053953603$&  $0.0053936359$\\
$4.0$& $-0.0232965652$   &$-0.0233051621$               & $0.0046781433$&  $0.0046773527$\\
$4.4$& $-0.0184555864$   &$-0.0184592682$               & $0.0040022089$&  $0.0040018495$\\
$4.8$& $-0.0144627380$   &$-0.0144642771$               & $0.0033820666$&  $0.0033819434$\\
$5.2$& $-0.0112267836$   &$-0.0112274146$               & $0.0028261291$&  $0.0028260937$\\
$5.6$& $-0.0086423718$   &$-0.0086426264$               & $0.0023375652$&  $0.0023374955$\\
$6.0$& $-0.0066037504$   &$-0.0066038518$               & $0.0019153422$&  $0.0019153413$\\
$6.4$& $-0.0050126760$   &$-0.0050127159$               & $0.0015559997$&  $0.0015560333$\\
$6.8$& $-0.0037823190$   &$-0.0037823346$               & $0.0012542094$&  $0.0012542300$\\
$7.2$& $-0.0028385851$   &$-0.0028385911$               & $0.0010036775$&  $0.0010036790$\\
$7.6$& $-0.0021198835$   &$-0.0021198858$               & $0.0007978552$&  $0.0007978387$\\
$8.0$& $-0.0015760545$   &$-0.0015760554$               & $0.0006302476$&  $0.0006303100$\\
$8.4$& $-0.0011669145$   &$-0.0011669148$               & $0.0004950965$&  $0.0004951137$\\
$8.8$& $-0.0008607071$   &$-0.0008607072$               & $0.0003868423$&  $0.0003868473$\\
$9.2$& $-0.00063262156$&$-0.00063262160$              & $0.0003007252$&  $0.0003007547$\\
$9.6$& $-0.00046346254$&$-0.00046346255$              & $0.0002327455$&  $0.0002327354$\\ 
$10.0$& $-0.00033850528$ &$-0.00033850528$           & $0.0001793012$&  $0.0001793144$\\
\hline
\end{tabular}
\end{center}
\caption{
The columns ${\mathfrak I}_{2n-1}=\big(\frac{R}{2\pi}\big)^{2n-1}\ I_{2 n-1}$ $(n=2,\,3$)
 were obtained from the numerical  solution of the system\ \eqref{DDV},\,\eqref{In-def}
 for 
$\delta=\frac{17}{47}=0.36\ldots,\ k_1=\frac{47}{150},\ k_2=\frac{47}{640}$.
The values ${\mathfrak I}_3$ and ${\mathfrak I}_5$ at $r=0$ are exact and given by formulae\,\eqref{kaaaajssa},\,\eqref{asosisasaisaoi}
and \eqref{ahsahasj}.
The columns $({\mathfrak I}_{2n-1})_{\rm IR}$ represent the large-$r$  approximation  \eqref{i2nm1},\,\eqref{manssaj}.}
\label{tab2}
\end{table}

\begin{table}
[!ht]
\begin{center}
\begin{tabular}{| c || l| l || l|l |}
\hline \rule{0mm}{3.6mm}
$r=MR$&$\ \ \ \ \ \ \ {\mathfrak h}^{(1)}_1$&  $\ \ \ ({\mathfrak h}^{(1)}_1)_{\rm IR}$&
$ \ \ \ \ \ \ \ \  {\bar {\mathfrak  h}}^{(1)}_1$ & $\ \ \ ({\bar {\mathfrak h}}^{(1)}_1)_{\rm IR}$ \\
\hline
$0.0$&  $0.0439788467$   &$0.0439081576$               & $0.0226596394$ &$0.0226232174$\\
\hline
\hline
$0.1$&  $0.0439462396$      &$0.0438921972$               & $0.0226428389$ &$0.0226149941$\\
$0.2$&  $0.0438483936$      &$0.0438367700$               & $0.0225924248$ &$0.0225864358$\\
$0.3$&  $0.0436852472$      &$0.0437318507$               & $0.0225083653$ &$0.0225323773$\\
$0.4$&  $0.0434567372$      &$0.0435689774$               & $0.0223906279$ &$0.0224484585$\\
$0.5$&  $0.0431628547$      &$0.0433413243$               & $0.0222392081$ &$0.0223311626$\\
$0.6$&  $0.0428037181$      &$0.0430437850$               & $0.0220541667$ &$0.0221778586$\\
$0.7$&  $0.0423796552$      &$0.0426729965$               & $0.0218356727$ &$0.0219868137$\\
$0.8$&  $0.0418912910$      &$0.0422272954$               & $0.0215840482$ &$0.0217571709$\\
$0.9$&  $0.0413396303$      &$0.0417066139$               & $0.0212998108$ &$0.0214888953$\\
$1.0$&  $0.0407261309$      &$0.0411123310$               & $0.0209837117$ &$0.0211826972$\\
$1.1$&  $0.0400527596$      &$0.0404470935$               & $0.0206367642$ &$0.0208399406$\\
$1.2$&  $0.0393220252$      &$0.0397146210$               & $0.0202602609$ &$0.0204625418$\\
$1.3$&  $0.0385369878$      &$0.0389195065$               & $0.0198557786$ &$0.0200528674$\\
$1.4$&  $0.0377012408$      &$0.0380670209$               & $0.0194251688$ &$0.0196136331$\\
$1.5$&  $0.0368188685$      &$0.0371629288$               & $0.0189705357$ &$0.0191478092$\\
$1.6$&  $0.0358943821$      &$0.0362133197$               & $0.0184942038$ &$0.0186585330$\\
$1.7$&  $0.0349326394$      &$0.0352244569$               & $0.0179986759$ &$0.0181490319$\\
$1.8$&  $0.0339387519$      &$0.0342026462$               & $0.0174865858$ &$0.0176225546$\\
$1.9$&  $0.0329179883$      &$0.0331541240$               & $0.0169606481$ &$0.0170823145$\\
$2.0$&  $0.0318756771$      &$0.0320849645$               & $0.0164236082$ &$0.0165314413$\\
$2.4$&  $0.0275947047$      &$0.0277140087$               & $0.0142178821$ &$0.0142793522$\\
$2.8$&  $0.0233568900$      &$0.0234186097$               & $0.0120343925$ &$0.0120661929$\\
$3.2$&  $0.0193901700$      &$0.0194199761$               & $0.0099905817$ &$0.0100059389$\\
$3.6$&  $0.0158326740$      &$0.0158463640$               & $0.0081576191$ &$0.0081646728$\\
$4.0$&  $0.0127459447$      &$0.0127520006$                &$0.0065672143$ &$0.0065703345$\\
$4.4$&  $0.0101364865$      &$0.0101390888$                &$0.0052227183$ &$0.0052240591$\\
$4.8$&  $0.0079761617$      &$0.0079772546$                &$0.0041096336$ &$0.0041101967$\\
$5.2$&  $0.0062180889$      &$0.0062185393$                &$0.0032038050$ &$0.0032040371$\\
$5.6$&  $0.0048077400$      &$0.0048079228$                &$0.0024771376$ &$0.0024772318$\\
$6.0$&  $0.0036900458$      &$0.0036901190$                &$0.0019012574$ &$0.0019012951$\\
$6.4$&  $0.0028135385$      &$0.0028135676$                &$0.0014496462$ &$0.0014496611$\\
$6.8$&  $0.0021324492$      &$0.0021324606$                &$0.0010987221$ &$0.0010987279$\\
$7.2$&  $0.0016074749$      &$0.0016074793$                &$0.0008282345$ &$0.0008282368$\\
$7.6$&  $0.0012057366$      &$0.0012057383$                &$0.0006212431$ &$0.0006212440$\\
$8.0$&  $0.0009002856$      &$0.0009002863$                &$0.0004638627$ &$0.0004638630$\\
$8.4$&  $0.000669396112$  &$0.000669396359$            &$0.000344899317$&$0.000344899451$\\
$8.8$&  $0.000495790485$  &$0.000495790579$            &$0.000255450850$&$0.000255450894$\\
$9.2$&  $0.000365887290$  &$0.000365887323$            &$0.000188519584$&$0.000188519604$\\
$9.6$&  $0.000269116027$  &$0.000269116042$            &$0.000138659212$&$0.000138659217$\\
$10.0$&$0.000197320756$  &$0.000197320760$            &$0.000101667450$&$0.000101667452$\\
\hline
\end{tabular}
\end{center}
\caption{
The columns
 ${\mathfrak h}_1^{(1)}= 
\big({\textstyle \frac{R}{2\pi}}\big)^{\frac{2}{a_1}} \, H^{(1)}_{1}$ 
and ${\bar {\mathfrak h}}_1^{(1)}= 
\big({\textstyle \frac{R}{2\pi}}\big)^{\frac{2}{a_1}} \, {\bar H}^{(1)}_{1}$
were  obtained from the numerical solution of \eqref{DDV},\,\eqref{jassa}.
The values ${\mathfrak h}_1^{(1)}$ and $ {\bar {\mathfrak h}}_1^{(1)}$ at
$r=0$ are exact and given by formulae\,\eqref{kajsska},\,\eqref{lsakslsakl}.
The columns $({\mathfrak h}^{(1)}_{1})_{\rm IR}$ and 
$({\bar {\mathfrak h}}^{(1)}_{1})_{\rm IR}$ represent the large-$r$  approximation \eqref{manssaj},\,\eqref{jsaisaiu}.}
\label{tab5}
\end{table}

\begin{table}
[!ht]
\begin{center}
\begin{tabular}{| c || l| l || l|l |}
\hline \rule{0mm}{3.6mm}
$r=MR$&$\ \ \ \ \ \ \ {\mathfrak h}^{(2)}_1$&  $\ \ \ \ \ \ \ ({\mathfrak h}^{(2)}_1)_{\rm IR}$&
$\ \ \ \  \ \ \ \   {\bar {\mathfrak h}}^{(2)}_1$ & $\ \ \ ({\bar {\mathfrak h}}^{(2)}_1)_{\rm IR}$ \\
\hline
$0.0$      & $-0.0669753762$ & $\ \ \ \ \ \ \ \ \ \star\star\star$                      & $-0.0434635157$   &$\ \ \ \ \ \ \ \ \ \star\star\star$\\
\hline
\hline
$0.1$      & $-0.0668127237$ & $\ \ \ \ \ \ \ \ \ \star\star\star$                       & $-0.0433579628$   &$\ \ \ \ \ \ \ \ \ \star\star\star$\\      
$0.2$      & $-0.0663278174$ & $\ \ \ \ \ \ \ \ \ \star\star\star$                       & $-0.0430432839$   &$\ \ \ \ \ \ \ \ \ \star\star\star$\\
$0.3$      & $-0.0655297064$ & $\ \ \ \ \ \ \ \ \ \star\star\star$                       &$-0.0425253517$    &$\ \ \ \ \ \ \ \ \ \star\star\star$\\
$0.4$      & $-0.0644331292$ & $\ \ \ \ \ \ \ \ \ \star\star\star$                       & $-0.0418137304$   &$\ \ \ \ \ \ \ \ \ \star\star\star$\\
$0.5$      & $-0.0630580132$ & $\ \ \ \ \ \ \ \ \ \star\star\star$                       & $-0.0409213520$   &$\ \ \ \ \ \ \ \ \ \star\star\star$\\
$0.6$      & $-0.0614288027$ & $-0.0811789003$                                       & $-0.0398640796$   &$-0.0526808599$\\
$0.7$      & $-0.0595736457$ & $-0.0769924035$                                       & $-0.0386601798$   &$-0.0499640425$\\
$0.8$      & $-0.0575234743$ & $-0.0727117211$                                       & $-0.0373297258$   &$-0.0471861035$\\
$0.9$      & $-0.0553110232$ & $-0.0684177771$                                       & $-0.0358939608$   &$-0.0443995585$\\
$1.0$      & $-0.0529698322$ & $-0.0641736558$                                       & $-0.0343746503$   &$-0.0416453457$\\
$1.1$      & $-0.0505332805$ & $-0.0600274826$                                       & $-0.0327934557$   &$-0.0389546962$\\
$1.2$      & $-0.0480336985$ & $-0.0560149907$                                       & $-0.0311713577$   &$-0.0363507989$\\
$1.3$      & $-0.0455015942$ & $-0.0521617577$                                       & $-0.0295281545$   &$-0.0338502522$\\
$1.4$      & $-0.0429650249$ & $-0.0484851239$                                       & $-0.0278820537$   &$-0.0314643092$\\
$1.5$      & $-0.0404491302$ & $-0.0449958208$                                       & $-0.0262493696$   &$-0.0291999340$\\
$1.6$      & $-0.0379758318$ & $-0.0416993376$                                       & $-0.0246443284$   &$-0.0270606889$\\
$1.7$      & $-0.0355636930$ & $-0.0385970603$                                       & $-0.0230789765$   &$-0.0250474732$\\
$1.8$      & $-0.0332279215$ & $-0.0356872127$                                       & $-0.0215631829$   &$-0.0231591343$\\
$1.9$      & $-0.0309804926$ & $-0.0329656289$                                       & $-0.0201047191$   &$-0.0213929687$\\
$2.0$      & $-0.0288303652$ & $-0.0304263819$                                       & $-0.0187093989$   &$-0.0197451303$\\
$2.4$      & $-0.0212931589$ & $-0.0219383725$                                       & $-0.0138181463$   &$-0.0142368561$\\
$2.8$      & $-0.0154551346$ & $-0.0157059311$                                       & $-0.0100295739$   &$-0.0101923277$\\
$3.2$      & $-0.0111002701$ & $-0.0111952532$                                       & $-0.0072034947$   &$-0.0072651337$\\
$3.6$      & $-0.0079224731$ & $-0.0079578263$                                       & $-0.0051412706$   &$-0.0051642130$\\
$4.0$      & $-0.0056325377$ & $-0.0056455405$                                       & $-0.0036552223$   &$-0.0036636605$\\
$4.4$      & $-0.0039941428$ & $-0.0039988848$                                       & $-0.0025919897$   &$-0.0025950670$\\
$4.8$      & $-0.0028268092$ & $-0.0028285276$                                       & $-0.0018344513$   &$-0.0018355665$\\
$5.2$      & $-0.0019973159$ & $-0.0019979355$                                       & $-0.0012961535$   &$-0.0012965556$\\                
$5.6$      & $-0.0014090326$ & $-0.0014092551$                                       & $-0.0009143884$   &$-0.0009145328$\\
$6.0$      & $-0.0009925018$ & $-0.0009925815$                                       & $-0.0006440817$   &$-0.0006441334$\\
$6.4$      & $-0.0006980355$ & $-0.0006980639$                                       & $-0.0004529885$   &$-0.0004530070$\\
$6.8$      & $-0.0004901835$ & $-0.0004901936$                                       & $-0.0003181034$   &$-0.0003181100$\\
$7.2$      & $-0.0003436985$ & $-0.0003437021$                                       & $-0.0002230423$   &$-0.0002230447$\\
$7.6$      & $-0.0002406262$ & $-0.0002406275$                                       & $-0.0001561538$   &$-0.0001561546$\\
$8.0$      & $-0.0001682163$ & $-0.0001682167$                                       & $-0.0001091636$   &$-0.0001091638$\\
$8.4$      & $-0.000117427887$  & $-0.000117428046$   &$-0.000076204556$&$-0.000076204659$\\
$8.8$      & $-0.000081860357$  & $-0.000081860413$   &$-0.000053123089$&$-0.000053123126$\\
$9.2$      & $-0.000056990041$  & $-0.000056990060$   &$-0.000036983555$&$-0.000036983568$\\
$9.6$      & $-0.000039625270$  & $-0.000039625276$ & $-0.000025714727$& $-0.000025714732$\\
$10.0$    & $-0.000027518134$  & $-0.000027518136$   &$-0.000017857829$&$-0.000017857831$\\
\hline
\end{tabular}
\end{center}
\caption{
The columns
 ${\mathfrak h}_1^{(2)}= 
\big({\textstyle \frac{R}{2\pi}}\big)^{\frac{2}{a_2}} \, H^{(2)}_{1}$ 
and ${\bar {\mathfrak h}}_1^{(2)}= 
\big({\textstyle \frac{R}{2\pi}}\big)^{\frac{2}{a_2}} \, {\bar H}^{(2)}_{1}$
were  obtained from the numerical solution of \eqref{DDV},\,\eqref{jassa}.
The values ${\mathfrak h}_1^{(2)}$ and $ {\bar {\mathfrak  h}}_1^{(2)}$ 
at $r=0$ are exact and given by formulae\,\eqref{kajsska},\,\eqref{lsakslsakl}.
The columns $({\mathfrak h}^{(2)}_{1})_{\rm IR}$ and $({\bar{\mathfrak  h}}^{(2)}_{1})_{\rm IR}$ 
represent the large-$r$  approximation\ \eqref{manssaj},\,\eqref{jsaisaiu}.}
\label{tab6}
\end{table}


\begin{thebibliography}{10}


\bibitem{Bukhvostov:1980sn}
A.~P. Bukhvostov and L.~N. Lipatov, ``Instanton\,--\,anti-instanton interaction in
  the {$O(3)$} nonlinear $\sigma$-model and an exactly soluble fermion
  theory'',
\href{http://dx.doi.org/10.1016/0550-3213(81)90157-7}{{\em Nucl. Phys.}
  {\bfseries B180} (1981) 116}.
\bibitem{Bazhanov:2016glt}
V.~V. Bazhanov, S.~L. Lukyanov and B.~A. Runov, ``{Vacuum energy of the
  {B}ukhvostov-{L}ipatov model}'',
  \href{http://dx.doi.org/10.1016/j.nuclphysb.2016.08.031}{{\em Nucl. Phys.}
  {\bfseries B911} (2016) 863--889},
\href{http://arxiv.org/abs/1607.04839}{{\ttfamily arXiv:1607.04839 [hep-th]}}.


\bibitem{Polyakov:1975yp}
A.~A. Belavin and A.~M. Polyakov, ``Metastable states of two-dimensional
  isotropic ferromagnets'', {\em JETP Lett.} {\bfseries 22} (1975) 245--248.
[Pisma Zh. Eksp. Teor. Fiz. 22, 503 (1975)].

\bibitem{Fateev:1979dc}
V.~A. Fateev, I.~V. Frolov and A.~S. Schwarz, ``Quantum fluctuations of
  instantons in the nonlinear sigma model'',
\href{http://dx.doi.org/10.1016/0550-3213(79)90367-5}{{\em Nucl. Phys.}
  {\bfseries B154} (1979) 1--20}.

\bibitem{Vor94}
A.~Voros, ``Exact quantization condition for anharmonic oscillators (in one
  dimension)'', {\em J. Phys. A} {\bfseries 27} no.~13, (1994) 4653--4661.

\bibitem{DT99b}
P.~Dorey and R.~Tateo, ``Anharmonic oscillators, the thermodynamic {B}ethe
  ansatz and nonlinear integral equations'', {\em J. Phys. A} {\bfseries 32}
  no.~38, (1999) L419--L425. [{\tt arXiv:hep-th/9812211}].

\bibitem{Bazhanov:1998wj}
V.~V. Bazhanov, S.~L. Lukyanov and A.~B. Zamolodchikov, ``{Spectral
  determinants for Schrodinger equation and $Q$ operators of conformal field
  theory}'', \href{http://dx.doi.org/10.1023/A:1004838616921}{{\em J. Statist.
  Phys.} {\bfseries 102} (2001) 567--576},
\href{http://arxiv.org/abs/hep-th/9812247}{{\ttfamily arXiv:hep-th/9812247.
 }}

\bibitem{Suzuki:2000fc}
J.~Suzuki, ``{Functional relations in Stokes multipliers: Fun with $x^6 + \alpha\,
  x^2$ potential}'', \href{http://dx.doi.org/10.1023/A:1004823608260}{{\em J.
  Statist. Phys.} {\bfseries 102} (2001) 1029--1047},
\href{http://arxiv.org/abs/quant-ph/0003066}{{\ttfamily arXiv:quant-ph/0003066.}}

\bibitem{Bazhanov:2001xm}
V.~V. Bazhanov, A.~N. Hibberd and S.~M. Khoroshkin, ``{Integrable structure of
  $W(3)$ conformal field theory, quantum Boussinesq theory and boundary affine
  Toda theory}'', \href{http://dx.doi.org/10.1016/S0550-3213(01)00595-8}{{\em
  Nucl. Phys.} {\bfseries B622} (2002) 475--547},
\href{http://arxiv.org/abs/hep-th/0105177}{{\ttfamily arXiv:hep-th/0105177.}}

\bibitem{Bazhanov:2003ni}
V.~V. Bazhanov, S.~L. Lukyanov and A.~B. Zamolodchikov, ``{Higher level
  eigenvalues of $Q$ operators and Schroedinger equation}'',
  \href{http://dx.doi.org/10.4310/ATMP.2003.v7.n4.a4}{{\em Adv. Theor. Math.
  Phys.} {\bfseries 7} no.~4, (2003) 711--725},
\href{http://arxiv.org/abs/hep-th/0307108}{{\ttfamily arXiv:hep-th/0307108.}}

\bibitem{Fioravanti:2004cz}
D.~Fioravanti, ``{Geometrical loci and CFTs via the Virasoro symmetry of the
  mKdV-SG hierarchy: An Excursus}'',
  \href{http://dx.doi.org/10.1016/j.physletb.2005.01.037}{{\em Phys. Lett.}
  {\bfseries B609} (2005) 173--179},
\href{http://arxiv.org/abs/hep-th/0408079}{{\ttfamily arXiv:hep-th/0408079.}}

\bibitem{Dorey:2006an}
P.~Dorey, C.~Dunning, D.~Masoero, J.~Suzuki and R.~Tateo,
  ``{Pseudo-differential equations, and the Bethe ansatz for the classical Lie
  algebras}'', \href{http://dx.doi.org/10.1016/j.nuclphysb.2007.02.029}{{\em
  Nucl. Phys.} {\bfseries B772} (2007) 249--289},
\href{http://arxiv.org/abs/hep-th/0612298}{{\ttfamily arXiv:hep-th/0612298.}}

\bibitem{Feigin:2007mr}
B.~Feigin and E.~Frenkel, ``{Quantization of soliton systems and Langlands
  duality}'',
\href{http://arxiv.org/abs/0705.2486}{{\ttfamily arXiv:0705.2486 [math.QA]}}.

\bibitem{Lukyanov:2010rn}
S.~L. Lukyanov and A.~B. Zamolodchikov, ``{Quantum Sine(h)-Gordon Model and
  Classical Integrable Equations}'',
  \href{http://dx.doi.org/10.1007/JHEP07(2010)008}{{\em JHEP} {\bfseries 07}
  (2010) 008},
\href{http://arxiv.org/abs/1003.5333}{{\ttfamily arXiv:1003.5333 [math-ph]}}.

\bibitem{Dorey:2012bx}
P.~Dorey, S.~Faldella, S.~Negro and R.~Tateo, ``{The Bethe Ansatz and the
  Tzitzeica-Bullough-Dodd equation}'',
  \href{http://dx.doi.org/10.1098/rsta.2012.0052}{{\em Phil. Trans. Roy. Soc.
  Lond.} {\bfseries A371} (2013) 20120052},
\href{http://arxiv.org/abs/1209.5517}{{\ttfamily arXiv:1209.5517 [math-ph]}}.

\bibitem{Lukyanov:2013wra}
S.~L. Lukyanov, ``{ODE/IM} correspondence for the {F}ateev model'',
  \href{http://dx.doi.org/10.1007/JHEP12(2013)012}{{\em JHEP} {\bfseries 12}
  (2013) 012},
\href{http://arxiv.org/abs/1303.2566}{{\ttfamily arXiv:1303.2566 [hep-th]}}.

\bibitem{Bazhanov:2013cua}
V.~V. Bazhanov and S.~L. Lukyanov, ``{Integrable structure of Quantum Field
  Theory: Classical flat connections versus quantum stationary states}'',
  \href{http://dx.doi.org/10.1007/JHEP09(2014)147}{{\em JHEP} {\bfseries 09}
  (2014) 147},
\href{http://arxiv.org/abs/1310.4390}{{\ttfamily arXiv:1310.4390 [hep-th]}}.

\bibitem{Masoero:2015lga}
D.~Masoero, A.~Raimondo and D.~Valeri, ``{Bethe Ansatz and the Spectral Theory
  of Affine Lie Algebra-Valued Connections I. The simply-laced Case}'',
  \href{http://dx.doi.org/10.1007/s00220-016-2643-6}{{\em Commun. Math. Phys.}
  {\bfseries 344} no.~3, (2016) 719--750},
\href{http://arxiv.org/abs/1501.07421}{{\ttfamily arXiv:1501.07421 [math-ph]}}.

\bibitem{Ito:2015nla}
K.~Ito and C.~Locke, ``{ODE/IM correspondence and Bethe ansatz for affine Toda
  field equations}'',
  \href{http://dx.doi.org/10.1016/j.nuclphysb.2015.05.016}{{\em Nucl. Phys.}
  {\bfseries B896} (2015) 763--778},
\href{http://arxiv.org/abs/1502.00906}{{\ttfamily arXiv:1502.00906 [hep-th]}}.

\bibitem{Faddeev:1979}
L.~D. Faddeev, E.~K. Sklyanin and L.~A. Takhtajan, ``The quantum inverse
  problem method. 1'', {\em Theor. Math. Phys.} {\bfseries 40} (1980) 688.

\bibitem{FT87}
L.~D. Faddeev and L.~A. Takhtajan, {\em Hamiltonian {M}ethods in the {T}heory
  of {S}olitons}.
\newblock Springer-Verlag, Berlin, 1987.

\bibitem{Bazhanov:2014joa}
V.~V. Bazhanov, G.~A. Kotousov and S.~L. Lukyanov, ``Winding vacuum energies in
  a deformed {$O(4)$} sigma model'',
  \href{http://dx.doi.org/10.1016/j.nuclphysb.2014.11.005}{{\em Nucl. Phys.}
  {\bfseries B889} (2014) 817--826},
\href{http://arxiv.org/abs/1409.0449}{{\ttfamily arXiv:1409.0449 [hep-th]}}.

\bibitem{Fateev:1996ea}
V.~A. Fateev, ``The sigma model (dual) representation for a two-parameter
  family of integrable quantum field theories'',
\href{http://dx.doi.org/10.1016/0550-3213(96)00256-8}{{\em Nucl. Phys.}
  {\bfseries B473} (1996) 509--538}.

\bibitem{Fateev:1992tk}
V.~A. Fateev, E.~Onofri and A.~B. Zamolodchikov, ``The sausage model
  (integrable deformations of {$O(3)$} sigma model)'',
\href{http://dx.doi.org/10.1016/0550-3213(93)90001-6}{{\em Nucl. Phys.}
  {\bfseries B406} (1993) 521--565}.

\bibitem{Saleur:1998wa}
H.~Saleur, ``The long delayed solution of the {B}ukhvostov-{L}ipatov model'',
  \href{http://dx.doi.org/10.1088/0305-4470/32/18/102}{{\em J. Phys.}
  {\bfseries A32} (1999) L207},
\href{http://arxiv.org/abs/hep-th/9811023}{{\ttfamily arXiv:hep-th/9811023.}}

\bibitem{Baxter:book:1982}
R.~J. Baxter, {\em Exactly solved models in statistical mechanics}.
\newblock Academic, London, 1982.

\bibitem{Lieb66}
E.~H. Lieb, ``Exact solution of the problem of the entropy of two-dimensional
  ice'', \href{http://dx.doi.org/10.1103/PhysRevLett.18.692}{{\em Phys. Rev.
  Lett.} {\bfseries 18} no.~17, (Apr, 1967) 692--694}.

\bibitem{Baxter:1971sam}
R.~J. Baxter, ``Generalized ferroelectric model on a square lattice'',
{\em Stud. Appl. Math.} {\bfseries 1} (1971) 51--69.

\bibitem{Baxterbook}
R.~J. Baxter, {\em Exactly {S}olved {M}odels in {S}tatistical {M}echanics}.
\newblock Academic, London, 1982.

\bibitem{TF79}
L.~A. Takhtajan and L.~D. Faddeev, ``The quantum method for the inverse problem
  and the {$XYZ$} {H}eisenberg model'', {\em Uspekhi Mat. Nauk} {\bfseries
  34(5)} no.~5, (1979) 13--63. (English translation: Russian Math. Surveys {\bf
  34(5)} (1979) 11--68).

\bibitem{Destri:1994bv}
C.~Destri and H.~J. De~Vega, ``Unified approach to thermodynamic {B}ethe
  {A}nsatz and finite size corrections for lattice models and field theories'',
  \href{http://dx.doi.org/10.1016/0550-3213(94)00547-R}{{\em Nucl. Phys.}
  {\bfseries B438} (1995) 413--454},
\href{http://arxiv.org/abs/hep-th/9407117}{{\ttfamily arXiv:hep-th/9407117}}.

\bibitem{Essler:1992uc}
F.~H.~L. Essler, V.~E. Korepin and K.~Schoutens, ``Exact solution of an
  electronic model of superconductivity in (1+1)-dimensions. 1.'',
  \href{http://dx.doi.org/10.1142/S0217979294001354}{{\em Int. J. Mod. Phys.}
  {\bfseries B8} (1994) 3205},
\href{http://arxiv.org/abs/cond-mat/9211001}{{\ttfamily arXiv:cond-mat/921100.}}

\bibitem{Kulish:1985bj}
P.~P. Kulish, ``Integrable graded magnets'',
  \href{http://dx.doi.org/10.1007/BF01083770}{{\em J. Sov. Math.} {\bfseries
  35} (1986) 2648--2662}.
[Zap. Nauchn. Semin.145,140(1985)].

\bibitem{Deguchi:1990}
T.~Deguchi, A.~Fujii and K.~Ito, ``Quantum superalgebra {$U_q(osp(2,2))$}'',
  {\em Phys. Lett. B} {\bfseries 238} no.~2, (1990) 242 -- 246.

\bibitem{Gould:1997}
M.~D. Gould, J.~R. Links, Y.-Z. Zhang and I.~Tsohantjis, ``Twisted quantum
  affine superalgebra {$U_q(sl(2|2)^{(2)})$}, {$U_q(osp(2|2))$} invariant
  {$R$}-matrices and a new integrable electronic model'', {\em J. Phys. A}
  {\bfseries 30} (1997) 4313.

\bibitem{Martins:1997ex}
M.~J. Martins and P.~B. Ramos, ``{On the solution of a supersymmetric model of
  correlated electrons}'',
  \href{http://dx.doi.org/10.1103/PhysRevB.56.6376}{{\em Phys. Rev.} {\bfseries
  B56} (1997) 6376},
\href{http://arxiv.org/abs/hep-th/9704152}{{\ttfamily arXiv:hep-th/9704152}}.

\bibitem{deVega:1989pj}
H.~J. de~Vega and F.~Woynarovich, ``{Solution of the {B}ethe {A}nsatz
  {E}quations with {C}omplex {R}oots for {F}inite {S}ize: {T}he {S}pin $S\ge1$
  Isotropic and Anisotropic Chains}'',
\href{http://dx.doi.org/10.1088/0305-4470/23/9/022}{{\em J. Phys.} {\bfseries
  A23} (1990) 1613}.

\bibitem{Lukyanov:2011wd}
S.~L. Lukyanov, ``Critical values of the {Y}ang-{Y}ang functional in the
  quantum sine-{G}ordon model'',
  \href{http://dx.doi.org/10.1016/j.nuclphysb.2011.07.028}{{\em Nucl. Phys.}
  {\bfseries B853} (2011) 475--507},
\href{http://arxiv.org/abs/1105.2836}{{\ttfamily arXiv:1105.2836 [hep-th]}}.

\bibitem{Lukyanov:2000jp}
S.~L. Lukyanov, ``{Finite temperature expectation values of local fields in the
  sinh-Gordon model}'',
  \href{http://dx.doi.org/10.1016/S0550-3213(01)00365-0}{{\em Nucl. Phys.}
  {\bfseries B612} (2001) 391--412},
\href{http://arxiv.org/abs/hep-th/0005027}{{\ttfamily arXiv:hep-th/0005027.}}

\bibitem{Klumper:1991}
A.~{Kl\"umper}, M.~T. {Batchelor} and P.~A. {Pearce}, ``Central charges of the
  6- and 19-vertex models with twisted boundary conditions'',
  \href{http://dx.doi.org/10.1088/0305-4470/24/13/025}{{\em J. Phys. A.}
  {\bfseries 24} (1991) 3111--3133}.

\bibitem{Destri:1992qk}
C.~Destri and H.~J. de~Vega, ``New thermodynamic {B}ethe ansatz equations
  without strings'',
\href{http://dx.doi.org/10.1103/PhysRevLett.69.2313}{{\em Phys. Rev. Lett.}
  {\bfseries 69} (1992) 2313--2317}.

\bibitem{Bazhanov:2017nzh} 
  V.~V.~Bazhanov, G.~A.~Kotousov and S.~L.~Lukyanov,
  ``{Quantum transfer-matrices for the sausage model}'',
\href{http://arxiv.org/abs/1706.09941}{{\ttfamily arXiv:1706.09941 [hep-th]}}.



\bibitem{Gorsky:2013xba}
A.~Gorsky, A.~Zabrodin and A.~Zotov, ``{Spectrum of Quantum Transfer Matrices
  via Classical Many-Body Systems}'',
  \href{http://dx.doi.org/10.1007/JHEP01(2014)070}{{\em JHEP} {\bfseries 01}
  (2014) 070},
\href{http://arxiv.org/abs/1310.6958}{{\ttfamily arXiv:1310.6958 [hep-th]}}.

\bibitem{Lukyanov:2003nj}
S.~L. Lukyanov, E.~S. Vitchev and A.~B. Zamolodchikov, ``{Integrable model of
  boundary interaction: The Paperclip}'',
  \href{http://dx.doi.org/10.1016/j.nuclphysb.2004.02.010}{{\em Nucl. Phys.}
  {\bfseries B683} (2004) 423--454},
\href{http://arxiv.org/abs/hep-th/0312168}{{\ttfamily arXiv:hep-th/0312168.}}

\end{thebibliography}


\newcommand\oneletter[1]{#1}\def\cprime{$'$}
\providecommand{\href}[2]{#2}\begingroup\raggedright\endgroup

\end{document}